\documentclass[utf8]{FrontiersinHarvard} 

\usepackage{url,hyperref,subcaption}
\usepackage[final, nopatch=footnote]{microtype}
\usepackage[onehalfspacing]{setspace}
\RequirePackage{fix-cm}

\usepackage{soul}
\usepackage{amsmath}
\usepackage{comment}
\usepackage{subcaption}
\usepackage{multirow}
\usepackage{array, makecell}
\newcommand{\veryshortarrow}[1][3pt]{\mathrel{%
   \hbox{\rule[\dimexpr\fontdimen22\textfont2-.2pt\relax]{#1}{.4pt}}%
   \mkern-4mu\hbox{\usefont{U}{lasy}{m}{n}\symbol{41}}}}

\def\keyFont{\fontsize{8}{11}\helveticabold }
\def\firstAuthorLast{Di Felice {et~al.}} 
\def\Authors{Louisa Jane Di Felice\,$^{1}$, Ada Diaconescu\,$^{2,*}$, Payam Zahadat\,$^{3}$ and Patricia Mellodge\,$^{4}$}
\def\Address{$^{1}$Department of Economic History, Institutions, Politics and World Economy, Universitat de Barcelona, Barcelona, Spain \\
$^{2}$Department of Computer Science and Networks, LTCI, Télécom Paris, Institut Polytechnique de Paris, Palaiseau, France  \\
$^{3}$Robotics, Evolution and Artificial Life lab, IT University of Copenhagen, Copenhagen, Denmark\\
$^{4}$Department of Electrical and Computer Engineering, University of Hartford, Hartford, CT, USA
}

\def\corrAuthor{Ada Diaconescu}
\def\corrEmail{ada.diaconescu@telecom-paris.fr}

\begin{document}
\onecolumn
\firstpage{1}

\title[The Value of Information]{The Value of Information in Multi-Scale Feedback Systems} 

\author[\firstAuthorLast ]{\Authors} 
\address{} 
\correspondance{} 

\extraAuth{}

\maketitle
\begin{abstract}
Complex adaptive systems (CAS) can be described as systems of information flows dynamically interacting across scales in order to adapt and survive. CAS often consist of many components that work towards a shared goal, and interact across different informational scales through feedback loops, leading to their adaptation. In this context, understanding how information is transmitted among system components and across scales becomes crucial
for understanding the behavior of CAS. Shannon entropy, a measure of syntactic information, is often used to quantify the size and rarity of messages transmitted between objects and observers, but it does not measure the value that information has for each specific observer. For this, semantic and pragmatic information have been conceptualized as describing the influence on an observer's knowledge and actions. Building on this distinction, we describe the architecture of multi-scale information flows in CAS through the concept of Multi-Scale Feedback Systems, and propose a series of syntactic, semantic and pragmatic information measures to quantify the value of information flows. While the measurement of values is necessarily context-dependent, we provide general guidelines on how to calculate semantic and pragmatic measures, and concrete examples of their calculation through four case studies: a robotic collective model, a collective decision-making model, a task distribution model, and a hierarchical oscillator model. Our results contribute to an informational theory of complexity, aiming to better understand the role played by information in the behavior of Multi-Scale Feedback Systems.

\tiny
 \keyFont{ \section{Keywords:} adaptation, syntactic, semantic, pragmatic, complexity} 
\end{abstract}

\section{Introduction}

Complex Adaptive Systems (CAS), such as organisms, societies, and socio-cyber-physical systems, self-organize and adapt to survive and achieve goals. Their viability and behavior depend on their ability to perceive and adapt to their internal states and external environment. Perception and adaptation are often modulated by feedback loops \cite{kephartAC2003},  \cite{müller2011organic}, \cite{lalanda2013autonomic} \cite{haken2016information}, which are based on the exchange of matter, energy, and information. When material and energy flows are perceived by system entities, we refer to these entities as \textit{agents}. Material and energy flows become information flows when agents perceive them, and use that perception to change their knowledge or behavior. Perception does not require consciousness, only the capacity to receive information and adapt to it -- a part of an engineered system, an ant, and a cell can all be described as agents. In this view, all material and energy flows are potential information flows, and all parts of CAS become informational. This is in line with informational structural realism \cite{floridi2008defence}, which considers the world as the “totality of informational objects dynamically interacting with each other” (p.219). The dynamic interactions turn material flows into informational ones, with agents transforming sources into informational inputs \cite{jablonka2002information}, \cite{von2013foray}. The presence of a goal is what distinguishes physical patterns from informational ones: while physical interactions can have no set purpose \cite{roederer2005information}, the agent extracting information from a source does so with a goal. Still, information flows are not decoupled from energy and matter, as perception, communication, and adaptation require the storage of information onto a physical substrate. Information can be tightly or loosely coupled to its physical substrate; in the latter case, it can be encoded onto different substrates \cite{feistel2016entropy}. 

The content and dynamics of information flows are central to CAS behavior \cite{atmanspacher1991information}. The environmental information needed for a CAS to maintain its viability can be measured \cite{kolchinsky2018semantic}, determining the subset of system-environment information used for its survival \cite{sowinski2023semantic}. Depending on the granularity selected to perceive a CAS, collections of entities can also be described as systems (e.g., a flock of birds, not just an individual bird) \cite{krakauer2020information}. In this case, considering the information flows generated and circulated within the system is necessary to understand its adaptive behavior -- a CAS using the same amount of environmental information may behave differently depending on how that information is circulated and used among its parts, and on how new information is generated within the system. Collective CAS with many components tend to operate across different scales, and to circulate information through feedback cycles \cite{Simon1991}, \cite{ahl1996hierarchy}, \cite{Pattee1973}. Our goal is to conceptualize what it means for information to flow through a collective CAS, and to measure how that information is used for the system's adaptation. For this, we focus on CAS formed by interacting components that operate under a collective goal and across different scales. 

The term \textit{scale} usually refers to spatial scales, temporal scales, or organizational levels. These different scales can be unified from an information perspective, defining an \textit{informational scale} as the chosen granularity for observing a system \cite{diaconescu2021information}. Within a collective CAS, information from the local scale can be abstracted to generate global information about the collective state, and individual agents can observe this coarse-grained information locally and adapt to it. 
We refer to CAS that contain such multi-scale feedback cycles as Multi-Scale Feedback Systems (MSFS) \cite{diaconescu2019multi} \cite{diaconescu2021exogenous}. Multi-scale feedback cycles help avoid scalability issues by allowing agents to coordinate without exchanging detailed information, relying instead on the local availability of global information \cite{flack2013timescales}. While at a higher scale information is less granular, this does not make higher scales less complex \cite{flack2021complexity}, but reduces the amount of information that the micro-scale has to process for its own adaptation. Bottom-up information abstraction is often referred to as coarse-graining, while top-down reification is referred to as downward causation \cite{flack2017coarse}. 

In this paper, we expand on previous conceptualizations of MSFS and propose a series of information measures that can be used to understand the behavior of such systems. Given the complex information dynamics in CAS, as well as the plurality of the concept of information \cite{floridi2005semantic}, measuring information flows and their impacts is a non-trivial task. As information circulates through a MSFS and its environment, agents perceive it and act upon it by adapting their knowledge and behavior. These three processes – information communication and perception, knowledge update, and action update – can be mapped onto three categories of information measures: syntactic, semantic, and pragmatic \cite{morris1938foundations}. Syntactic information measures quantify the amount of information transmitted to an agent. Among these, Shannon entropy quantifies the amount of information in a source, and can be interpreted as a measure of compressibility, uncertainty or resource use \cite{shannon1948mathematical} \cite{tribus1971energy} \cite{timpson2013quantum}. While relevant to message transmission, syntactic measures provide no indication about the actual meaning or usefulness of the message to each specific agent. These are described by semantic and pragmatic information measures. Semantic information measures quantify changes in the agent's state, and pragmatic information measures quantify changes in its behavior \cite{haken2016information}. Although pragmatic information measures are sometimes included in the semantic category \cite{kolchinsky2018semantic}, and vice versa \cite{gernert2006pragmatic}, we consider them as separate and interrelated. As semantic and pragmatic categories consider the \textit{value} of information to an agent, they are necessarily relative \cite{von2013foray}, requiring a case-dependent approach for their measurement. 

Building on previous work \cite{diaconescu2019multi} \cite{Mellodge2021} \cite{diaconescu2021exogenous}, we propose a series of syntactic, semantic, and pragmatic information measures for CAS, focusing on MSFS. We provide generic guidelines to calculate them, and offer examples through four case studies: a robotic collective model, a collective decision-making model, a task distribution model, and a hierarchical oscillator model. The breadth of these case studies allows generalizing the role of the proposed information measures in MSFS, building towards a better understanding of the role of information in CAS.

\section{Background: Information Measures}

While agents turn material flows into information, information is still tied to a physical substrate \cite{walker2014top}. The information perceived by the agent can be more or less decoupled from its substrate. When it is tightly coupled, it can be referred to as structural information. When it becomes less dependent on it, and can be encoded differently onto alternative substrates, it can be referred to as symbolic information \cite{feistel2016entropy} \cite{bellman-lang&mov1984}. Beyond the physicality of the information carrier, resources are also needed when information is stored, interpreted or transformed. Shannon entropy applies to the communication of messages \cite{shannon1948mathematical}. It is observer-dependent, since (i) it requires the existence of an observer perceiving the message; and (ii) probability distributions depend on the observer’s knowledge of the system \cite{lewis1930symmetry}. Alternatively, Algorithmic Information Theory \cite{AIT_Vitanyi2008} proposes a universal measure for the irreducible information content of an object. The object's Algorithmic Complexity, or Kolmogorov complexity \cite{kolmogorovIntro2019}, is its most compressed, self-contained representation. It can be defined as the length of the shortest binary computer program that generates the object. Kolmogorov complexity only depends on the description language, or universal Turing machine running the program. While Shannon entropy measures the quantity of information within an average object from a probabilistic set, Kolmogorov complexity measures the absolute information within an individual object, without requiring prior knowledge of its probabilistic distribution. 

Applying these concepts to human cognition, \cite{jldRelevance2013}
considers the importance of information to human observers, in terms of its interest, surprise, memorability, or relevance. This is assessed in terms of the difference between the perceived and expected algorithmic complexity of an observed object. Similarly, \cite{jldBadLuckEmotion2010} measures an observer's emotion about an event depending on the difference between its stakes and the causal complexity of its occurrence. Despite their richness and wide application range, the above information measures exclusively focus on the content of the informational object(s), taken in isolation, irrespective of their actual usage by an agent. This makes them insufficient for assessing the importance of information flows to an agent that employs them to self-adapt.The limitations of syntactic information measures are argued by many authors \cite{brillouin1962science} \cite{atmanspacher1991information} \cite{nehaniv1999meaning} \cite{gernert2006pragmatic}, leading to different approaches to measure semantic and pragmatic information. Semantic information is particularly relevant for biology \cite{jablonka2002information}. \cite{kolchinsky2018semantic} focus on system viability, and measure the environmental information needed for a system to survive, referred to as the \textit{viability value of information}. For pragmatic information measures, \cite{weizsacker1972wiederaufnahme} first noted that any measure of pragmatic information should be zero when novelty or confirmation are zero. Novelty refers to whether the message contains any new information for the receiver, and confirmation to whether the message is understandable. \cite{weinberger2002theory} put forward a formal definition of pragmatic information that also tends to zero when novelty or confirmation are zero, noting that this idea was shared by \cite{atmanspacher1991information}, and is in-line with Crutchfield's complexity. \cite{frank2003pragmatic} argues that pragmatic information can only be measured with respect to an action. They argue that if someone uses two different maps to go from one place to another (one map more detailed than the other), resulting in the same route, then the pragmatic information of those two maps is the same. This ignores the amount of energy needed to extract information from a message -- also referred to as \textit{negentropy} \cite{brillouin1953negentropy}. In the MSFS context, we find it useful to distinguish between semantic and pragmatic information (impact on knowledge and impact on action). This helps to assess the importance of information for a CAS that acquires knowledge to adapt and achieve goals. In engineering, this applies to autonomic computing \cite{kephartAC2003}, \cite{autonomicLalanda2013},
 organic computing \cite{müller2011organic}, \cite{OrganicCMllerSchloer2017}, self-adpative systems \cite{sasWeynsBook2021},\cite{SASreview2022}, self-integrating systems \cite{selfIntegrationMastering2021}, and self-aware systems \cite{selfAwareNotion2017}. 

\section{Multi-Scale Feedback Systems}
\subsection{Overview}

\setcounter{figure}{0}
\begin{subfigure}
    \centering
    \begin{minipage}[b]{0.47\textwidth}
        \includegraphics[scale=0.12]{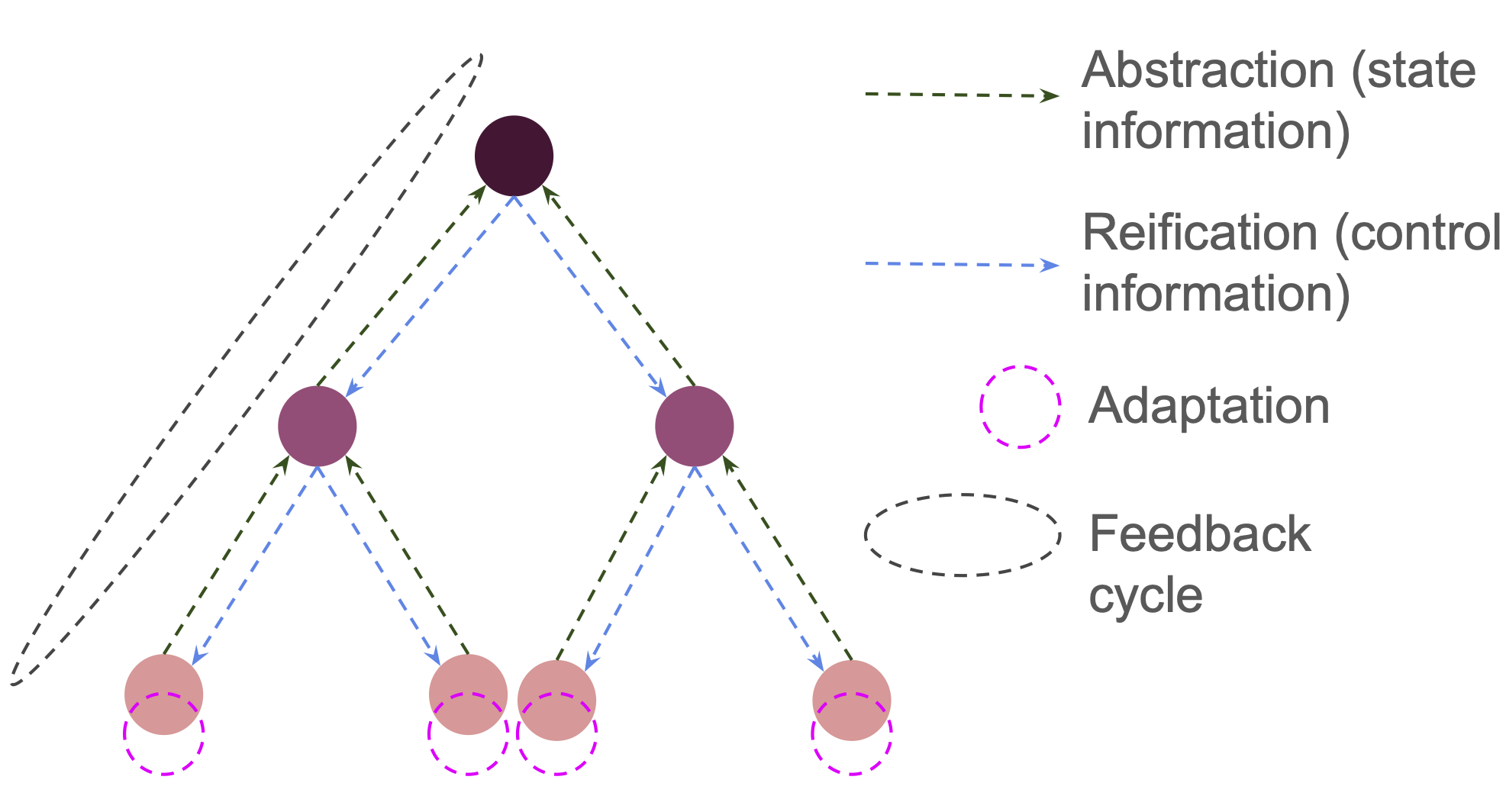}
        \caption{MSFS across three scales, \\ with a single feedback cycle}
        \label{fig:generic_cycle}
    \end{minipage}  
    \begin{minipage}[b]{0.47\textwidth}
        \includegraphics[scale=0.12]{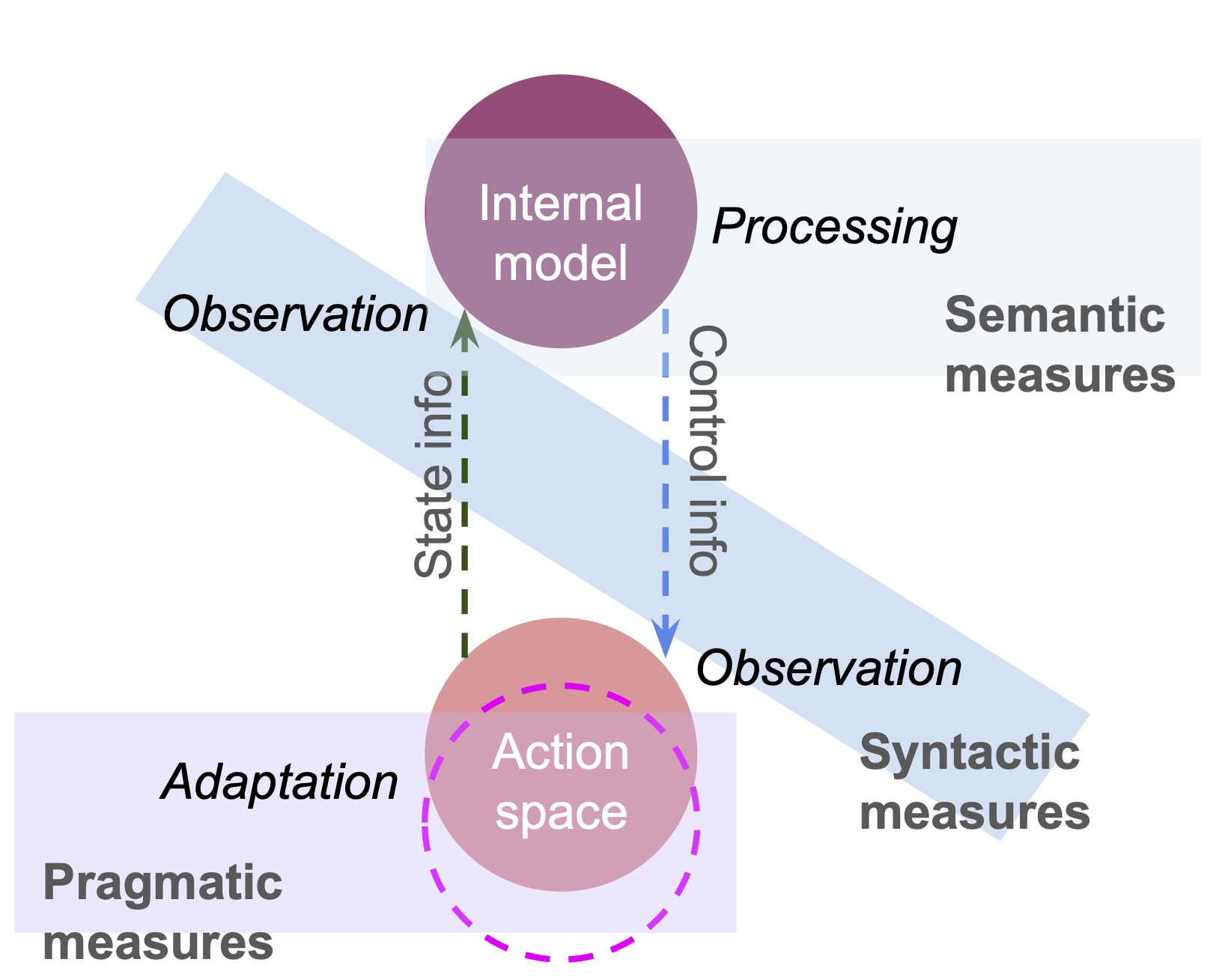}
        \caption{Syntactic, semantic, and pragmatic information measure domains}
        \label{fig:three_measures}
    \end{minipage}

\setcounter{subfigure}{-1}
    \caption{Main MSFS elements}
    \label{fig:abstraction_reification}
\end{subfigure}

In MSFS, information observed at the micro-scale merges into macro-information. Information at the macro-scale has a coarser granularity than at the micro-scale, with a loss of information that can be due to an abstraction function (e.g., the information from the micro-scale being averaged out) or to the sampling frequency of micro-information. 
Macro-information has a larger scope than each information source at the micro-scale. 
This limits resource requirements for the micro-scale to adapt to collective information. We refer to information abstracted from micro to macro as \textit{state information}; and to information that flows back from macro to micro as \textit{control information}. When control information is reified from macro to micro, it can become more detailed (e.g., adding new information flows specific to the micro-scale) or more abstracted (e.g., several decisions at the macro-scale resulting in the same action at the micro-scale). Figure \ref{fig:generic_cycle} shows a multi-scale feedback cycle, including information abstraction, information reification, and adaptation. Processing may also take place at different scales of the feedback cycle, whenever an agent observes information and updates its knowledge. 
 
Macro-information can take different forms depending on how it is coupled to its physical substrate:

\begin{itemize}
    \item \textbf{Exogenous or endogenous}. \textit{Exogenous} macro-information uses a substrate that is external to the substrate of the micro-information sources. The macro-substrate can be another agent -- e.g., a manager collecting information from its workers; or it can be an entity within the environment -- e.g., a public blackboard showing relevant information to investors, or an anthill providing information to ants. \textit{Endogenous} macro-information uses the same substrate as the micro-information sources. Micro-agents act based on their internal model of an abstract entity in the world -- e.g., members of a society acting based on informal norms governing the group. 
    \item \textbf{Centralized or distributed}. \textit{Centralized} macro-information uses a substrate that is monolithic, based on tightly coupled components, which are either static or change together as a whole -- e.g., managers, pheromone trails, or anthills. \textit{Distributed} substrates consist of components that are loosely coupled or that change or move independently. 
    \item \textbf{Structural or symbolic}. We use the distinction put forth by \cite{feistel2016entropy}. \textit{Structural} macro-information is tightly coupled to the structure of its substrate -- e.g., the size of a bird flock or a bee hive that may impact individual bird or bee behavior. \textit{Symbolic} macro-information is loosely coupled to the structure of its substrate. It may be encoded differently onto various substrates and decoded by observers accordingly -- e.g., managers issuing orders verbally, via email or written documents. 
\end{itemize}

The above macro-categories allow for a broad range of CAS to be described as MSFS, with the underlying mechanism being a feedback cycle connecting a minimum of two scales, and with the micro-scale adapting based on macro-information. Figure \ref{fig:three_measures} shows the adaptation of a micro-agent with respect to macro-information. The micro-agent is connected to the scales above and below by a flow of state information going upward, and control information going downward. Additional flows may exist (e.g., communication among micro-agents). The transmission and observation of these information flows are described by syntactic information measures. When an agent processes information, it may subsequently change its knowledge (its internal model or the output generated by that model) and the action it generates. Processing can  
happen at different scales. Figure {\ref{fig:three_measures} shows a macro-agent processing information and sending down the ensuing control information. The micro-agent could also observe the macro-information and process it for its 
adaptation. The impact of the information flow on an agent's knowledge is measured by semantic information measures. Changes in the resulting action are measured by pragmatic information measures. 

\subsection{Information Measures}

Information values are observer-dependent \cite{brillouin1953negentropy}. To measure them, the purpose of the information flow needs to be specified.
This can be system viability \cite{kolchinsky2018semantic}, or utility towards a goal. In general, value is measured by comparing how the system performs towards a goal state 
with and without information \cite{gould1974risk}. As all elements in a CAS are informational, various comparison scenarios can be developed. For MSFS, the impact of  feedback cycles on the agents' knowledge and actions is calculated by comparing the system at 
different times. The relevant gap for comparison, depending on the application, can be a single time-step, a single feedback cycle, or multiple cycles. We define the time at which the information measure is calculated as $t$, the previous time used for comparison as $t-\theta$, and the information flowing between $t$ and $t-\theta$ as $I^{t-\theta\rightarrow t}$. 

There are three steps in the construction and interpretation of information values. First, the system state at time $t$ is assigned a \textit{state value} with respect to a goal state. These may be knowledge or action states. The state value measures how valuable the state at time $t$ is, with respect to the goal state. Depending on the system, this may include how likely it is for the system to reach its goal and/or stay at that goal within a certain time-frame. Then, the state values of the system at times $t$ and $t-\theta$ are compared, to measure the impact of $I^{t-\theta\rightarrow t}$. Finally, information values for different system configurations are compared, to contextualize their dependencies -- e.g., varying the system architecture, algorithms, or adding errors to $I^{t-\theta\rightarrow t}$. In addition to information values, descriptive information measures can be calculated, not taking state values into account. We propose the following information values and descriptive measures:

\begin{itemize}
    \item The \textbf{syntactic information content ($C_{syn}$)} measures the amount of information in $I^{t-\theta\rightarrow t}$, During the feedback cycle, information flows are transmitted, stored and processed, and $C_{syn}$ estimates the amount of resources needed for this. It can be calculated using Shannon entropy $H$, if probability distributions are known. It can also be calculated using Kolmogorov complexity, the amount of memory used to hold information, or other meaningful proxies of resource use. $C_{syn}$ can be calculated for individual variables, a whole scale or  a full feedback cycle ($C_{syn, cycle}$).
    \item The \textbf{semantic information content ($C_{sm}$)} is the subset of $C_{syn}$ that has meaning for the agent(s) observing the information. If all $C_{syn}$ can be interpreted, $C_{sm} = C_{syn}$. 
    \item The \textbf{semantic delta ($\Delta_{sm}$)}is a descriptive measure that quantifies the magnitude of change in the knowledge due to $I^{t-\theta\rightarrow t}$, irrespective of the goal. It can be split into sub-measures depending on which aspects of the knowledge are relevant. 
    \item The \textbf{semantic truth value ($V_{sm,th}$)} measures whether $I^{t-\theta\rightarrow t}$ brings the knowledge closer to an observable ground truth ($th$). First, ground truth state values $SV_{th}$ are assigned at times $t$ and $t-\theta$, considering how valuable the knowledge is with respect to a ground truth $th$. The calculation of $SV_{th}$ depends on the system, but usually includes a \textit{delta of truth} $\Delta_{th}^t$, i.e., the distance between the knowledge and $th$ at $t$. Then, $V_{sm,th}$ is calculated based on the distance between $SV_{th}$ at times $t$ and $t-\theta$:
    \begin{equation}
        V_{sm,th}^{t-\theta\rightarrow t} = SV_{th}^{t} - SV_{th}^{t-\theta}
    \end{equation}
    \item The \textbf{semantic goal value ($V_{sm,gl}$)} measures whether $I^{t-\theta\rightarrow t}$ brings the knowledge closer to the optimal knowledge $opt$ needed to reach the goal. This may or may not be the same as the ground truth, depending on the CAS. First, optimal knowledge state values $SV_{opt}$ are calculated at $t$ and $t-\theta$. This includes $\Delta_{opt}^t$, i.e., the distance at $t$ between knowledge and $opt$. Then, $V_{sm,gl}$ is calculated as the 
    difference between $SV_{opt}$ at $t$ and $t-\theta$:
    \begin{equation}
        V_{sm,gl}^{t-\theta\rightarrow t} = SV_{opt}^{t} - SV_{opt}^{t-\theta}
    \end{equation}
    \item The \textbf{pragmatic delta} ($\Delta_{pr}$) is a descriptive measure that quantifies the magnitude of change that $I^{t-\theta\rightarrow t}$ has generated on the adaptation. Similar to $\Delta_{sm}$, it can be 
    divided into sub-measures depending on the relevant adaptation aspects.
    \item The \textbf{pragmatic goal value ($V_{pr,gl}$)} measures whether the adaptation following $I^{t-\theta\rightarrow t}$ brings
    brought the system closer to its goal. Goal state values $SV_{gl}$ are calculated at $t$ and $t-\theta$, usually requiring the calculation of a $\Delta_{gl}^t$, i.e., the distance at $t$ between the system state and the goal. $V_{pr,gl}$ represents the change in $SV_{gl}$ between $t$ and $t-\theta$:
    \begin{equation}
        V_{pr,gl}^{t-\theta\rightarrow t} = SV_{gl}^{t} - SV_{gl}^{t-\theta}
    \end{equation}
    \item \textbf{Efficiencies} of each measure can be calculated by dividing it by the $C_{syn}$ of $I^{t-\theta\rightarrow t}$. 
\end{itemize}

The exact calculations of each measure and its units depend on the system, and not all measures are relevant to every system -- e.g., it may not be possible to identify a ground truth for $V_{sm,th}$. Our measures differ in certain cases from the existing literature. While some authors argue that data need to be contingently true to be considered information \cite{floridi2005semantic} \cite{fetzer2004information}, $V_{sm,th}$ measures whether the information flow has brought knowledge closer or further away from the truth, considering it to have value in both cases. We do not measure pragmatic value as the product of novelty and confirmation, since in our framework both of these refer to the semantic category. If confirmation is zero then the $C_{sm,cycle}$ of the information is zero, leading to all other semantic and pragmatic measures also being zero. If novelty is zero, however, the information flow may still carry semantic value, since it can lead to an update of the agent's internal model. Efficiencies, then, connect the values to the physical substrates. Considering the example proposed by \cite{frank2003pragmatic}, while the pragmatic measures would be the same for two different maps leading to the same route, their efficiencies would differ, as they would require different amounts of resources to be processed. 

\section{Case studies}

\subsection{Overview}

Figure \ref{fig:case_studies} shows a multi-scale representation of the four case studies: robotic collective (RC) (\ref{fig:cases_robotic}), collective decision-making (CD) (\ref{fig:cases_collective}), task distribution (TD) (\ref{fig:cases_task-distr}), and hierarchical oscillators (HO) (\ref{fig:cases_hierarchical}).

\stepcounter{figure}

\begin{subfigure}
    \centering
    \begin{minipage}[b]{0.55\textwidth}
        \includegraphics[scale=0.71]{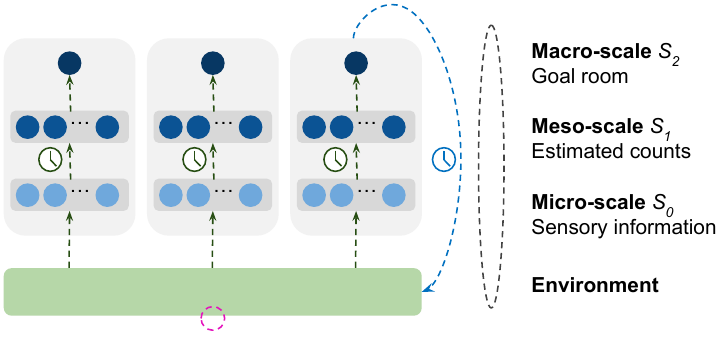}
        \caption{Robotic collective}
        \label{fig:cases_robotic}
        
        \includegraphics[scale=0.71]{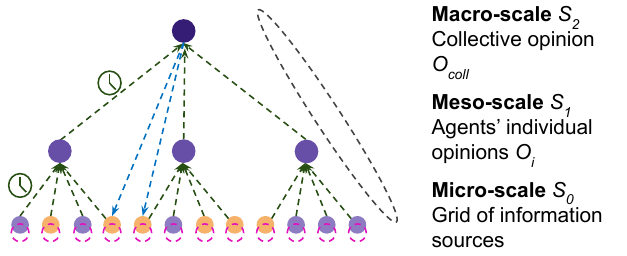}
        \caption{Collective decision-making}
        \label{fig:cases_collective}
    \end{minipage}
    \begin{minipage}[b]{0.44\textwidth}
        \includegraphics[scale=0.71]{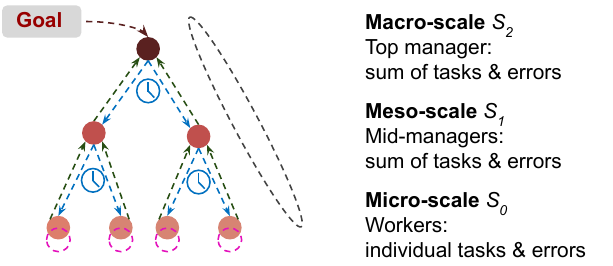}
        \caption{Task distribution}
        \label{fig:cases_task-distr}

        \includegraphics[scale=0.71]{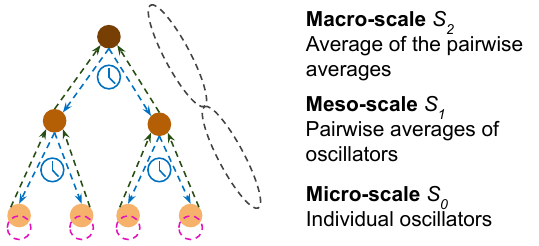}
        \caption{Hierarchical oscillators}
        \label{fig:cases_hierarchical}
    \end{minipage}

\setcounter{subfigure}{-1}
    \caption{Multi-scale representation of the four case studies. Green arrows show abstracted flows of state information, green arrows flows of control information, and pink dotted circles show the adaptation. Clocks (blue or green) denote which flows have a delay. In \ref{fig:cases_collective}, the macro-scale sends control information to each micro-scale agent -- only two flows are shown for simplicity. Similarly, in \ref{fig:cases_robotic} , where each gray box represents an individual robot, only one flow connecting the robot to the environment is shown.}
    \label{fig:case_studies}
\end{subfigure}

\subsubsection{Robotic Collective Multi-Scale Overview}
The RC model (Fig.~\ref{fig:cases_robotic}) is based on the simulation of a swarm of robots that self-distribute across multiple rooms in proportion to the amount of collectible objects in each room, as in \cite{Zahadat2023ACSOS}. 
Each robot acts as a micro-agent equipped with sensors that provide it with the lowest scale of information, related to the number of robots and objects in the rooms based on the sensing of the environment. This forms the micro-scale ($S_0$). Over time, robots abstract their micro-scale information into meso-scale information ($S_1$), consisting of estimations correlated with the current number of robots and objects in the rooms. This is further abstracted into robots' estimations of each room's demand for robots, constituting the system's macro-scale $S_2$. Based on that, the robots decide their motion actions 
, forming a feedback loop.
The system dynamics are simulated in discrete time steps, and the robots are synchronized: at each time step, all robots update their information at every scale, starting with $S_0$ and ending with $S_2$, followed by a motion action. 
The process of abstracting micro-information into meso-information occurs over time.
In contrast, the abstraction of meso-information into macro-information happens in a single step. Robots' motion actions also unfold over time.
As a result, feedback cycles are inseparable from each other, extended in time, and occur at variable intervals depending on the system's state. We present our measurements either at consecutive timesteps or over a series of intervals.

\subsubsection{Collective Decision-Making Multi-Scale Overview}
The CD model simulates a system in which a group of agents makes a collective decision about the size of two tasks by observing their environment. The environment is composed of a fixed set of information sources, each showing one of the two tasks, $A$ and $B$. This is the micro-scale ($S_0$). Agents are placed on the grid of information sources and each agent forms an individual opinion $O_i$ about the relative size of $A$, by observing the environment around them and averaging out what they see. The meso-scale $S_1$ is formed by the set of individual agent opinions. In the second abstraction step, agents go through a consensus process to form a collective opinion $O_{coll}$ of the size of $A$, constituting the macro-scale $S_2$. Then, $O_{coll}$ generates feedback on the micro-scale -- i.e., the environment of information sources. If agents underestimate the size of $A$, the task will grow until the next feedback cycle, and vice versa. If task $A$ becomes too big or too small the environment collapses. The goal of the system is to avoid collapse for as long as possible. Timing is discrete: the environment of information sources ($S_0$) changes at each timestep, and the length of the feedback cycle depends on the system variables and algorithms. We focus on measuring the value of the amount of information used by agents to observe their immediate environment; the number of agents scanning the environment; and the algorithm used to reach consensus.

\subsubsection{Task Distribution Multi-Scale Overview}
This system aims to distribute a set of tasks of two types
across a set of workers. We consider a small system with three scales, for traceability reasons (Fig. \ref{fig:cases_task-distr}). Larger instances were presented elsewhere \cite{diaconescu2021exogenous}. At the micro-scale ($S_0$), workers select an active task of one of two types. The initial choice is  random, then it is driven by feedback from the upper scale ($S_1$). Managers at the meso-scale ($S_1$) collect, aggregate and forward the workers' task choices to the macro-scale ($S_2$) and then retrieve, split and distribute the feedback from the macro-scale ($S_2$) to the workers ($S_0$). The top manager 
(at $S_2$) receives the task distribution goal as initial input and provides feedback to the meso-scale ($S_1$) about the difference between the goal and the workers' current task choices. Information processing and transmission at each scale depend on the task distribution strategy employed in the system. 
This forms a feedback cycle $c$, taking two simulation steps. The workers' state information is sent to the top in one step, then feedback returns one scale at each step. As soon as meso-managers detain new control information, workers adapt to it and send their new states up. Hence, there is a two-step delay between workers sending their state ($t_i$) and adapting to feedback obtained from that state ($t_{i+2}$). 
We calculate information measures at each step $t_i$ considering the information flows over the preceding feedback cycle from $t_{i-2}$ to $t_i$. 

\subsubsection{Hierarchical Oscillators Multi-Scale Overview}
The HO case study simulates a system of communicating oscillators whose goal is to achieve synchronized oscillation.  This system is based on the differential equation model for coupled biochemical oscillators presented in \cite{Kim2010} which has been extended to a hierarchical structure \cite{Mellodge2021}.  A single oscillator consists of two interacting components $X$ and $Y$ (e.g., mRNA and protein) in which $X$ inhibits its own synthesis while promoting that of $Y$ and $Y$ inhibits its own and $X$'s synthesis, resulting in a feedback relationship within a single oscillator that induces oscillations in the concentrations of both $X$ and $Y$.  In the HO system, multiple scales of oscillators communicate by means of their $X$ concentrations. The oscillators at the micro-scale $S_0$ send their $X$ concentrations to the oscillators at the meso-scale $S_1$ which average that information.
These oscillators in turn send their $X$ concentrations to the next highest scale until the macro-scale $S_{M-1}$ is reached, which contains the average of all micro-scale $X$ concentrations.  Information is also passed down the system one scale at a time through feedback.  The 
control information contains time-delayed $X$ concentration averages.  At each scale $S_m$, oscillators update their own $X$ and $Y$ concentrations based on a combination of current averages from the lower scale $S_{m-1}$ and time-delayed averages from the higher scale $S_{m+1}$.  Information measures are used to compare the behavior and performance of two system structures, a two-scale hierarchy and a three-scale hierarchy. 

\subsection{Robotic Collective}

This case study is based on \cite{Zahadat2023ACSOS} where a homogeneous robotic swarm is designed to self-distribute across four interconnected rooms proportional to the quantity of collectible objects in each room. The four-room layout forms a ring, with each room connected to two others. Here, 200 collectible objects are distributed across four rooms, with 70 objects in each of the first two rooms and 30 in each of the remaining two rooms. The robotic swarm, consisting of 100 robots, is initially distributed uniformly across all rooms. The robots can only detect objects beneath their circular bodies, and their communication range with other robots is twice their diameter. The robots are autonomous and use a variation of the Response Threshold algorithm~\cite{Theraulaz98} to independently decide which room they want to belong to: their \textbf{goal} room. 
The decision is regularly updated based on the robot's internal model of the environment's state. %
Each robot models only the state of the current room and the two rooms connected to that, collectively called the robot's \textbf{relevant rooms}.
The \textbf{locomotion behavior} of the robots is a collision-avoiding random walk, modulated by the current and the goal room: robots can cross room boundaries freely unless they are in their goal room.

We consider the multi-scale feedback generated by the information flowing through the robots' internal processing structures and feeding back to them via the environment (Fig.\ref{fig:cases_robotic}).
The $S_0$ consists of information collected through sensory inputs from physical interactions with the outside world. This includes detecting other robots, detecting objects, and receiving communications from other robots.
The $S_1$ information is an abstraction of the micro-information into \textbf{estimated counts}. These estimated counts are approximations designed to correlate with actual counts and are represented as a vector $\hat{v}^{t}(i)$ for each robot $i$ at time $t$.
The vector consists of six components: the first three components, $\hat{v}_j^{t}(i)$ for $j=1,2,3$, are the estimated object counts in the robot's three relevant rooms; and the last three components, $\hat{v}_{j+3}^{t}(i)$ for $j=1,2,3$, are the estimated robot counts in those same rooms.
To compute the estimated counts, micro-information is first integrated into sense-based and communication-based quantities over time (\cite{Zahadat2023ACSOS}).
Sense-based quantities are calculated as a simple moving average 
of the number of detections of a robot (or object) over the past $M$ time-steps in each relevant room, where $M$ is the \textbf{sensing horizon}.
Communication-based quantities are computed as exponential moving averages 
of the estimated counts received from other robots. 
These quantities are then averaged pairwise and normalized across rooms to form $\hat{v}^{t}(i)$.
In $S_2$, the $\hat{v}^{t}(i)$ is processed to decide on the robot's goal room.
First, a representation of \textbf{estimated demand} for each relevant room is computed:
\begin{equation}
    \hat{\varphi}_j^{t}(i)=\hat{v}_j^{t}(i)/(\hat{v}_j^{t}(i)+\hat{v}_{j+3}^{t}(i)), \text{ for } j=1,2,3.
\end{equation}
Then the robot selects the room with the largest estimated demand as its goal: $g^{t}(i) = \arg\max_{j=1,2,3} \hat{\varphi}_j^{t}(i)
$.
A robot's $g^{t}(i)$, and whether it matches its current room, influence its locomotion behavior. This can affect the rate of robot exchange between rooms, thereby altering the actual demand in the rooms, which in turn feeds back into the robot's knowledge system through its sensors.

\subsubsection{Comparison strategies}
The information processing described above forms our main strategy in this study. We use two configurations for this strategy: \textbf{main} strategy with sensing horizon of $M=100$, and \textbf{main (short memory)} strategy with $M=10$.
The two configurations reflect the effects  of memory resource usage on the various information measures. 
In addition to these, two other strategies are implemented: \textbf{ground truth} and \textbf{random} strategies.
In both of them, all the processing that leads to the estimated counts is omitted. 
In the ground truth strategy, the actual counts of robots and objects, normalized across the relevant rooms, are directly copied into the estimated counts. 
Comparing the measures between the main and ground truth strategies reflects the effects of the estimation algorithm, particularly regarding sub-optimal accuracy and time delays 
.
In the random strategy, estimated counts are generated by sampling from a uniform distribution, normalized across rooms. Since this strategy lacks feedback loops, it allows us to evaluate the benefits of feedback in achieving the system's goal.

\subsubsection{Information measures}
\paragraph{Calculations}

\subparagraph{Syntactic information}
The amount of memory resources that a robot needs to store information at each scale is used as a measure of $C_{syn}$ (detailed in the Supplementary Material).
We use 10, $6M+24$, and $4$ memory units in $S_0$, $S_1$ and $S_2$, respectively. 
The syntactic information content is therefore $C_{syn}^{main}=6M+38$ memory units for the main strategies:
$C_{syn}^{main}(M=100) = 638$ and $C_{syn}^{main}(M=10) = 98$. 
Since in both ground truth and random strategies the estimated counts are set directly, their syntactic information contents are
$C_{syn}^{gt} = C_{syn}^{rn} = 10$, i.e., 6 memory units for the estimated counts in $S_1$ and 4 units used in $S_2$.

\subparagraph{Semantic information}

For the \textbf{semantic delta} $\Delta_{sm}$, the change in the knowledge between subsequent time steps is calculated. 
We first find the absolute change over time for each of the six components of the estimated counts.
Since the estimated counts are normalized over the relevant rooms, their values are interrelated. Therefore, we select the largest of the six absolute changes, which is then averaged across all robots to obtain the $\Delta_{sm}$ at timestep $t$.

\begin{equation}
\label{eq:robotic_semanticdelta}
\Delta_{sm}^{(t-1) \rightarrow t}  = \frac{1}{N}\sum_{i=1}^N\max_{j=1\dots 6} |\hat{v}^{t-1}_j(i) - \hat{v}^{t}_j(i)|,
\end{equation}
where $\hat{v}^t(i)$ and $\hat{v}^{t-1}(i)$ represent the current and previous estimated counts by robot $i$,
and $N$ is the number of robots.

To calculate the \textbf{semantic truth value} $V_{sm,th}$, 
we first calculate the \textit{delta of truth} $\Delta_{th}$ after each step, then measure its reduction. 
%
The robots' internal models consist of several sub-models. We consider three: one related to estimated counts at the meso-scale, 
and two related to macro-scale modeling of room demands. 

\textit{Estimated counts sub-model:} 
We calculate a measure of discrepancy between the estimated and actual counts over time. The estimated count vector $\hat{v}_j^{t}(i)$, has six elements for each robot $i$ at time $t$. At each timestep, the average of the absolute differences between the estimated and actual counts is taken across the six elements and all robots:
\begin{equation}
\label{eq:robotic_deltatruthcounts}
     \Delta_{th,counts}^{t} = \frac{1}{6N}\sum_{i=1}^N \sum_{j=1}^6 {\mid \hat{v}_j^{t}(i) - v_j^{t}\mid }
\end{equation}
where $v^{t}$ represents the vector of actual counts. 

\textit{Highest demand full sub-model}:
To measure the discrepancy between the internal models 
and the actual demands, we check whether the demand estimation for the relevant rooms is sufficient to select the room with the highest actual demand as their goal. To quantify this, we calculate the normalized count of robots whose goal room does not match the room with the highest actual demand among the relevant rooms:
\begin{equation}
\label{eq:robotic_deltatruthfull}
    \Delta_{th,full}^{t} = \frac{1}{N}\sum_{i=1}^N\delta_i^{t}
\end{equation}
\begin{equation}
    \delta_i^{t} =
    \begin{cases}
    1, & \text{if }  
    g^{t}(i) \neq \arg\max_{j=1,2,3}\varphi_j^{t}(i) \\
    0, & \text{else}
    \end{cases}
\end{equation}
where $\varphi^{t}(i)$ represents the actual demands in the relevant rooms for robot $i$, and $g^t(i)$ is the goal of the robot $i$ at time step $t$.

\textit{Highest demand partial sub-model:}
When examining the robots’ internal model for estimated demands (data not shown),
we observed a strong tendency to identify the local room as having the highest demand. 
In practice, this means that the robots often choose their local room as their goal and only occasionally target one of the neighboring rooms.
The measure here is the same as the one for the full model but only considers cases where the model does not identify the local room as having the highest demand:
\begin{equation}
\label{eq:robotic_deltatruthpartial}
    \Delta_{th,partial}^{t} = \frac{1}{N}\sum_{i=1}^N\delta_i^{t}
\end{equation}
\begin{equation}
    \delta_i^{t} =
    \begin{cases}
    1, & \text{if }  
    g^{t}(i) \neq L^{t}(i) \text{ and}\\
     &
    g^{t}(i) \neq \arg\max_{j=1,2,3}\varphi_j^{t}(i) \\
    0, & \text{else}
    \end{cases}
\end{equation}
where $L^{t}(i)$ represents the local room, 
$\varphi^{t}(i)$ represents the actual demands in the relevant rooms for robot $i$, and $g^t(i)$ is the goal of the robot $i$ at time step $t$.

$V_{sm,th}$ for each of the three sub-models ($V_{sm,th,model}^{t-1\rightarrow t}\in[-1,1]$) and their efficiencies are calculated as 
\begin{equation}
\label{eq:robotic_semantictruth}
V_{sm,th,model}^{t-1\rightarrow t} = 
\Delta_{th}^{t-1} - \Delta_{th}^{t}
\end{equation}

\begin{equation}
\label{eq:robotic_semantictruth_efficiency}
E_{sm,th,model}^{t-1\rightarrow t} = \frac{V_{sm,th,model}^{t-1\rightarrow t}}{C_{syn}}
\end{equation}

\subparagraph{Pragmatic information}

The \textbf{pragmatic goal value} $V_{pr,gl}$ measures whether, at each time step $t$, the distribution of robots across the four rooms, denoted by $r^{t}$, moves closer to or further away from the desired distribution, 
denoted by vector $o$.
We first calculate the distance between $r^{t}$ and $o$:
\begin{equation}
\label{eq:robotic_pragmaticdeltaofgoal}
    \Delta_{gl}^t = \frac{1}{4}\sum_{k=1}^4 \mid o_k - r^{t}_k \mid, \quad\text{where } k\in \{1,2,3,4\} \text{ is the room number} 
\end{equation}

The values $\Delta_{gl}^t \in [0,1]$ are then used to calculate $V_{pr,gl}$ and efficiencies for a given period, $\theta$:
\begin{equation}
\label{eq:robotic_pragmaticgoalvalue}
    V_{pr,gl}^{t-\theta\rightarrow t} = \Delta_{gl}^{t-\theta} - {\Delta}_{gl}^{t}
\end{equation}

\begin{equation}
\label{eq:robotic_pragmaticgoal_efficiency}
    E_{pr,gl}^{t-\theta\rightarrow t}  = \frac{V_{pr,gl}^{t-\theta\rightarrow t} }{C_{syn}}
\end{equation}

\paragraph{Results}
The results are pooled from 100 repetitions of independent experiment runs.
Figure~\ref{fig:semantic_output_value} shows $\Delta_{sm}$ for the two configurations of the main strategy. 
In both configurations the knowledge initially undergoes significant changes, driven by the information received at the beginning, which adapts the default knowledge system. After this abrupt initial change, subsequent changes are minor. 
%
\stepcounter{figure}
\begin{subfigure}
\centering
    \begin{minipage}[b]{0.49\textwidth}
        \includegraphics[scale = 0.71]{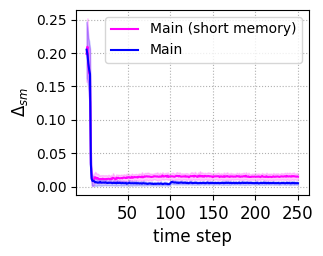}
            \caption{$\Delta_{sm}^{(t-1) \rightarrow t}$}
            \label{fig:semantic_output_value}
        \end{minipage}
\centering
     \begin{minipage}[b]{0.49\textwidth}
        \includegraphics[scale = 0.71]{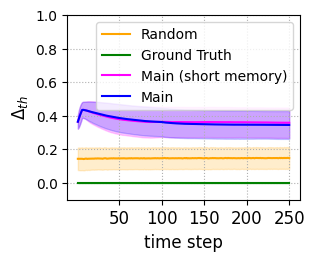}
            \caption{$\Delta_{th,counts}^{t}$}
            \label{fig:est_count_error}
        \end{minipage}
\centering
     \begin{minipage}[b]{0.48\textwidth}
        \includegraphics[scale = 0.72]{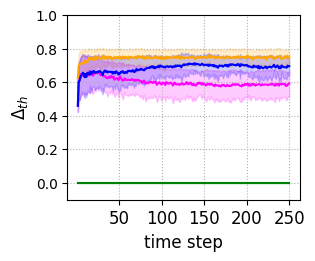}
        \caption{$\Delta_{th,full}^{t}$}
        \label{fig:decision_error}
        \end{minipage}
\centering
    \begin{minipage}[b]{0.49\textwidth}
        \includegraphics[scale = 0.71]{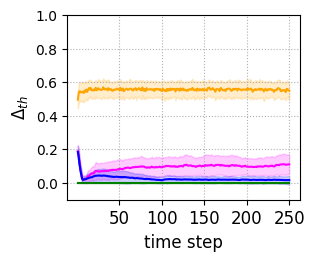}
    \caption{$\Delta_{th,partial}^{t}$}
    \label{fig:decision_error_neig}
        \end{minipage}

\setcounter{subfigure}{-1}
    \caption{Semantic delta calculated using  eq.~\ref{eq:robotic_semanticdelta} (a) and delta to truth for the three model components calculated using eq.~\ref{eq:robotic_deltatruthcounts}, \ref{eq:robotic_deltatruthfull}, \ref{eq:robotic_deltatruthpartial} (b-d).}
    \label{fig:semantic_outputdelta_deltatotruth}
\end{subfigure}
We also compare the four strategies to determine whether changes in different parts of the knowledge move toward or away from the ground truth they aim to model.
Figure~\ref{fig:est_count_error}-\ref{fig:decision_error_neig} show 
the discrepancy between the sub-models and the actual states.
As expected, there is no discrepancy between the models in the ground truth strategy. 
For the random strategy, estimated counts are assigned uniformly at random, and the delta is averaged over the three relevant rooms for each robot, resulting in a constant average discrepancy between the actual and the sub-model  counts (Fig.~\ref{fig:est_count_error}). 
In contrast, the 
counts for the main strategies initially show an increase in discrepancy, followed by a decrease and eventual stabilization.
This initial increase occurs because these strategies cause robots to move between rooms, changing the number of robots in each room. Since these changes take time to be captured in the model, the discrepancy temporarily rises before stabilizing.
However, the short memory strategy converges to a slightly higher level of discrepancy. 
Figures~\ref{fig:decision_error} and~\ref{fig:decision_error_neig} show the discrepancies in identifying the room with the highest demand. In both cases, the random strategy performs the worst. After the initial adaptation, the short memory strategy outperforms the long memory strategy for the full model of the relevant rooms (Fig.~\ref{fig:decision_error}) but performs worse when focusing only on the neighboring rooms (Fig.~\ref{fig:decision_error_neig}). 
For both of the main strategies, the partial models are significantly closer to the ground truth strategy and farther from the random strategy when compared with the full models.
This suggests that robots often assume the current room has the highest demand; otherwise they select the neighboring room with the actual highest demand.
Choosing the current room enhances system stability, while selecting the correct neighboring room drives the system toward the desired distribution of robots across rooms.
Figure~\ref{fig:rob_semantic_truth_and_efficiency} shows the semantic truth values and efficiencies for the three sub-models. 
It highlights the high fluctuation in the random models across all three cases. 
A comparison of Fig.~\ref{fig:rob_semantic_efficiency_full} and Fig.~\ref{fig:rob_semantic_efficiency_neig} shows higher fluctuations in the full sub-models for the main strategies.

\stepcounter{figure}
\begin{subfigure}[ht]
\centering
    \begin{minipage}[b]{0.32\textwidth}
        \includegraphics[scale = 0.65]{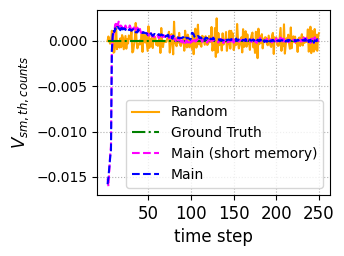}
            \caption{$V_{sm,th,counts}^{t-1\rightarrow t}$}
            \label{fig:rob_semantic_truth_counts}
        \end{minipage}
\centering
     \begin{minipage}[b]{0.32\textwidth}
        \includegraphics[scale = 0.65]{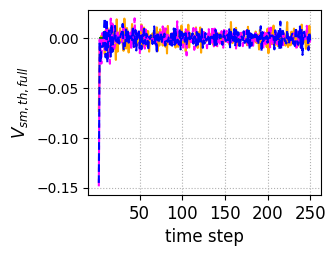}
            \caption{$V_{sm,th,full}^{t-1\rightarrow t}$}
            \label{fig:rob_semantic_truth_full}
        \end{minipage}
\centering
     \begin{minipage}[b]{0.32\textwidth}
        \includegraphics[scale = 0.65]{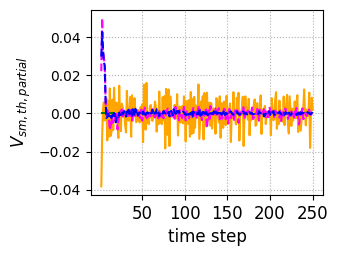}
        \caption{$V_{sm,th,partial}^{t-1\rightarrow t}$}
        \label{fig:rob_semantic_truth_neig}
        \end{minipage}

\centering
    \begin{minipage}[b]{0.32\textwidth}
        \includegraphics[scale = 0.65]{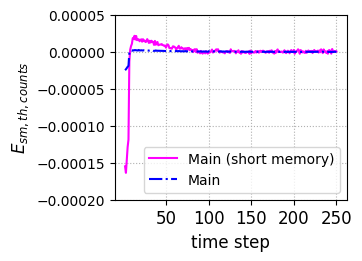}
    \caption{$E_{th,counts}^{t-1\rightarrow t}$}
    \label{fig:rob_semantic_efficiency_counts}
        \end{minipage}
\centering
    \begin{minipage}[b]{0.32\textwidth}
        \includegraphics[scale = 0.65]{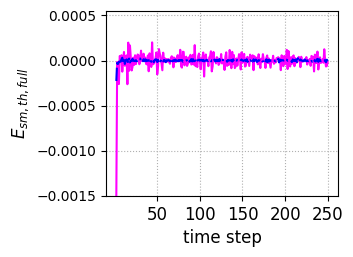}
    \caption{$E_{th,full}^{t-1\rightarrow t}$}
    \label{fig:rob_semantic_efficiency_full}
        \end{minipage}
\centering
    \begin{minipage}[b]{0.32\textwidth}
        \includegraphics[scale = 0.65]{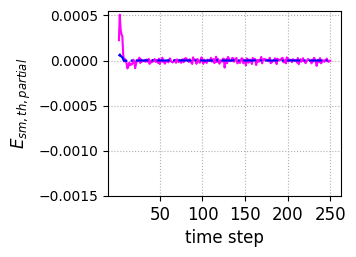}
    \caption{$E_{th,partial}^{t-1\rightarrow t}$}
    \label{fig:rob_semantic_efficiency_neig}
        \end{minipage}
\setcounter{subfigure}{-1}
    \caption{Semantic truth values (eq.~\ref{eq:robotic_semantictruth}) and efficiencies (eq.~\ref{eq:robotic_semantictruth_efficiency}) for the three model components.}
    \label{fig:rob_semantic_truth_and_efficiency}
\end{subfigure}

Figures~\ref{fig:rob_prag_delta_of_goal} and \ref{fig:rob_parg} compare the pragmatic measures. Figure~\ref{fig:rob_prag_delta_of_goal} shows the distance between the actual and desired distributions of robots across the rooms. 
The random strategy performs the worst, while the main strategy with the long memory achieves the best performance, followed by the short memory strategy. 
The ground truth strategy performs worse than the main strategies, despite relying on precise models. This is likely due to movement delays, causing late overreactions to perceived issues. This suggests a lack of critical models, possibly related to the spatial organization of robots and their interactions within rooms, which are essential for effective coordination but remain inaccessible to the robot. In contrast, the main strategies demonstrate that partial and imprecise information distributed across robots, rather than attempting an unattainable complete model, leads to better collective behavior.

Figures~\ref{fig:rob_prag_goalvalue_c10}-~\ref{fig:rob_prag_goalvalue_c500} 
depict the extent to which the robots' distribution 
in the rooms approaches the desired distribution over periods of 10, 100, and 500 timesteps. 
This indicates a pattern similar to that in Fig.~\ref{fig:rob_prag_delta_of_goal}, 
with large initial adaptations that subsequently oscillate around zero, stabilizing the system. 
Figures~\ref{fig:rob_prag_eff_c10}-\ref{fig:rob_prag_eff_c500} 
illustrate the efficiencies of the main strategies, 
highlighting the higher resource efficiency of the short memory strategy, particularly over longer feedback cycles.

\stepcounter{figure}
\begin{subfigure}
\centering
    \begin{minipage}[b]{0.96\textwidth}
        \includegraphics[scale = 0.75]{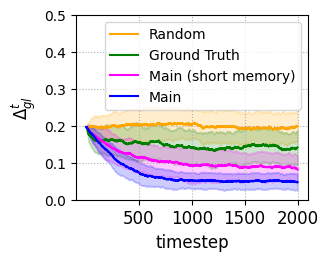}
        \end{minipage}
\setcounter{subfigure}{-1}
    \caption{Pragmatic delta (eq.~\ref{eq:robotic_pragmaticdeltaofgoal})  of the various strategies.}
    \label{fig:rob_prag_delta_of_goal}
\end{subfigure}

\stepcounter{figure}
\begin{subfigure}
\centering
    \begin{minipage}[b]{0.32\textwidth}
        \includegraphics[scale = 0.65]{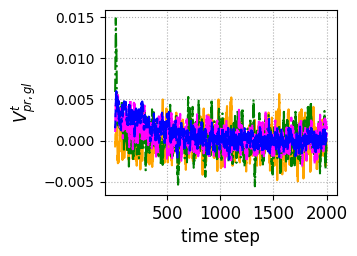}
    \caption{$V_{pr,gl}^{t-10\rightarrow t}$}
    \label{fig:rob_prag_goalvalue_c10}
        \end{minipage}
\centering
    \begin{minipage}[b]{0.32\textwidth}
        \includegraphics[scale = 0.65]{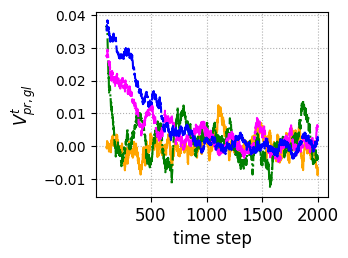}
    \caption{$V_{pr,gl}^{t-100\rightarrow t}$}
    \label{fig:rob_prag_goalvalue_c100}
        \end{minipage}
\centering
    \begin{minipage}[b]{0.32\textwidth}
        \includegraphics[scale = 0.65]{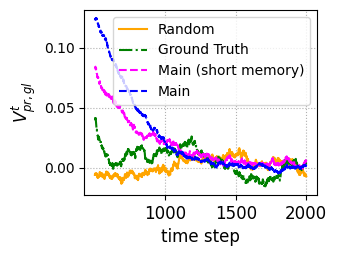}
    \caption{$V_{pr,gl}^{t-500\rightarrow t}$}
    \label{fig:rob_prag_goalvalue_c500}
        \end{minipage}
\centering
    \begin{minipage}[b]{0.32\textwidth}
        \includegraphics[scale = 0.65]{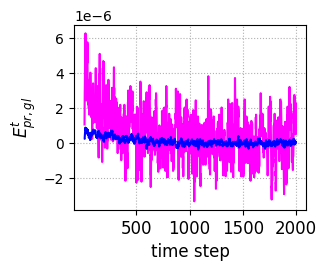}
    \caption{$E_{pr,gl}^{t-10\rightarrow t}$}
    \label{fig:rob_prag_eff_c10}
        \end{minipage}
\centering
    \begin{minipage}[b]{0.32\textwidth}
        \includegraphics[scale = 0.65]{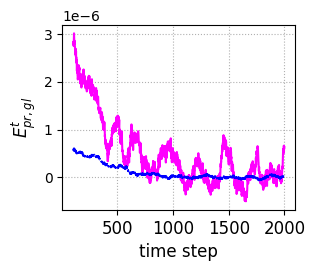}
    \caption{$E_{pr,gl}^{t-100\rightarrow t}$}
    \label{fig:rob_prag_eff_c100}
        \end{minipage}
\centering
     \begin{minipage}[b]{0.32\textwidth}
        \includegraphics[scale = 0.65]{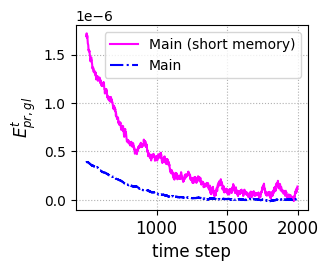}
            \caption{$E_{pr,gl}^{t-500\rightarrow t}$}
            \label{fig:rob_prag_eff_c500}
        \end{minipage}
\setcounter{subfigure}{-1}
    \caption{
    Pragmatic goal values (eq.~\ref{eq:robotic_pragmaticgoalvalue}), and efficiencies (eq.~\ref{eq:robotic_pragmaticgoal_efficiency}) of the various strategies.}
    \label{fig:rob_parg}
\end{subfigure}

\subsection{Collective Decision-Making}

\subsubsection{Model description}

The CD model is partially based on the work presented in \cite{difelice2022agent}. It is a simplified representation of decision-making in small, consensus-based groups ($N<30$, where $N$ is the number of agents), where agents react to their environment to avoid collapse. Agents estimate the size of two tasks, $A$ and $B$. The environment is modeled as a polarized grid of $58$x$58$ information sources, each showing task $A$ or task $B$, with the sum of $A$ and $B$ remaining fixed. The size of each task is measured by the weight of $A$ $W_{A} \in [0.0,1.0]$, determining the share of total information sources showing task $A$. The feedback cycle includes two abstraction steps. First, agents scan $R$ information sources around them, each showing either $0$ (task $B$) or $1$ (task $A$) and average them out to form an individual opinion $O_{i} \in [0.00,1.00]$ of $W_{A}$, with a memory component between subsequent feedback cycles. Second, agents follow a consensus process to reach a collective opinion $O_{coll} \in [0.00,1.00]$, representing the average of $O_i$. The maximum opinion divergence between agents and the number of agents $N$ determine the consensus time $T_{CN}$. For full details on the decision-making algorithm, see the Supplementary Material. 

After $T_{CN}$, $O_{coll}$ generates feedback on the environment $W_{A}$, which grows at each time step until the next feedback cycle if $O_{coll} < W_A$, and shrinks if $O_{coll} > W_A$. If $W_A$ reaches either $0.10$ or $0.90$, the environment collapses. 

\subsubsection{Comparison strategies}
We refer to the decision-making strategy described above as the \textit{consensus} strategy. To evaluate it, we implement three reference strategies, randomizing different abstraction steps -- 
opinion formation, consensus formation, or both: 

\begin{itemize}
    \item The \textit{\textbf{random\textsubscript{OP}}} strategy randomizes opinion formation: agents form a random $O_{i}{[0.00,1.00]}$ without perceiving the environment, within one time-step. They then form $O_{coll}$ normally, reaching consensus among those random $O_i$. This reflects the extreme situation where opinions are completely disconnected from the environment;
    \item The \textit{\textbf{random\textsubscript{CN}}} strategy randomizes consensus: opinions are formed normally, and a random $O_{i}$ is chosen as $O_{coll}$, with the consensus process taking one time step. This reflects the extreme situation where one agent, who is not better informed than others, holds an opinion with maximum power;
    \item The \textit{\textbf{random\textsubscript{TOT}}} strategy randomizes $O_{coll}$, taking one time-step.
\end{itemize}

Comparing the \textit{consensus} strategy with \textit{random\textsubscript{OP}} highlights the value of evidence-based opinion-formation; comparing with \textit{random\textsubscript{CN}} shows the advantage of the consensus process once opinions have been formed; and comparing with \textit{random\textsubscript{TOT}} evaluates the usefulness of the entire feedback cycle.

\subsubsection{\label{subsec:collective_experiments}Experiments}

Parameters $N$ and $R$ are both varied between $[1,30]$; the starting $W_{A}$ is set at $0.6$; each simulation is capped at $5000$ timesteps; and $40$ simulations are ran for each $N$x$R$ combination, leading to $36000$ simulations per strategy, and a total of $144000$ simulations. For each simulation, agents are placed randomly on the inner $50$x$50$ information grid at $t=0$, with their position not changing, and the environment $W_{A}$ changing at each time step. Multiple feedback cycles are executed sequentially until the environment collapses, or the maximum time is reached.

\stepcounter{figure}
\begin{subfigure}
    \centering
    \begin{minipage}[b]{0.23\textwidth}
        \includegraphics[scale=0.27]{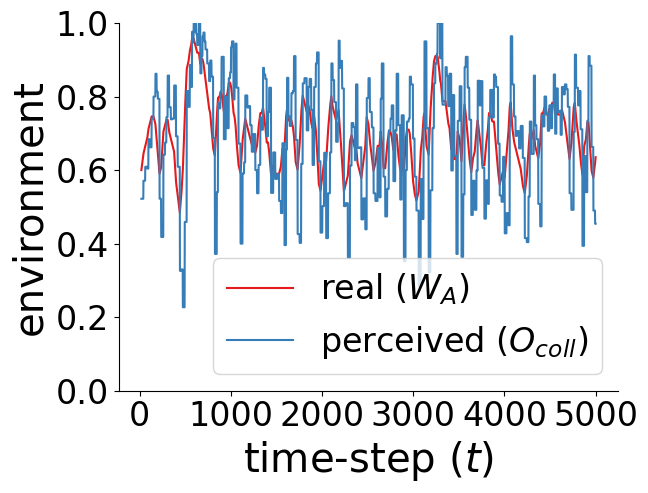}
        \caption{\textit{consensus} strategy, $N$=$R$=$5$}
        \label{subfig:gen_5_5}
    \end{minipage}
    \hfill
    \begin{minipage}[b]{0.23\textwidth}
        \includegraphics[scale=0.27]{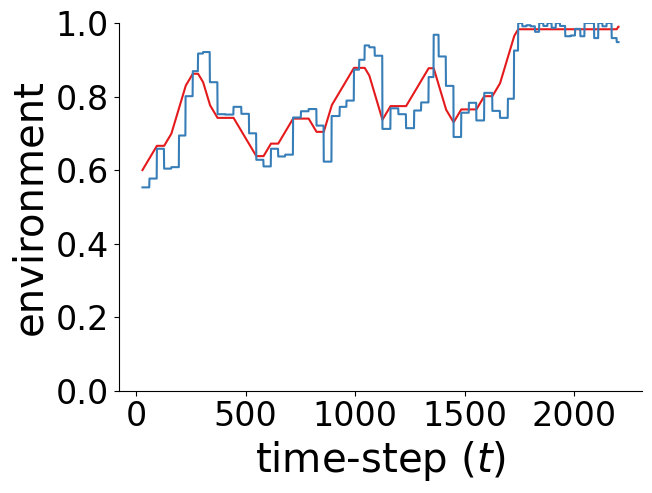}
        \caption{\textit{consensus} strategy, $N$=$R$=$10$}
        \label{subfig:gen_10_10}
    \end{minipage}
    \hfill
    \begin{minipage}[b]{0.23\textwidth}
        \includegraphics[scale=0.27]{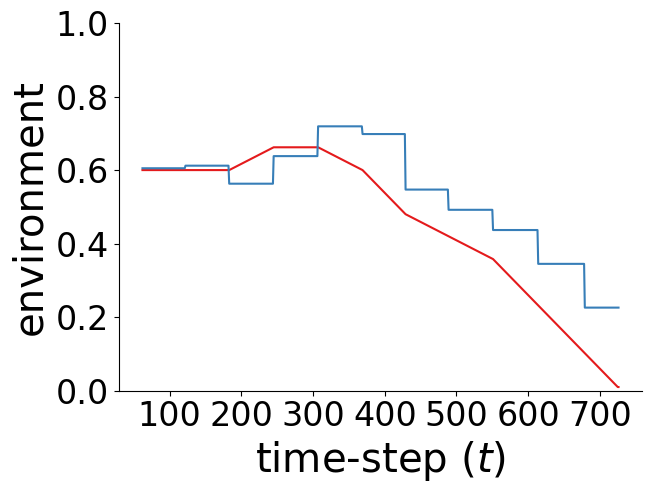}
        \caption{\textit{consensus} strategy, $N$=$R$=$25$}
        \label{subfig:gen_25_25}
    \end{minipage}
    \hfill
    \begin{minipage}[b]{0.23\textwidth}
        \includegraphics[scale=0.27]{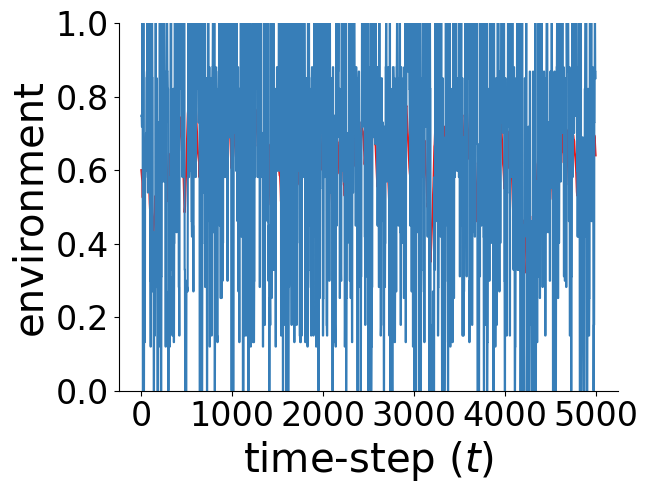}
        \caption{\textit{random\textsubscript{CN}} strategy, $N$=$R$=$5$}
        \label{subfig:rand2_5_5}
    \end{minipage}
    \hfill
   \begin{minipage}[b]{0.23\textwidth}
        \includegraphics[scale=0.27]{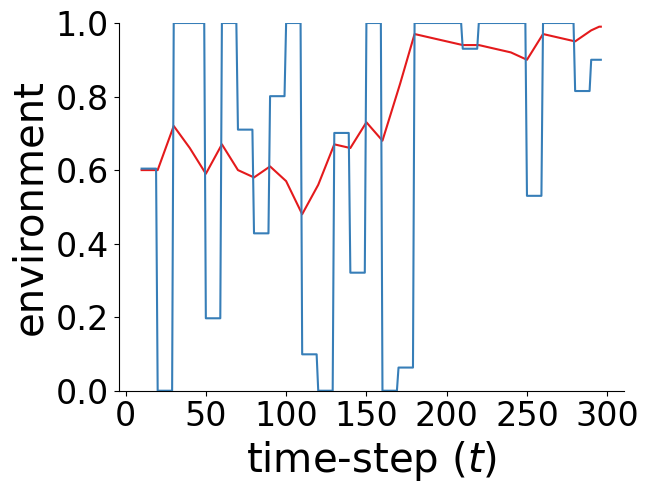}
        \caption{\textit{random\textsubscript{CN}} strategy, $N$=$R$=$10$}
        \label{subfig:rand2_10_10}
    \end{minipage}
    \hfill
    \begin{minipage}[b]{0.23\textwidth}
        \includegraphics[scale=0.27]{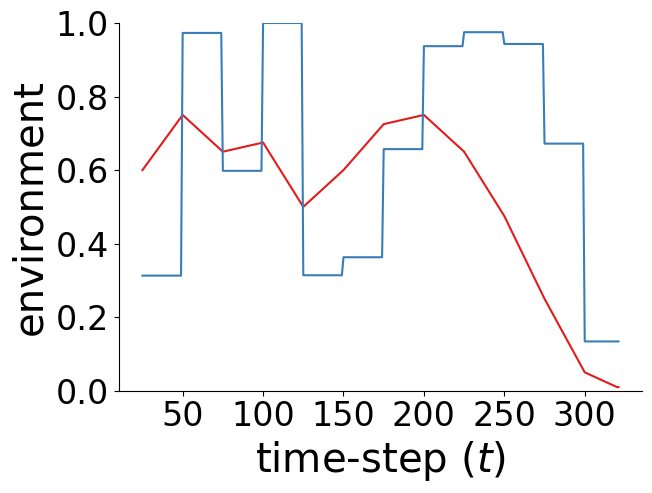}
        \caption{\textit{random\textsubscript{CN}} strategy, $N$=$R$=$25$}
        \label{subfig:rand2_25_25}
    \end{minipage}
    \hfill
    \begin{minipage}[b]{0.23\textwidth}
        \includegraphics[scale=0.27]{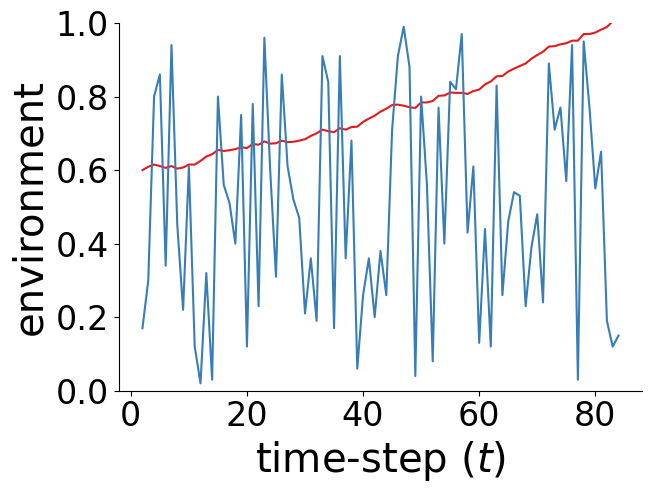}
        \caption{\textit{random\textsubscript{TOT}} strategy, $N$=$R$=$10$}
        \label{subfig:rand1_10_10}
    \end{minipage}
    \hfill
    \begin{minipage}[b]{0.23\textwidth}
        \includegraphics[scale=0.27]{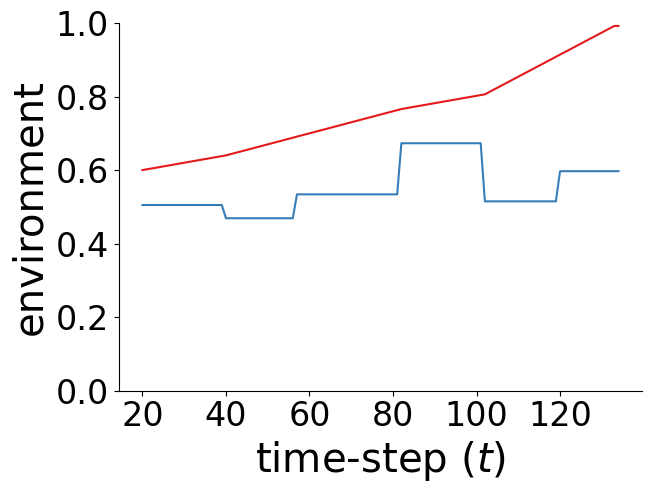}
        \caption{\textit{random\textsubscript{OP}} strategy, $N$=$R$=$10$}
        \label{subfig:rand3_10_10}
    \end{minipage}
    \hfill

\setcounter{subfigure}{-1}
    \caption{Examples of individual simulations, showing the evolution of $W_{A}$ and $O_{coll}$ at each time step $t$. The pink line shows the real environment ($W_{A}$); the blue line shows the perceived environment ($O_{coll}$).}
    \label{fig:individual_runs}
\end{subfigure}

\stepcounter{figure}
\begin{subfigure}
    \centering
    \begin{minipage}[b]{0.32\textwidth}
        \includegraphics[scale=0.32]{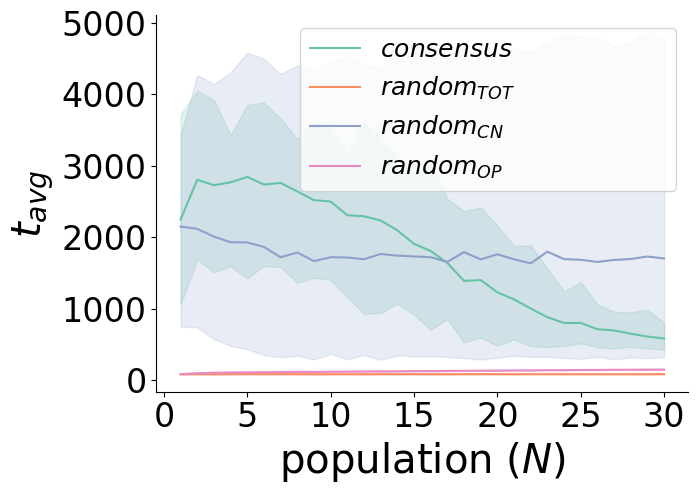}
        \caption{Average simulation length ($t_{avg}$) as a function of $N$ for all strategies}
        \label{subfig:general_beh_N}
    \end{minipage}
    \hfill
    \begin{minipage}[b]{0.32\textwidth}
        \includegraphics[scale=0.32]{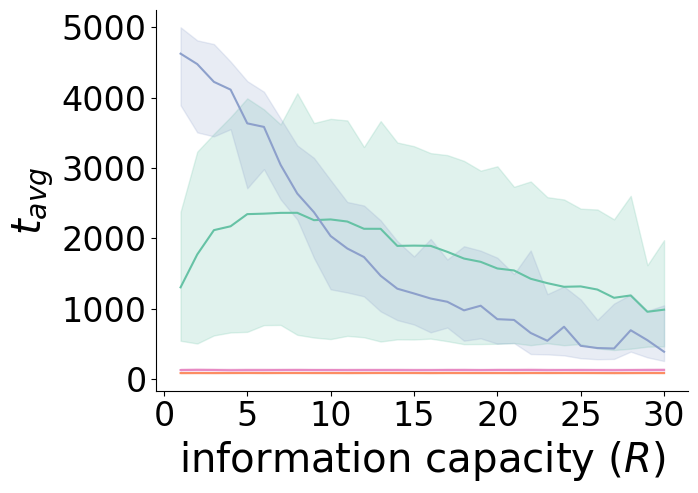}
        \caption{Average simulation length ($t_{avg}$) as a function of $R$ for all strategies}
        \label{subfig:general_beh_S}
    \end{minipage}
    \hfill
    \begin{minipage}[b]{0.32\textwidth}
        \includegraphics[scale=0.32]{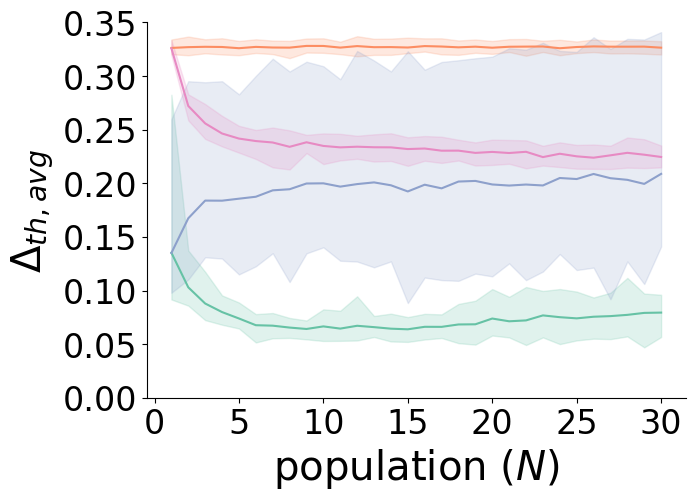}
        \caption{Average delta of truth ($\Delta_{th,avg}$) as a function of $N$ for all strategies}
        \label{subfig:perception_delta_N}
    \end{minipage}
    \hfill
    \begin{minipage}[b]{0.32\textwidth}
        \includegraphics[scale=0.32]{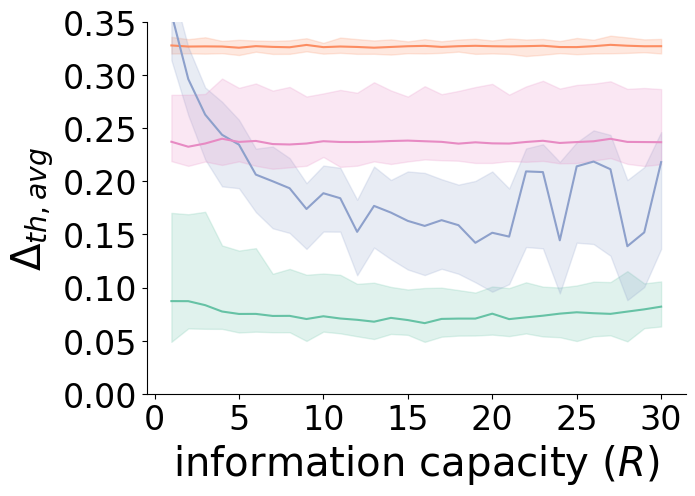}
        \caption{Average delta of truth ($\Delta_{th,avg}$) as a function of $R$ for all strategies}
        \label{subfig:perception_delta_S}
    \end{minipage}
    \hfill
   \begin{minipage}[b]{0.32\textwidth}
        \includegraphics[scale=0.2]{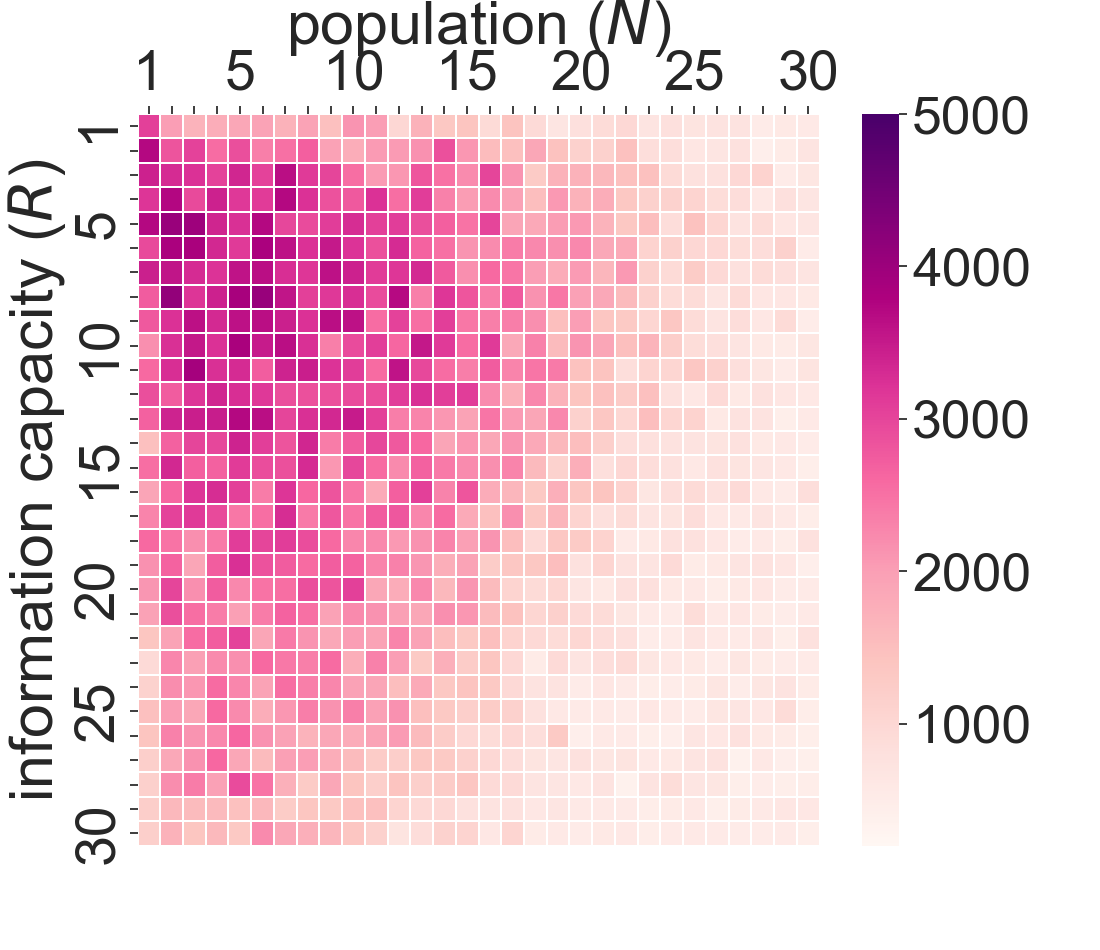}
        \caption{Average simulation length ($t_{avg}$) as a function of $N$ and $R$ for the \textit{consensus} strategy}
        \label{subfig:heat_general}
    \end{minipage}
    \hfill
    \begin{minipage}[b]{0.32\textwidth}
        \includegraphics[scale=0.2]{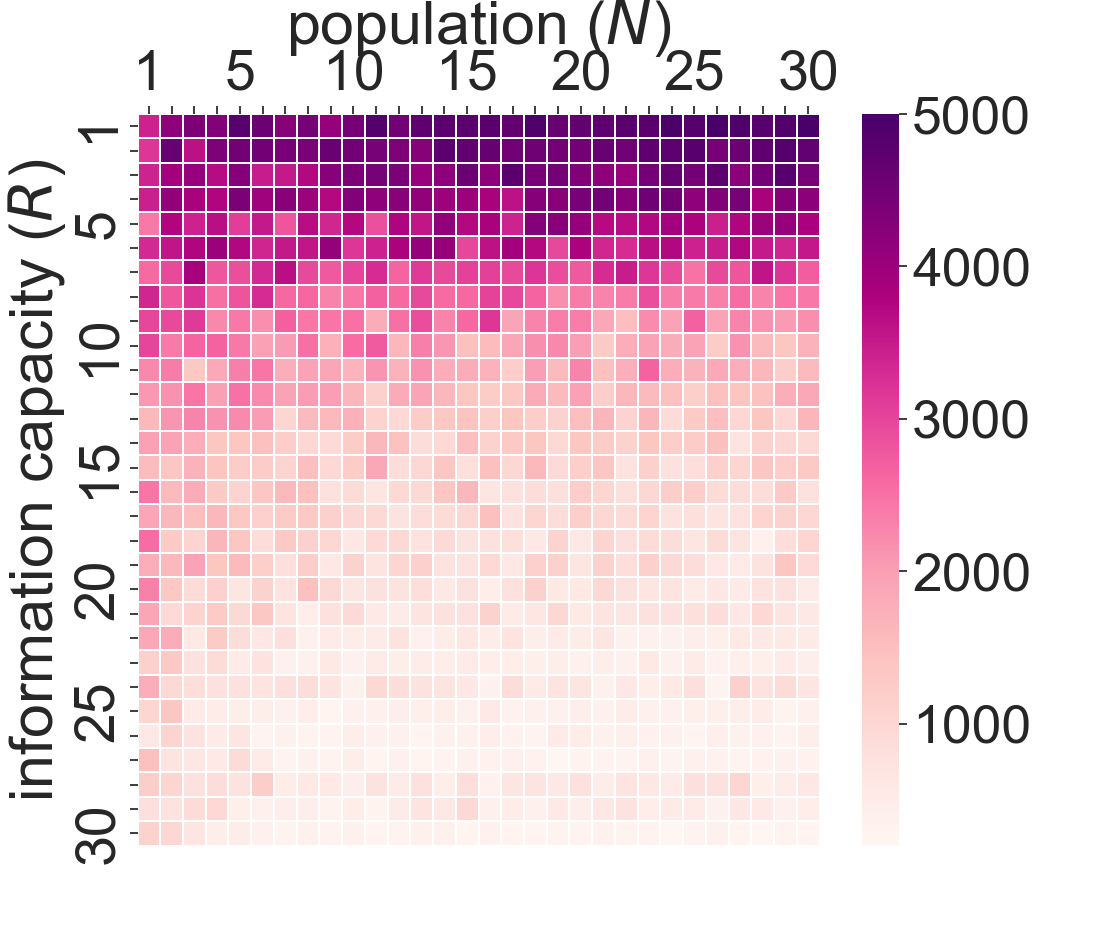}
        \caption{Average simulation length ($t_{avg}$) as a function of $N$ and $R$ for the \textit{random\textsubscript{CN} strategy}}
        \label{subfig:heat_random2}
    \end{minipage}
    \hfill

\setcounter{subfigure}{-1}
    \caption{Comparison of average behavior across strategies.}
    \label{fig:general_behaviour_coll}
\end{subfigure}

Figure \ref{fig:individual_runs} shows examples of the main types of behaviors that can be observed in each simulation. Figure \ref{fig:general_behaviour_coll} shows average behaviors over the $36000$ simulations for each strategy. Long-lasting simulations (with $t_{avg}$ = $5000$ timesteps) show two patterns. In the first one, $O_{coll}$ follows $W_{A}$, with some delay, leading to an oscillation of both values (Figure \ref{subfig:gen_5_5}). The general strategy behaves this way when $N<15$ and $3<R<15$} (see the darker area in Figure \ref{subfig:heat_general}), surviving for $5000$ timesteps.
In the second pattern, $O_{coll}$ rapidly oscillates between polarized values, forcing $W_{A}$ to remain stable with tight oscillations around its starting point of $0.6$ (Figure \ref{subfig:rand2_5_5}). This is what is observed in the high performance area of \textit{random\textsubscript{CN}} (Figure  \ref{subfig:heat_random2}). Here, $T_{CN}$ is reduced to one time step, the $O_{i}$ randomly chosen as $O_{coll}$ is either $0$ or $1$ for $R$=$1$, and likely to remain so for low $R$, leading to a very fast feedback cycle with drastic fluctuations in $O_{coll}$. In other instances, $W_{A}$ and $O_{coll}$ both enter an oscillatory behavior but $W_{A}$ crashes before $5000$ $t$ (Figure \ref{subfig:gen_10_10}). This can happen in the \textit{consensus} strategy when a slower feedback cycle cannot keep up with changes in $W_{A}$. Finally, if $W_{A}$ does not oscillate, it crashes before $t$ reaches $400$ timesteps (Figures \ref{subfig:rand2_10_10} - \ref{subfig:rand3_10_10}); this is what is observed in both \textit{random\textsubscript{TOT}} and \textit{random\textsubscript{OP}}. 

Figures \ref{subfig:perception_delta_N} and \ref{subfig:perception_delta_S} show the average delta of truth $\Delta_{th, avg}$ -- i.e., the average difference between $O_{coll}$ and $W_{A}$ -- for each parameter combination. A lower $\Delta_{th}$ shows that agents are better at perceiving the environment, collectively. The relation between $\Delta_{th}$ and $t_{avg}$ highlights the weight of the timing mechanisms in the model. The \textit{consensus} strategy has the lowest $\Delta_{th,avg}$, and it decreases and stabilizes as $N$ increases (Figure \ref{subfig:perception_delta_N}): poor performance at high $N$ is not due to incorrect $O_{coll}$, but to a long feedback cycle that prevents $W_A$ oscillations. 
For \textit{random\textsubscript{CN}}, $\Delta_{th,avg}$ is highest for low $R$, even higher than \textit{random\textsubscript{TOT}} and \textit{random\textsubscript{OP}} (Figure \ref{subfig:perception_delta_S}). In this range, high $t_{avg}$ is not the result of a good $O_{coll}$, but of an oscillating $O_{coll}$ that stabilizes $W_{A}$.

\subsubsection{\label{subsec:decision_info}Information measures}

\paragraph{Calculations}

We consider $O_{coll}$ as the knowledge for the semantic measures (which could also be calculated at the scale of individual agents' opinions $O_{i}$). For the pragmatic measures, the action taken by the system is the change in the environment $W_{A}$, and we consider the system goal to be furthest from the points of collapse $W_{A} = 0.10$ or $0.90$. Agents are unaware of how far they are from collapse, aiming instead for their local goal of perceiving their immediate environment. 
Information values are calculated at the granularity of single feedback cycles, comparing the system at time $t$ with $t$-$c$, where $c$ is the number of time steps in the feedback cycle ending at $t$. Values are averaged across feedback cycles within a simulation, and then across simulations within each set of $N$x$R$ parameters. Averages are weighted by the number of feedback cycles in each simulation. 

\subparagraph{Syntactic information}

We calculate the amount of resources needed to store information at the meso-scale $S_{1}$ of agent's individual opinions $O_{i}$ and at the macro-scale $S_{2}$ of the group's collective opinion $O_{coll}$. We do not consider the amount of information in the environment at $S_{0}$ since this is the same for all strategies. We employ Shannon entropy as a measure of resource use at each scale. The syntactic content is summed across scales for the total $C_{syn,cycle}$. Probabilities are simplified by assuming that each information source is equally likely to show one of the two tasks. For the \textit{consensus} strategy, this leads to a Shannon entropy of each agent's individual opinion $H(O_i)$ of:

\begin{equation}
    H(O_{i}) = -\sum_{k} p(O_{i,k}) \log_2 p(O_{i,k})
\end{equation}

Where $O_{i, k}$ in $K$ is the set of all possible opinions, and $p(O_{i,k})$ is the probability of an opinion being $O_{i,k}$. For the entropy at the scale $S_{1}$ this is multiplied by the number of agents $N$:

\begin{equation}
    H(S_{1}) = N*H(O_{i})
\end{equation}

Similarly, the entropy of $S_2$ is:

\begin{equation}
    H(S_2) = -\sum_{k} p(O_{coll,k}) \log_2 p(O_{coll,k})
\end{equation}

Where $O_{coll,k}$ represents all possible values of $O_{coll}$, given a set of $O_{i, k}$, and $p(O_{coll,k})$ the probability that $O_{coll}$ is $O_{coll,k}$. This leads to a final $C_{syn,cycle}$ of:

\begin{equation}
    C_{syn,cycle} = H(S_1) + H(S_2)
\end{equation}

Similar calculations are carried out for the three random strategies -- full details are included in the Supplementary Material, as well as graphs showing how $C_{syn,cycle}$ varies with $N$ and $R$ for all strategies. 

\subparagraph{Semantic information}

For the \textbf{semantic delta} $\Delta_{sm}$, the change in $O_{coll}$ between feedback cycles is calculated. This is a proxy of how stable $O_{coll}$ is:

\begin{equation}
    \Delta_{sm}^{t-c \rightarrow t} = \mid O_{coll}^{t-c} - O_{coll}^{t} \mid
\end{equation}

where $O_{coll}^t$ is $O_{coll}$ at the end of the feedback cycle of length $c$, and $O_{coll}^{t-c}$ at the end of the previous feedback cycle. $\Delta_{sm} \in [0.00,1.00]$, with a larger value reflecting higher change in $O_{coll}$ across feedback cycles. The \textbf{semantic truth value} $V_{sm,th}$ measures whether the information used at each feedback cycle has brought $O_{coll}$ closer or further away from the environment $W_{A}$, with respect to the $O_{coll}$ of the previous cycle. First, the \textit{delta of truth} $\Delta_{th}$ is calculated:

\begin{equation}
    \Delta_{th}^{t} = \mid O_{coll}^{t} - W_{A}^{t} \mid 
\end{equation}

Then, $\Delta_{th}^{t}$ is subtracted across feedback cycles to calculate $V_{sm,th} \in [-1,1]$:

\begin{equation}
    V_{sm,th}^{t-c \rightarrow t} = \Delta_{th}^{t-c} - \Delta_{th}^{t}
\end{equation}

$V_{sm,th}$ is positive when $\Delta_{th}$ decreases between cycles, and negative if it increases. Its efficiency is calculated by dividing the value by $C_{syn,cycle}$.

\subparagraph{Pragmatic information}

The \textbf{pragmatic delta} $\Delta_{pr} \in [0,1]$ measures changes in $W_{A}$ between feedback cycles:

\begin{equation}
    \Delta_{pr}^{t-c \rightarrow t} = \mid W_{A}^{t-c} - W_{A}^{t} \mid
\end{equation}

It is a proxy of how stable the environment is, with a higher value reflecting larger environmental change. The \textbf{pragmatic goal value} $V_{pr,gl}$ measures whether the system has moved closer or further away from the goal between a feedback cycle and the next. For this, the \textit{delta of goal} $\Delta_{gl} \in [0,1]$ is calculated at each $t$, with higher values for system states that are further from the point of collapse:

\begin{equation}
    \Delta_{gl}^t = 1 - 2*\mid 0.5 - W_{A}^t \mid
\end{equation}

Then we calculate $V_{pr,gl}^{t-c \rightarrow t} \in [-1,1]$ as:

\begin{equation}
    V_{pr,gl}^{t-c \rightarrow t} = \Delta_{gl}^{t-c} - \Delta_{gl}^{t}
\end{equation}

Similar to $V_{sm, th}$, $V_{pr,gl}$ takes negative values when the system moves closer to collapse relative to the previous feedback cycle, and vice versa. 


\paragraph{Results}

\stepcounter{figure}
\begin{subfigure}
\centering
    \begin{minipage}[b]{0.23\textwidth}
        \includegraphics[scale=0.16]{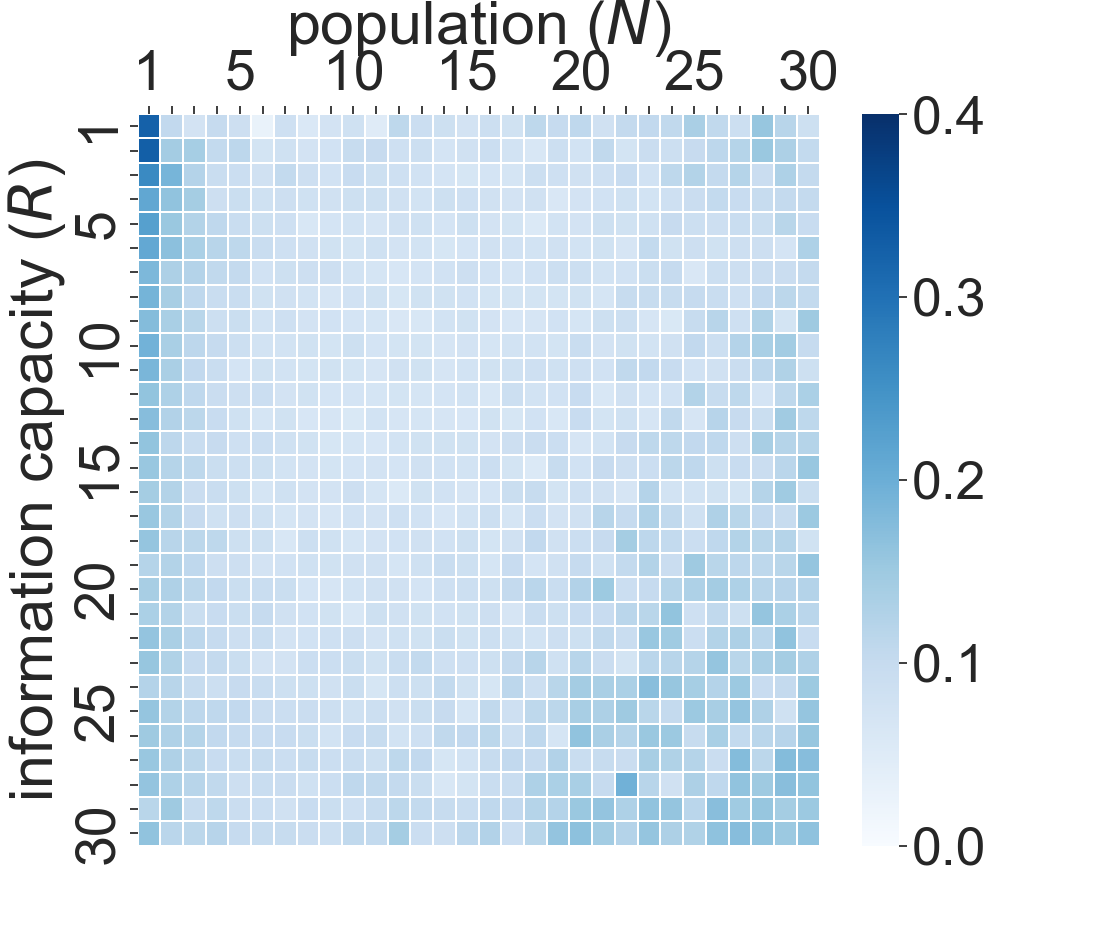}
        \caption{Semantic delta $\Delta_{sm}$ of the \textit{consensus} strategy}
        \label{subfig:sem_delta_gen}
        \end{minipage}
        \hfill
\centering
     \begin{minipage}[b]{0.23\textwidth}
        \includegraphics[scale=0.16]{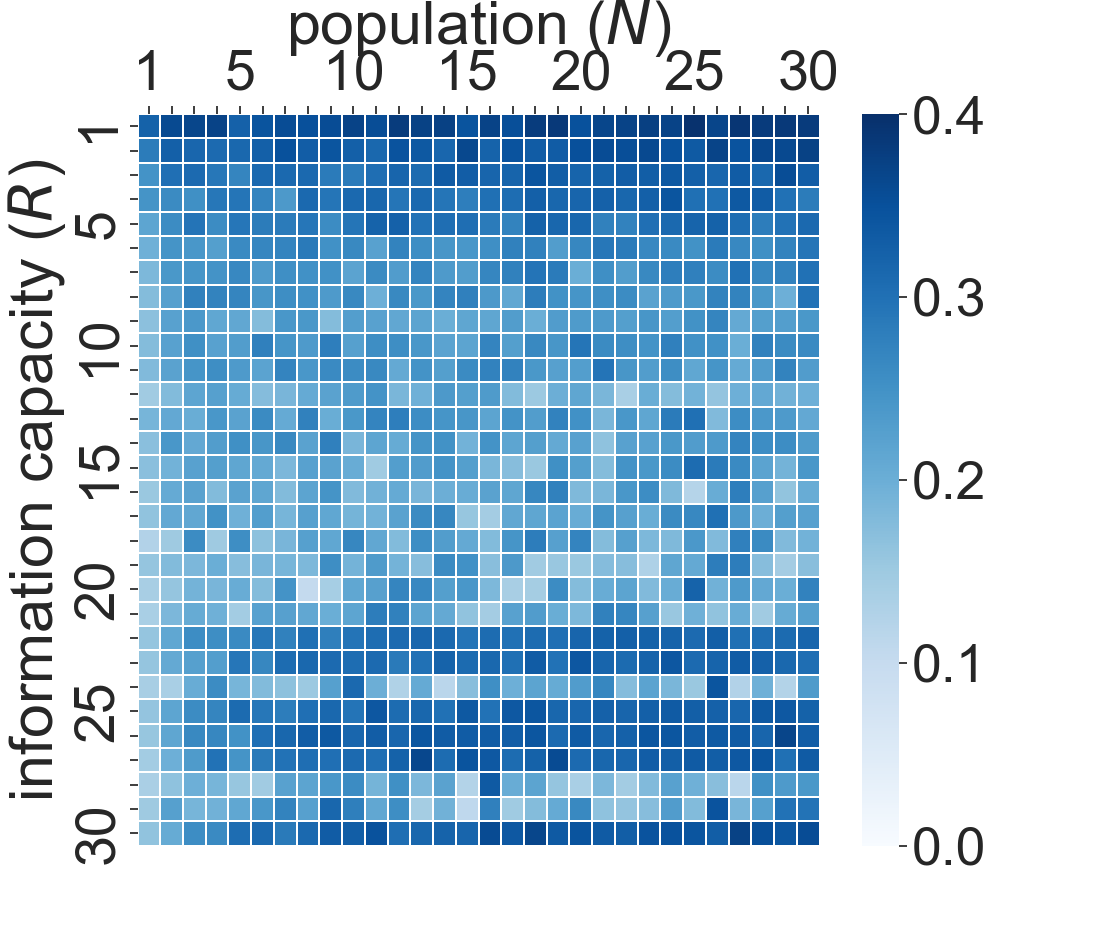}
        \caption{Semantic delta $\Delta_{sm}$ of the \textit{random\textsubscript{CN}} strategy}
        \label{subfig:sem_delta_rand2}
        \end{minipage}
        \hfill
\centering
     \begin{minipage}[b]{0.23\textwidth}
        \includegraphics[scale=0.16]{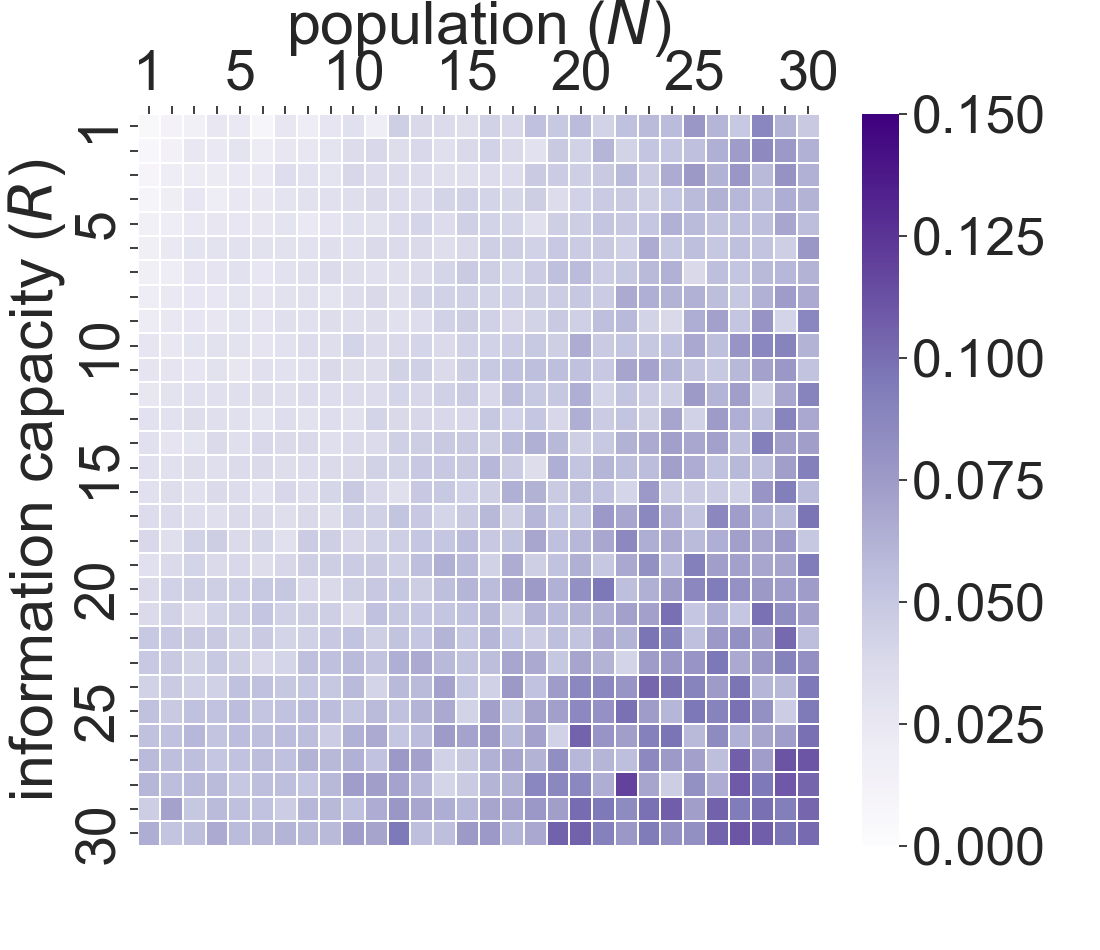}
        \caption{Pragmatic delta $\Delta_{pr}$ of the \textit{consensus} strategy}
        \label{subfig:prag_delta_gen}
        \end{minipage}
        \hfill
\centering
    \begin{minipage}[b]{0.23\textwidth}
        \includegraphics[scale=0.16]{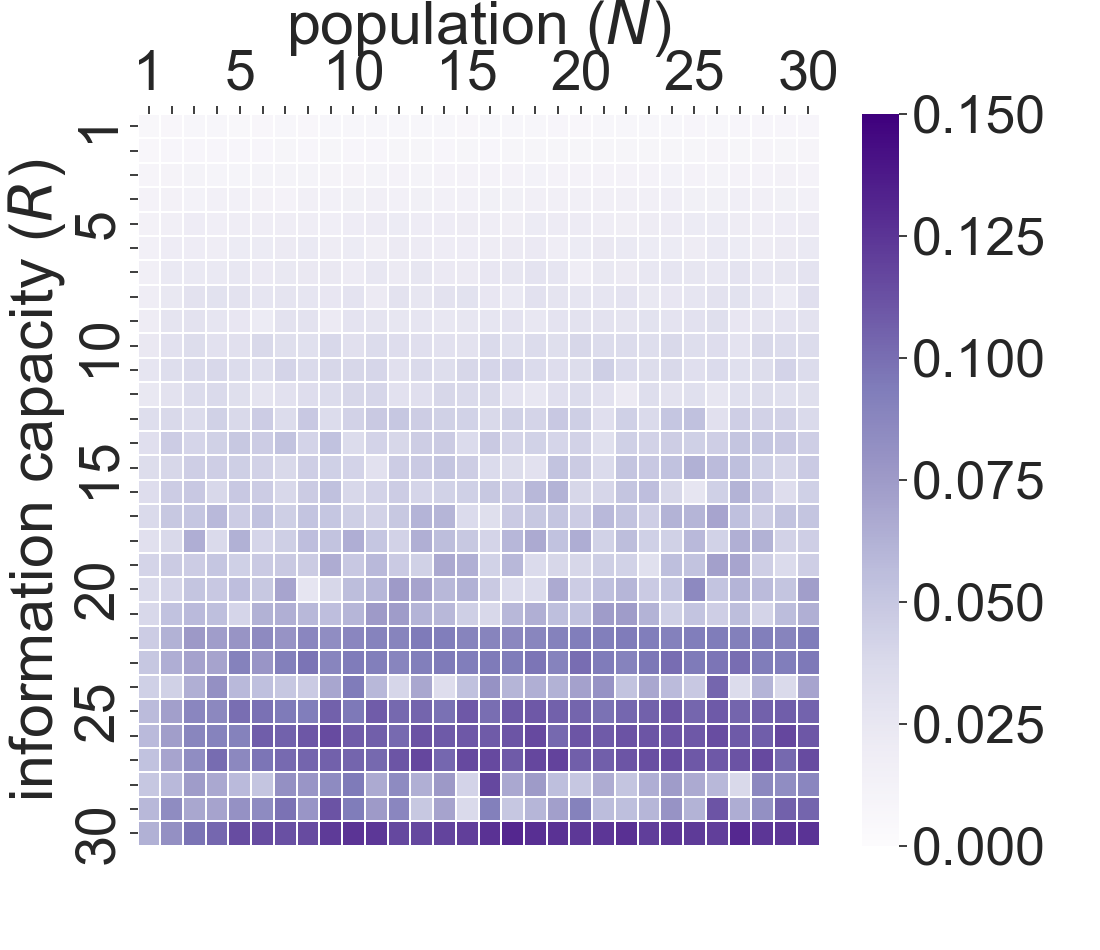}
        \caption{Pragmatic delta $\Delta_{pr}$ of the \textit{random\textsubscript{CN}} strategy}
        \label{subfig:prag_delta_rand2}
        \end{minipage}
        \hfill
\centering
     \begin{minipage}[b]{0.32\textwidth}
        \includegraphics[scale=0.32]{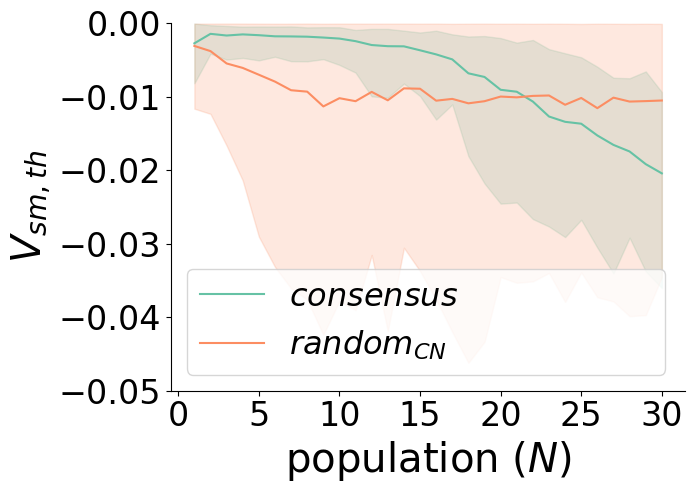}
        \caption{Semantic truth value $V_{sm,th}$ for the \textit{consensus} and \textit{random\textsubscript{CN}} strategies, as a function of $N$}
        \label{subfig:semantic_truth_pop_lim}
        \end{minipage}
        \hfill
\centering
    \begin{minipage}[b]{0.32\textwidth}
        \includegraphics[scale=0.32]{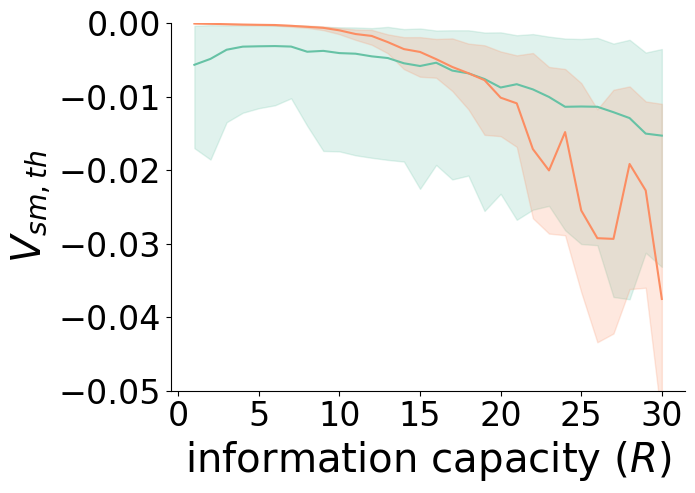}
        \caption{Semantic truth value $V_{sm,th}$ for the \textit{consensus} and \textit{random\textsubscript{CN}} strategies, as a function of $R$}
        \label{subfig:semantic_truth_info_lim}
        \end{minipage}
        \hfill
\centering
    \begin{minipage}[b]{0.27\textwidth}
        \includegraphics[scale=0.32]{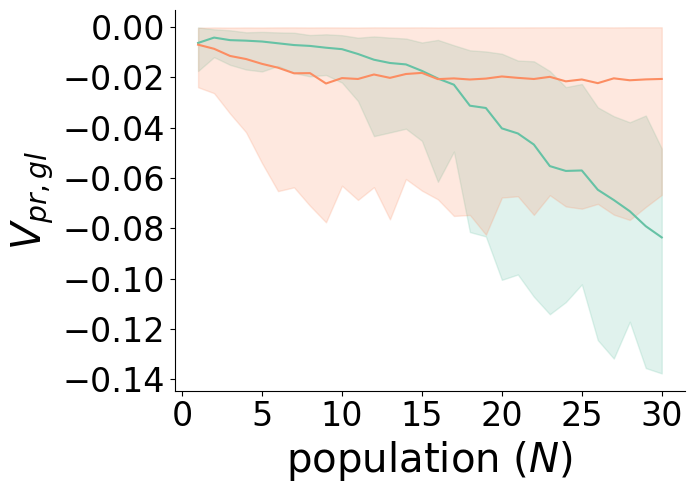}
        \caption{Pragmatic goal value $V_{pr,gl}$ for the \textit{consensus} and \textit{random\textsubscript{CN}} strategies, as a function of $N$}
        \label{subfig:pragmatic_goal_pop_lim}
        \end{minipage}
        \hfill
\centering
    \begin{minipage}[b]{0.27\textwidth}
        \includegraphics[scale=0.32]{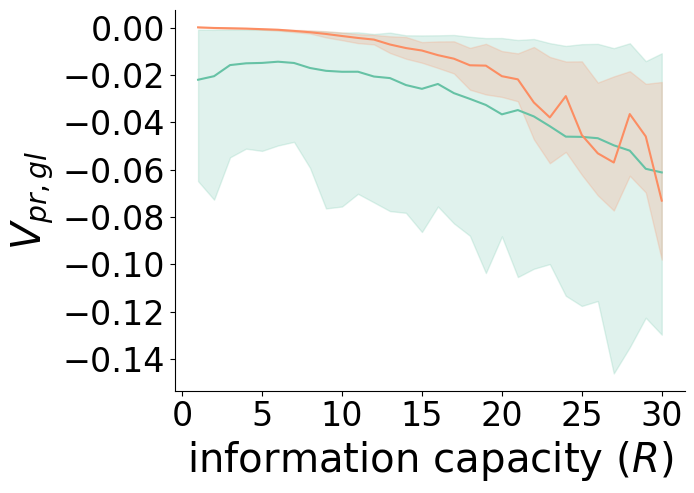}
        \caption{Pragmatic goal value $V_{pr,gl}$ for the \textit{consensus} and \textit{random\textsubscript{CN}} strategies, as a function of $R$}
        \label{subfig:pragmatic_goal_info_lim}
        \end{minipage}
        \hfill
\centering
    \begin{minipage}[b]{0.27\textwidth}
        \includegraphics[scale=0.32]{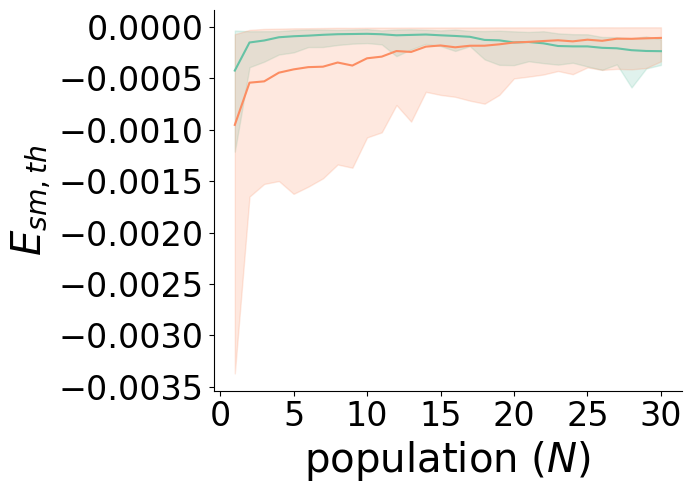}
        \caption{Efficiency of $V_{sm,th}$ for the \textit{consensus} and \textit{random\textsubscript{CN}} strategies, as a function of $N$}
        \label{subfig:semantic_truth_eff_pop_lim}
        \end{minipage}
        \hfill
\centering
    \begin{minipage}[b]{0.27\textwidth}
        \includegraphics[scale=0.32]{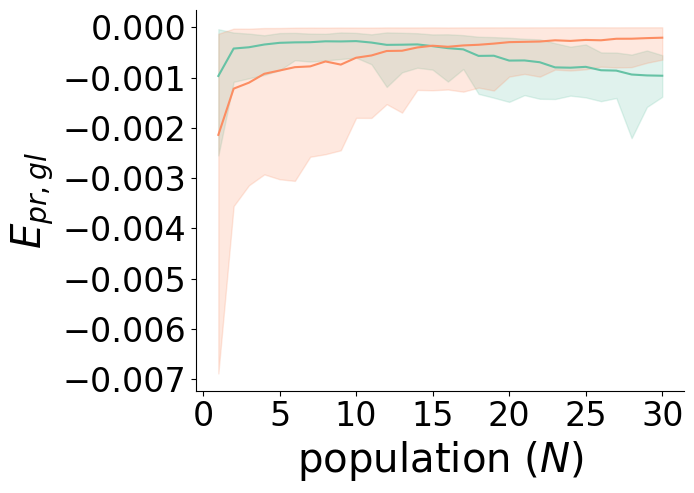}
        \caption{Efficiency of $V_{pr,gl}$ for the \textit{consensus} and \textit{random\textsubscript{CN}} strategies, as a function of $N$}
        \label{subfig:pragmatic_goal_eff_pop_lim}
        \end{minipage}
        \hfill

\setcounter{subfigure}{-1}
    \caption{Semantic and pragmatic information measures for the collective decision-making case study.}
    \label{fig:sem_prag_info_collective}
\end{subfigure}

Figure \ref{fig:sem_prag_info_collective} shows the average information measures as a function of $N$ and $R$. We focus on the behavior of the \textit{consensus} and \textit{random\textsubscript{CN}} strategies, that are comparable in terms of $t_{avg}$. $\Delta_{sm}$ and $\Delta_{pr}$ show the evolution of variations in $O_{coll}$ and $W_{A}$, respectively. For the \textit{consensus} strategy, there is an overall correlation between $t_{avg}$ (Figure \ref{subfig:heat_general}), $\Delta_{sm}$ and $\Delta_{pr}$: higher fluctuations in $W_{A}$ are associated with a lower $t_{avg}$, and caused by higher fluctuations in $O_{coll}$. There is an exception for $N$=$1$ (Figure \ref{subfig:sem_delta_gen}), where $O_{coll}$ oscillates between $0$ and $1$, leading to a stable environment (Figure \ref{subfig:prag_delta_gen}). For \textit{random\textsubscript{CN}}, $\Delta_{sm}$ is considerably higher ($O_{coll}$ is unstable), and there are two possible relations between $\Delta_{sm}$, $\Delta_{pr}$, and $t_{avg}$. In the area of low $R$, $\Delta_{sm}$ is high, $\Delta_{pr}$ is low, and $t_{avg}$ is high: $O_{coll}$ fluctuates strongly between a feedback cycle and the next, forcing the environment to stabilize. From $R=23$ a different pattern emerges, with bands of high $\Delta_{sm}$ (Figure \ref{subfig:sem_delta_rand2}) leading also to a high $\Delta_{pr}$, and to a low $t_{avg}$. In this range \textit{random\textsubscript{CN}} acts like the \textit{consensus} strategy, with changes in $O_{coll}$ coupled to a slower feedback cycle, leading to a faster collapse. The discrepancy in values in the range of $R>23$ for \textit{random\textsubscript{CN}} is likely due to noisy results as a consequence of the limited sample size, and needs further exploration. 
Depending on the system, specific combinations of $\Delta_{sm}$, $\Delta_{pr}$ and $t_{avg}$ may be more or less desirable. E.g., an $O_{coll}$ that changes in line with $W_{A}$, leading to stable oscillations (such as when the \textit{consensus} strategy performs well), may be more desirable than a highly varying $O_{coll}$ that flattens changes in $W_{A}$, but that may quickly lead to collapse if conditions change.

$V_{sm,th}$ and $V_{pr,gl}$ show whether each feedback cycle on average brings $O_{coll}$ closer or further away from $W_{A}$, and whether $W_{A}$ moves closer or further away from a desirable state between subsequent cycles. Values are almost entirely negative for both \textit{consensus} and \textit{random\textsubscript{CN}} (Figures \ref{subfig:semantic_truth_pop_lim} - \ref{subfig:pragmatic_goal_info_lim}), meaning that each feedback cycle performs on average worse than the previous one.
The exception is for low $R$ in \textit{random\textsubscript{CN}}: in this case, strong fluctuations of $O_{coll}$ keep $W_{A}$ stable. Stability can also be considered across parameter ranges: while \textit{random\textsubscript{CN}} leads to a stable $W_{A}$ for low $R$, increasing the range of $R$ quickly leads to instability, while parameter transitions are smoother for the \textit{consensus} strategy. 
Efficiencies (\ref{subfig:semantic_truth_eff_pop_lim} and \ref{subfig:pragmatic_goal_eff_pop_lim}) 
show how for both strategies (and more strongly for \textit{random\textsubscript{CN}}), at low $N$, less information is needed to produce larger (negative) values. 
Overall, comparison with \textit{random\textsubscript{CN}} provides insights into the value of the consensus process as formalized in this model, with its polarized environment, opinion formation and consensus algorithms. While for low $N$ it might be better to pick a random agents' opinion as the prevailing one in terms of $t_{avg}$, the \textit{consensus} strategy maintains more stability in $O_{coll}$, and a higher $V_{sm,th}$ of the average feedback cycle. The comparison with \textit{random\textsubscript{TOT}} and \textit{random\textsubscript{OP}}, while not explored in detail here, shows that the opinion formation step is essential to the functioning of the model. 

\subsection{Task Distribution}
\label{subsec:t-d}

\subsubsection{Model Description}
\label{subsubsec:t-d:model-descr}

This model is a simplified version of the ABM in \cite{diaconescu2021exogenous}, aiming to achieve a given task distribution among a set of agents. We exemplify a small system  to facilitate the formal analysis of information flows. The system contains $N=3$ scales $S_s$, $s=0..2$, populated by agents $a_{s,j}$, at scale $S_s$, index $j$. $S_0$ includes four workers ($|S_0|=4$) that must select two task types, $k_0$ and $k_1$ (sometimes noted as $0$ and $1$) according to given proportions: $g_0$, $g_1$, $g_0 + g_1 = |S_0|$.  Workers coordinate their task selection via managers, each handling $C=2$ children (Fig. \ref{fig:cases_task-distr}). At $t_0$, the top-manager $a_{2,0}$ receives the goal $v_g = g_1$, and workers $a_{0,j}$, $j=0..4$, are assigned an active task $k_a \in \{k_0,k_1\}$.  
Time is modeled in discrete steps $t_i$. 
For simplicity, we only consider information about $k_1$, $k_0$ being complementary to it. 

Each manager $a_{s,j}$, $s>0$, holds two variables concerning its children:   the number of workers that execute $k_1$ (abstraction $v_{s,j}^A$) and that should switch to $k_1$ (control-error $v_{s,j}^{\Delta}$).
For workers, $v_{0,j}^A=v_{0,j}^{k_a}$ (their active task) and $v_{0,j}^{\Delta} = v_{1,l}^{\Delta}$ (control from parent $a_{1,l}$).  

The system uses the Blackboard (BB) strategy (simplified from \cite{diaconescu2021exogenous}, see the Supplementary Material for further details), forming a multi-scale feedback cycle via the following information flows: 
\begin{itemize}
    \item \textbf{Abstraction:} managers sum up their children states (active tasks), $v_{s,j}^A = \sum_{c=0..1} v_{s-1,c}^A$
    \item \textbf{Processing:} the top-manager calculates the error between its abstract state and the goal, $v_{2,0}^{\Delta}=v_{2,0}^{A} - v_g$, where negative errors indicate that more workers should pick $k_1$.
    \item \textbf{Reification:} mid-managers fetch the control error $v_{2,0}^{\Delta}$ from the top, split it equally between them and round up the remainder; then send the result to workers $v_{1,j}^{\Delta}$ 
    \item \textbf{Adaptation:} workers that perform $k_a=k_0$ and receive a negative error $v_{1,l}^{\Delta} < 0$ switch to $k_1$ with probability $p_{ch}=0.15$; otherwise they do nothing.    
\end{itemize}

The BB simulation repeats this cycle recursively until the goal is reached, and maintained for $N=3$ steps.

\subsubsection{Comparison Strategies}

We define several alternative strategies for reference: 
\begin{itemize}
    \item \textbf{Random-switch (RS)}: at each step, workers switch  states with probability $p_{ch}=0.15$ (same as BB). 
    \item \textbf{Random-blind (RB)}: at each step, workers switch  states randomly, $p_{ch}=0.5$.  
    \item \textbf{Static (St)}: workers keep their initial states. 
    \item \textbf{Model (Md)}: managers collect detailed information about which worker performs which task and feedback exactly which workers should switch tasks -- hence reaching the goal in $N=3$ steps.   
\end{itemize}

\subsubsection{Experiments}

In the initial state, all workers perform $k_0$ ($v_{S_0}^k=0000$) whereas the goal $v_g=4$ indicates they should perform $k_1$ ($v_{g,S_0}^k=1111$).  
Experiments rely on an analytic simulation to facilitate information tracking.  We employ two experimental alternatives: 
\begin{itemize}
    \item \textbf{Probabilistic analysis}, $Ex_{all}$, for all strategies: considers all possible system behaviors over $50$ steps. We use this to estimate probabilities for the syntactic content and pragmatic values.  
    \item \textbf{Concrete scenario}, $Ex_{scenario}$, for BB and Md: illustrates a particular behavior over $6$ steps (see Fig. \ref{fig:t-d:exo-steps} for BB; Supplementary Material for Md). We use this to track semantic and pragmatic delta measures, step by step. 
\end{itemize}

\stepcounter{figure}
\begin{subfigure}
\begin{minipage}{0.33\textwidth}
\centering
\includegraphics[scale=0.25]{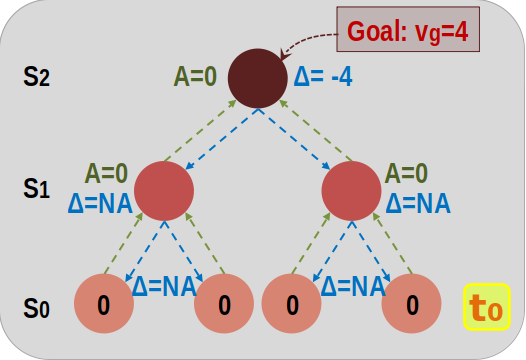}
\caption{$t_0$}
\label{fig:t-d:ex-t0}
\end{minipage}
\begin{minipage}{0.33\textwidth}
\centering
\includegraphics[scale = 0.25]{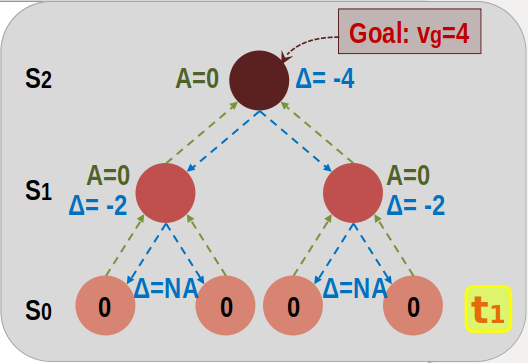}
\caption{$t_1$}
\label{fig:t-d:ex-t1}
\end{minipage}
\begin{minipage}{0.33\textwidth}
\centering
\includegraphics[scale=0.25]{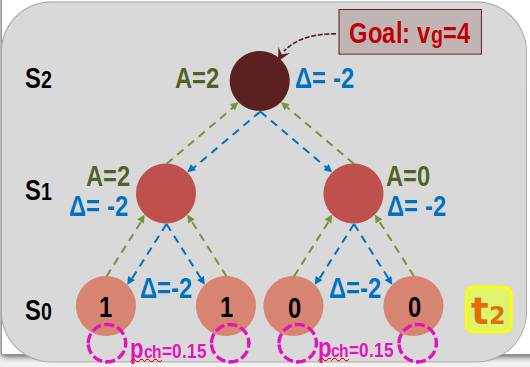}
\caption{$t_2$}
\label{fig:t-d:ex-t2}
\end{minipage}
\newline
\begin{minipage}{0.33\textwidth}
\centering
\includegraphics[scale = 0.25]{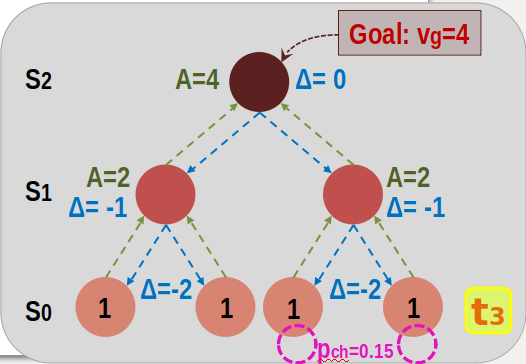}
\caption{$t_3$}
\label{fig:t-d:ex-t3}
\end{minipage}
\begin{minipage}{0.33\textwidth}
\centering
\includegraphics[scale = 0.25]{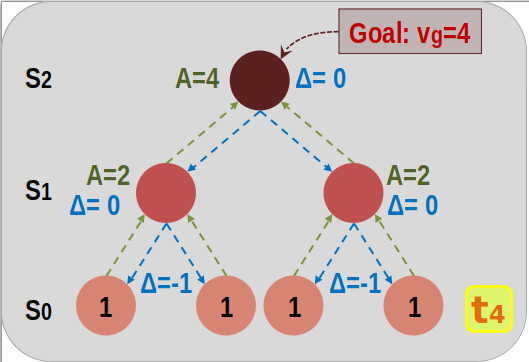}
\caption{$t_4$}
\label{fig:t-d:ex-t4}
\end{minipage}
\begin{minipage}{0.33\textwidth}
\centering
\includegraphics[scale = 0.25]{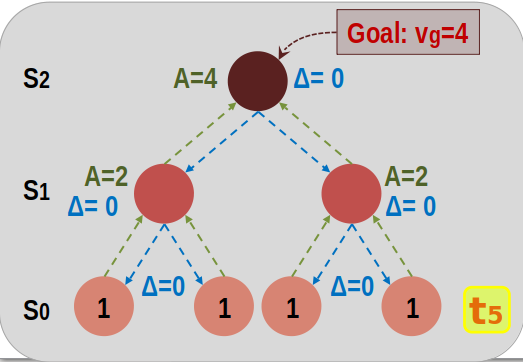}
\caption{$t_5$}
\label{fig:t-d:ex-t5}
\end{minipage}
\setcounter{subfigure}{-1}
\caption{Task-distribution scenario $Ex_{scenario}$ using the Blackboard (BB) strategy, with 4 workers, 2 mid-managers and 1 top-manager; reaching goal $v_{g,S_0}=1111$ from initial state $v_{S_0}^{t_0}=0000$, in $4$ steps}
\label{fig:t-d:exo-steps}
\end{subfigure}

\subsubsection{Information measures}

We define information measures through generic equations, then illustrate them in specific cases. 
We consider all measures for BB and Md; and only syntactic content and pragmatic values for RS, RB and St, as they lack multi-scale coordination. 
Table \ref{tab:t-d:results} summarizes results for syntactic ($Ex_{all}$), semantic ($Ex_{scenario}$) and pragmatic delta ($Ex_{scenario}$) measures. Fig. \ref{fig:t-d:PV-goal-all} compares   pragmatic values of all strategies ($Ex_{all}$) (see the Supplementary Material for further results and calculations).

\paragraph{Syntactic Information}

We estimate the \textbf{syntactic information content $C_{syn}$} using Shannon's entropy $H(\cdot)$ \cite{diaconescu2021information}) for each agent variable.  
To provide an upper limit for $C_{syn}$, we consider the workers' task selections and the agents' variables $v_{s,j}^A$, $v_{s,j}^\Delta$ as independent.  
Furthermore, we only distinguish between collective states (at each scale) featuring different task distributions: i.e., the same number $z$ of workers performing $k_a=k_1$, regardless of their arrangements.   
E.g., for $z=1$, $v_{S_0}^k \in \{0001$, $0010$, $0100$, $1000$\} are equivalent, leading to $v_2^A=1$, $v_2^{\Delta}=-3$. 

For any variable $v_{s,j}$ taking numerical values $val_{s} \in V_{s}$ with probabilities $p_{val_{s}}$, we have:

\begin{equation}
    C_{syn}(v_{s,j}) = H(v_{s,j}) = \sum_{val_{s} \in V_s} -p_{val_{s}}*log_2(p_{val_{s}})
    \label{eq:t-d:entropy-agent}
\end{equation}

For an entire scale $S_s$, the entropy is the sum of individual entropies of variables $H(v_{s,j})$, iff $v_{s,j}$ have different information sources. 
Otherwise, $H(S_s)$ is the same as $H(v_{s,j})$, for all $j$. This gives: \\

\begin{equation}
H(S_{s}) = 
\begin{cases} 
\sum_{j=0}^{|S_s|-1} H(v_{s,j}) \mbox{, if } v_{s,j} \mbox{ has different info sources for each } j \\ 
\mbox{(e.g. abstraction from different micro-sources)}\\
\hline 
H(v_{s,j}) \mbox{, for any } j \mbox{, if all } v_{s,j} \mbox{ have the same information source}\\ 
\mbox{(e.g. error broadcast from the same macro-source)}
\end{cases}
\label{eq:t-d:entropy-scale}
\end{equation}

We can then estimate the amount of information `lost' or `gained' via inter-scale abstraction or reification: 

\begin{equation}
\Delta H_{S_s \rightarrow S_{s+i}} = H(S_{s+i}) - H(S_s)
\label{eq:t-d:entropy-between-scales}
\end{equation}

E.g., BB's abstraction flow loses $25\%$ from $S_0$ to $S_1$ ($\Delta H_{S_0 \rightarrow S_1}^{BB,A}=-1$); and further $\approx32\%$ from $S_1$ to $S_2$ ($\Delta H_{S_1 \rightarrow S_2}^{BB,A}=-0.97$).   
BB's control flow also loses $41\%$ of information from $S_2$ to $S_1$ ($\Delta H_{S_2 \rightarrow S_1}^{\Delta} \approx -0.8$), as  different error indicators from the top-manager map to the same indicator from mid-managers (e.g. both $v_{2,0}^{\Delta}=4$ and $3$ are split into $v_{1,0}^{\Delta}=v_{1,1}^{\Delta}=2$).   
No information is lost from $S_1$ to $S_0$ as mid-managers broadcast to workers.
In contrast, Md loses no information in either state and control flows.

\begin{table}
    \centering
    \begin{tabular}{|c|c|c|c|c|c|c|}
    \hline
        \multicolumn{7}{|c|}{\textbf{BB: Syntactic information content $C_{syn}^{BB}$}, using $Ex_{all}$}  \\
        \hline
        \hline
        Var. $v_{s,j}$ 
        & \multicolumn{3}{|c|}{$C_{syn}(v_{s,j}^A)$} 
        & \multicolumn{3}{|c|}{$C_{syn}(v_{s,j}^{\Delta})$}\\
        \hline
        $v_{0,j}$ 
        & \multicolumn{3}{|c|}{$H(v_{0,j}^k)=1$} 
        & \multicolumn{3}{|c|}{$H(v_{0,j}^{\Delta})=1.198$} \\
        \hline
        $v_{1,j}$ 
        & \multicolumn{3}{|c|}{$H(v_{1,j}^k)=1.5$} 
        & \multicolumn{3}{|c|}{$H(v_{1,j}^{\Delta})=1.198$} \\
        \hline
        $v_{2}$ 
        & \multicolumn{3}{|c|}{$H(v_{2}^k)=2.03$} 
        & \multicolumn{3}{|c|}{$H(v_{2}^{\Delta})=2.03$} \\
        \hline
        Scale $S_s$ 
        & \multicolumn{3}{|c|}{$C_{syn}(S_s^A)$} 
        & \multicolumn{3}{|c|}{$C_{syn}(S_s^{\Delta})$}\\ 
        \hline
        $S_0$ 
        & 
        \multicolumn{3}{|c|}{$H(S_0^A) = 4$} & 
        \multicolumn{3}{|c|}{$4 * H(S_0^{\Delta}) = 4.79$} \\ 
        \hline 
        $S_1$ 
        & \multicolumn{3}{|c|}{$H(S_1^A) = 3$} & 
        \multicolumn{3}{|c|}{$2 * H(S_1^{\Delta}) = 2.396$} \\ 
        \hline
        $S_2$ 
        & \multicolumn{3}{|c|}{$H(S_2^A) = 2.03$} 
        & \multicolumn{3}{|c|}{$H(S_2^{\Delta}) = 2.03$}\\
        \hline
        \hline
          \multicolumn{7}{|c|}{\textbf{Syntactic information content for one cycle, $C_{syn,cycle}^{Strategy}$}, using $Ex_{all}$}  \\
        \hline
        \hline 
        Strategy & BB & Md & RS & RB & St & .\\
        \hline
        $C_{syn,cycle}^{strategy}$ & 18.248 & 24 & 4 & NA & NA & .\\
        \hline\hline 
        \multicolumn{7}{|c|}{\textbf{Semantic information}, using $Ex_{scenario}$}  \\
        \hline
        \hline 
        \textbf{Measure} 
          & $t_0\veryshortarrow t_1$  & $t_2\veryshortarrow t_2$ 
          & $t_2\veryshortarrow t_3$  & $t_3\veryshortarrow t_4$ 
          & $t_4 \veryshortarrow t_5$  & $t_5\veryshortarrow t_6$ \\
        \hline
        \makecell{Semantic delta BB: $\Delta_{sm,sys}^{BB}$} 
        & $4$   & $12$ & $10$  & $4$ & $2$  & $0$ \\
        \hline
        \makecell{Semantic delta Md: $\Delta_{sm,sys}^{Md}$} 
        & $4$   & $16$ & $0$  & $0$ & $0$  & $0$ \\
        \hline
        \textbf{Measure} 
        & $t_0$   & $t_1$ & $t_2$  & $t_3$ & $t_4$  & $t_5$\\
        \hline
        \makecell{Delta truth BB: $\Delta_{th}^{BB}(Est_1^{\Delta,t},Obs_0^{\Delta,t})$} 
        & $4$   & $0$ & $-2$  & $-2$ & $0$  & $0$\\
        \hline
        \makecell{Delta truth Md: $\Delta_{th}^{Md}(Est_1^{\Delta,t},Obs_0^{\Delta,t})$} 
        & $-4$   & $0$ & $0$  & $0$ & $0$  & $0$ \\
        \hline
        \makecell{Sem. truth Val. BB: $V_{sm,th}^{BB}(Est_1^{\Delta},Obs_0^{\Delta})$} 
        & $NA$   & $1$ & $-0.3$  & $0$ & $0.3$  & $0$ \\
        \hline
        \makecell{Sem. truth Val. Md: $V_{sm,th}^{Md}(Est_1^{\Delta},Obs_0^{\Delta})$} 
        & $NA$   & $1$ & $0$  & $0$ & $0$  & $0$ \\
        \hline
        \makecell{Sem. Effi. BB: $E_{sm,th}^{BB}(S_1^\Delta)$} 
        & $NA$   & $0.54$ & $0.018$  & $0.27$ & $0.36$  & $0.27$ \\
        \hline
        \makecell{Sem. Effi. Md: $E_{sm,th}^{Md}(S_1^\Delta)$} 
        & $NA$   & $0.41$ & $0.208$ & $0.208$ & $0.208$ & $0.208$ \\
        \hline\hline 
        \multicolumn{7}{|c|}{\textbf{Pragmatic information}, using $Ex_{scenario}$}  \\
        \hline
        \hline
        \textbf{Measure} 
        & $t_0$   & $t_1$ & $t_2$  & $t_3$ & $t_4$  & $t_5$\\
        \hline
        \makecell{Prag. Scope Delta BB: $\Delta_{pr,sp}^{BB}(S_0)$} 
        & $0$   & $0$ & $4$  & $2$ & $0$  & $0$\\
        \hline
        \makecell{Prag. Scope Delta Md: $\Delta_{pr,sp}^{Md}(S_0)$} 
        & $0$   & $0$ & $4$  & $0$ & $0$  & $0$\\
        \hline
        \makecell{Prag. Adp. Delta BB: $\Delta_{pr,ad}^{BB}(S_0)$} 
        & $NA$   & $NA$ & $2$  & $0$ & $-2$  & $0$ \\
        \hline
         \makecell{Prag. Adp. Delta Md: $\Delta_{pr,ad}^{Md}(S_0)$} 
        & $NA$   & $NA$ & $4$  & $-4$ & $0$  & $0$\\
        \hline  
    \end{tabular}
    \caption{Information measures for the task distribution case. Syntactic content and pragmatic values rely on the probabilistic  analysis of all possible behaviors via $Ex_{all}$. Semantic value and pragmatic delta measures are exemplified via $Ex_{scenario}$.}
    \label{tab:t-d:results}
\end{table}

In all cases, resource use depends on the number of variables at each scale:

\begin{equation}
    C_{syn}(S_s)=\sum_{j=0}^{|S_s|-1} C_{syn}(v_{s,j})
    \label{eq:t-d:resources-scale}
\end{equation}

 For BB, the control flow's resource usage is larger than its information content: $C_{syn}^{BB}(S_s^{\Delta}) > H^{BB}(S_s^{\Delta})$ (see Table \ref{tab:t-d:results}).  
 It gains $18\%$ in syntactic content from $S_2$ to $S_1$ 
 ; and $99.9\%$ from $S_1$ to $S_0$. 

Finally, we consider the syntactic content of an entire 
feedback cycle (see Table \ref{tab:t-d:results}):

\begin{equation}
    C_{syn,cycle}=\sum_{s=0..2} C_{syn}(S_s^A) + \sum_{s=0..2} C_{syn}(S_s^{\Delta}) 
    \label{eq:t-d:resources-cycle}
\end{equation}


\paragraph{Semantic Information} 

The \textbf{semantic delta} $\Delta_{sm}$ of an agent variable $v_{s,j}$ measures changes in its numerical values. For one step, we have: 

\begin{equation}
\Delta_{sm}^{t_i \rightarrow t_{i+1}}(v_{s,j}) = |v_{s,j}^{t_{i+1}} - v_{s,j}^{t_i}|
\label{eq:t-s:semantic-delta-value}
\end{equation}

For the entire system: 

\begin{equation} 
\Delta_{sm,sys}^{t_i \rightarrow t_{i+1}} = \sum_s \sum_j \Delta_{sm}^{t_i \rightarrow t_{i+1}}(v_{s,j})
\label{eq:t-s:semantic-delta-value-full}
\end{equation}

Using $Ex_{scenario}^{BB}$ (Fig. \ref{fig:t-d:exo-steps}), $t=0..5$,  gives the subsequent measures: $\Delta_{sm,sys}^{BB,t \rightarrow t+1} = 4, 12, 10, 4, 2, 0$ (see full details in the Supplementary Material).  
Hence, BB's knowledge changes during adaptation ($t=0..4$), then stabilizes when the goal is reached ($t \geq 5$). 
Md features $\Delta_{sm,sys}^{Md, t \rightarrow t+1} = 4, 16, 0, 0, 0, 0$, indicating larger, quicker knowledge changes, as Md converges faster to the goal.

We calculate the \textbf{semantic truth value} $V_{sm,th}$ based on three steps. We focus on $V_{sm,th}$ for the managers' control flows, to highlight the difference between their knowledge (leading to workers' adaptation) and the workers' actual state, at $t$.

First, we estimate the \textit{delta of truth }$\Delta _{th}$ at $t$ as the difference between the managers' estimated error, $Est_{s}^{\Delta,t}$, and the workers' actual error to the goal $Obs_0^{\Delta,t}$:  

\begin{equation}
    Est_s^{\Delta,t} = \sum_{j=0}^{|S_s|-1} v_{s,j}^{\Delta,t} 
    \mbox{, $s>0$}
\end{equation}

\begin{equation}
Obs_0^{\Delta,t} = v_g - \sum_{j=0}^{|S_0|-1} v_{0,j}^{k,t}
\end{equation}

\begin{equation}
    \Delta_{th}(Est_{s}^{\Delta,t},Obs_{0}^{\Delta,t}) = 
    Est_{s}^{\Delta,t} -  Obs_{0}^{\Delta,t} 
    \label{eq:t-d:sem-truth-val}
\end{equation}

We consider $\Delta_{th}=0$ when $Est_s^{\Delta,t} = NA$, $t=0..1$.

Using $Ex_{scenario}$, $t=0..5$, we obtain for BB: $\Delta_{th}^{BB}(Est_1^{\Delta,t},Obs_0^{\Delta,t}) = 4,0,-2,-2,0,0$; and for Md: $\Delta_{th}^{Md}(Est_1^{Md,\Delta,t},Obs_0^{Md,\Delta,t}) = -4, 0,  0, 0, 0, 0$. 
 These differences are due to communication delays. As above, they occur during the adaptation period (shorter for Md). 

Second, we assign values $SV_{th}$ to different knowledge states, depending on their distance to the truth: 

\begin{equation}
    SV_{th}(Est_{s}^{\Delta,t},Obs_{0}^{\Delta,t}) = 
    1 - \lvert \frac{\Delta_{th}(Est_s^{\Delta,t},Obs_0^{\Delta,t})}{\Delta_{th,max}}\rvert
\end{equation}
with $\Delta_{th,max}=4$ the maximum $\Delta_{th}$ in our case.

Maximum $SV_{th}=1$ indicates no difference to truth, and minimum $SV_{th}=0$ the largest difference. 
This gives, for BB: $SV_{th}^{BB}= 0, 1, 0.5, 0.5, 1, 1$; and for Md: $SV_{th}^{Md} = 0, 1, 1, 1, 1, 1$. We note that, due to delays, $SV_{th}$ becomes $1$ (highest) when the system undergoes no adaptation; and $\leq 1$ (lower) otherwise.

Third, we calculate 
$V_{sm,th} \in [-1,1]$:

\begin{equation}
V_{sm,th}^{t}(Est_s^{\Delta,t},Obs_0^{\Delta,t}) =
\begin{cases}
    \textbf{NA} \mbox{\textit{, if no knowledge of $Est_s$ at $t$-$1$ or $t$}} \\
     SV_{kn}(Est_s^{\Delta,t-1},Obs_0^{\Delta,t}) == NA \lor 
     SV_{kn}(Est_s^{\Delta,t},Obs_0^{\Delta,t}) == NA \\
    \hline         
    \textbf{0} \mbox{\textit{, if no change from} } $t-1$ \mbox{ to } $t$\\
            SV_{kn}(Est_s^{\Delta,t-1},Obs_0^{\Delta,t}) 
             == SV_{kn}(Est_s^{\Delta,t},Obs_0^{\Delta,t})\\
    \hline            
    \frac{SV_{kn,s}^{\Delta,t} - SV_{kn,s}^{\Delta,t-1}}{SV_{kn,s}^{\Delta,t} + SV_{kn,s}^{\Delta,t-1}}
    \mbox{, \textit{otherwise} }
\end{cases}
\label{eq:t-s:STV}
\end{equation}

We have $V_{sm,th}^t = 0$ (neutral) if knowledge was unavailable at $t$-$1$ or remained unchanged from $t$-$1$ to $t$; $V_{sm,th}^t \in (0,1]$ (positive) if $SV_{kn}$ increased (closer to truth); and $V_{sm,th}^t \in [-1,0)$ (negative) if $SV_{kn}$ decreased (farther from truth). 
As before, fluctuations are due to estimation lags and neutral values to stable states.  

We estimate the \textbf{efficiency of the semantic truth value} $E_{sm,th} \in [0,1]$  
based on the resources required through a feedback cycle 
 $C_{syn,cycle}$ (eq. \ref{eq:t-d:resources-cycle}):  

\begin{equation}
    E_{sm,th}^{t}(S_1^\Delta) = 
    \frac{(V_{sm,th}(Est_{S_1^\Delta}^t,Obs_{S_0^\Delta}^t)+1)/2}{C_{syn,cycle}} * coef
    \label{eq:t-d:semantic-efficiency}
\end{equation}

We normalize $E_{sm,th}$ to fit the interval $[0,1]$ and to indicate better efficiency for higher $V_{sm,th}$ and lower $C_{syn,cycle}$; and worse efficiency for worse $V_{sm,th}$ and higher $C_{syn,cycle}$. 
We employ a scaling factor $coef=10$ for readability reasons. 
We note that BB features $\approx 31.5\%$ better efficiency than Md for the same $V_{sm,th}$ values, as Md uses $\approx 31.5\%$ more resources than BB. 

\subparagraph{Pragmatic Information} 

We focus on evaluating how mid-managers' control information $v_{S_1}^\Delta$ influence workers' adaptations $K_{act}(S_0)$ toward the goal $v_g$. 

We consider two types of \textbf{pragmatic delta} $\Delta_{pr}$ for worker changes: 
\begin{itemize}
    \item \textbf{pragmatic scope delta} $\Delta_{pr,sp}$: changes in the scope of possible adaptations $|K_{act}|$;
    \item \textbf{pragmatic adaptation delta} $\Delta_{pr,ad}$: changes in the magnitude of actual adaptations $|K_{ch}|$.
\end{itemize}

The \textbf{pragmatic scope delta} $\Delta_{pr,scp}$ measures how the number of adaptation actions $|K_{act}(S_0)|$ changes when workers receive $v_{S_1}^{\Delta}$ or not:  

\begin{equation}
    \Delta_{pr,sp}^t(S_0) = \sum_{j=0}^3 (|K_{act}^{t,inf}(a_{0,j})| - |K_{act}^{t,noInf}(a_{0,j})|)
\end{equation}

In BB and Md, workers may only switch tasks when receiving control information; otherwise they stay idle. Hence, no information implies $|K_{act}^{t,noInf}(a_{0,j})| = 0$; and received information implies $|K_{act}^{t,inf}(a_{0,j})| = 1$ if the worker switches 
and $|K_{act}^{t,inf}(a_{0,j})| = 0$ otherwise.  
Positive $\Delta_{pr,sp}$ means that information increases the number of potential adaptations; and negative $\Delta_{pr,sp}$ that it reduces them. 

In $Ex_{scenario}$, $t=0..5$, we have $\Delta_{pr,scp}^{BB}(S_0) = 0, 0, 4, 2, 0, 0$. These correspond to workers' potential adaptations at $t_2$ and $t_3$ (see the pink circles in Figs \ref{fig:t-d:ex-t2} and \ref{fig:t-d:ex-t3})
. For Md, we have $V_{pr,scp}^{Md}(S_0) = 0, 0, 4, 0, 0, 0$, indicating that all adaptations happen at $t=2$.  

The \textbf{pragmatic adaptation delta} $\Delta_{pr,ad}$ 
measures actual adaptation differences between $t-1$ and $t$:  

\begin{equation}
    \Delta_{pr,adp}^t(S_0) = \sum_{j=0}^3 (|K_{ch}^t(a_0,j)| -  |K_{ch}^{t-1}(a_0,j)|)
\end{equation}

Here, $|K_{ch}(a_{0,j})|=1$ if worker $a_{0,j}$ switches tasks; and $|K_{ch}(a_{0,j})|=0$ otherwise. 

In $Ex_{scenario}$, $t=0..5$, we have: $\Delta_{pr,adp}^{BB,t} = NA,NA,2,0,-2,0$; and: $\Delta_{pr,adp}^{Md,t} = NA,NA,4,-4,0,0$. This indicates that adaptation happens progressively for BB ($t_2$ and $t_3$) and all at once for Md ($t_2$).  

We estimate the \textbf{pragmatic goal value} ($V_{pr,gl}$) of information based on the likelihood that strategies bring the task distribution towards the goal $v_g$. 
We calculate $V_{pr,gl}$ in three steps:
\begin{itemize}
    \item Assign a \textbf{state value} $SV_{gl}(v_{S_0}^k)$ to each task-distribution state $v_{S_0}^k$;
    \item Assign a \textbf{pragmatic adaptation value} $V_{pr,ad}(v_{S_0}^k)$ to each state, based on its potential to adapt into more valuable states (closer to the goal); 
    \item Calculate the \textbf{pragmatic goal value} $V_{pr,gl}(v_{S_0}^k)$ based on the above adaptation value, current state state value and ability to maintain the goal once reached. 
\end{itemize}
 
We assign \textbf{state values} $SV_{gl} \in [0,1]$ to different task distribution groups $S_0^z$ ($z$ the number of workers executing $k_1$), depending on their distance from $v_g$. $SV_{gl}=1$ is the highest value (goal reached) and $0$ the lowest one. Concretely, we set: $SV_{gl}(S_0) = \{0,0.25,0.5,0.75,1\}$, meaning that $SV_{gl}(S_0^0) = 0$, $SV_{gl}(S_0^1) = 0.25$, $SV_{gl}(S_0^2) = 0.5$, $SV_{gl}(S_0^3) = 0.75$ 
and $SV_{gl}(S_0^4) = 1$. 

We evaluate the \textbf{pragmatic adaptation value} $V_{pr,ad} \in [0,1]$ based on the workers' probability to adapt from their present state $v_{S_0}^{k,t} \in V_{S_0}$ to all possible states $v_{S_0}^{w,t+m} \in V_{S_0}$, within $m$ steps. We weight this state-transition probability $p(v_{S_0}^{w,t+m}|v_{S_0}^{k,t})$ by the state value $SV_{gl}$ of each adapted state $v_{S_0}^w$. This gives: 

\begin{equation}
V_{pr,ad}^{t \rightarrow t+m}(v_{S_0}^k)=\sum_{v_{S_0}^w:V_{S_0}} SV_{gl}(v_{S_0}^w) * p(v_{S_0}^{w,t+m}|v_{S_0}^{k,t})
\label{eq:t-d:PVadapt}
\end{equation}
 
We start from state $v_{S_0}^0=0000$ and aim for $v_g=1111$. Thus, state-transition probabilities are the likelihood that workers switch from $k_0$ to $k_1$, at each step (see details in the Supplementary Material). For BB, this gives the stochastic matrix $M$, for one step; with states $S_0^z \in \{0000, 0001, 0011, 0111, 1111\}$ from left-to-right (columns) and top-down (rows):

\begin{equation}
M = 
\begin{bmatrix}
0.522 & 0.3684 & 0.0975 & 0.0114 & \textbf{0.0005} \\
0 & 0.6141 & 0.3251 & 0.0573 & 0.0034 \\
0 & 0 & 0.723 & 0.255 & 0.0225 \\
0 & 0 & 0 & 0.85 & 0.15 \\
0 & 0 & 0 & 0 & 1
\end{bmatrix}
\label{eq:t-d:matrix1}
\end{equation}

The likelihood that BB reaches the goal in one step is 0.0005. We can then obtain state-transition probabilities over $m$ steps via $M^m = \prod_{m} M$. Fig. \ref{fig:t-d:probas-toGoal-0to1-calculated} shows the probabilities of reaching the goal in $m=0..50$ steps (cf. Supplementary Material for corresponding $V_{pr,adp}$ values and calculations).

Finally, we assess the \textbf{pragmatic goal value} $V_{pr,gl} \in [-1,1]$ considering progress between the current state's value $SV_{ds}(v_{S_0}^k)$ and its pragmatic adaptation value $V_{pr,adp}(v_{S_0}^k)$; and the system's ability to maintain the goal $V_{pr,adp}(v_g)$:

\begin{equation}
\begin{split}
&V_{pr,gl}^{t \rightarrow t+m}(v_{S_0}^k) = 
\frac{V_{pr,adp}^{t \rightarrow t+m}(v_{S_0}^k) - SV_{gl}(v_{S_0}^k)}
{p_{goal-keep}} \\[10pt]
&p_{goal-keep} =
\begin{cases}
2 - V_{pr,adp}^{t \rightarrow t+m}(v_g) \qquad
\mbox{, if } V_{pr,adp}^{t \rightarrow t+m}(v_{S_0}^k) \geq SV_{gl}(v_{S_0}^k)\\[10pt]
V_{pr,adp}^{t \rightarrow t+m}(v_g) \qquad \qquad
\mbox{, if } V_{pr,adp}^{t \rightarrow t+m}(v_{S_0}^k) < SV_{gl}(v_{S_0}^k)
\end{cases}\\[10pt]
&\mbox{considering } V_{pr,gl}^{t \rightarrow t+m}(v_{S_0}^k) = -1 \qquad
\mbox{ , if } p_{goal-keep} == 0
\end{split}
\label{eq:t-d:pv-goal}
\end{equation}

The maximum value $V_{pr,gl}^{t \rightarrow t+m}(v_{S_0}^k)=1$ occurs when the system is sure to reach $v_g$, from the least valuable state, within $m$ steps; and then maintain it, $V_{pr,adp}^{t \rightarrow t+m}(v_g)=1$. 
The minimum value $-1$ occurs when the system is sure to regress from $v_g$ to the least valuable state. 

Using $Ex_{all}$, Fig. \ref{fig:t-d:pv-goal_0to1_0-025-075-1} compares $V_{pr,gl}$ results for different strategies, over $m=0..50$ steps. 
BB has a low pragmatic goal value $V_{pr,gl}^{BB}\approx 0.15$ for $m=1$, yet this increases to $\approx 0.72$ at $m=10$; $\approx 0.94$ at $m=20$; and $\approx 0.98$ at $m=30$. 
As BB maintains the goal and the initial state value is $SV_{ds}(0000)=0$, we have $V_{pr,gl}^{BB} = V_{pr,adp}^{BB}$, $\forall m$.  
Md reaches the goal in $N$-$1$ steps and maintains it, hence $V_{pr,gl}^{Md}=1$, for $m \geq 2$. St does not adapt, hence  $V_{pr,gl}^{St}=0$, $\forall m$. RB and RS adapt randomly and are further penalized 
for not maintaining the goal (i.e. $p_{goal-keep}^{RS} \approx p_{goal-keep}^{RS} \approx 1.5$). 
RB features a constant $V_{pr,gl}^{RB}=0.33$ and RS converges to the same value after $m \geq 20$. 

To emphasize the role of multi-scale information flows separately from their processing strategies, we introduce an error for BB and Md:  worker $a_{0,0}$ always reports state $v_{0,0}^{k_a}=k_1$, regardless of its actual state. Fig. \ref{fig:t-d:pv-goal_0to1_0-025-075-1_withError} shows the impact on $V_{pr,gl}$. RS, RB and St remain unaffected, as they do not employ multi-scale flows. Md loses $25\%$ of its $V_{pr,gl}$; and BB $\approx 23\%$. This is proportional to the $25\%$ information error from $V_{S_0}^{k_a}$, with BB slightly less sensitive than Md.     

We assess the \textbf{efficiency of the pragmatic goal value} $E_{pr,gl}$ considering required resources over $m$ steps:

\begin{equation}
    E_{pr,gl}^{t \rightarrow t+m} = 
    \frac{\sum_{i=1}^m V_{pm,gl}^{t_i}}{m*C_{syn,cycle}}
    \label{eq:t-d:eff-prag}
\end{equation}

This gives the pragmatic efficiency values in Fig. \ref{fig:t-d:pv-effi-res-nrml}. 
Md surpasses BB in efficiency as it converges faster, yet as BB consumes less resources to maintain the goal it will become slightly more efficient than Md over the longer term ($\approx 16\%$ more by step 50).  

\stepcounter{figure}
\begin{subfigure}[ht!]
\begin{minipage}{0.5\textwidth}
\centering
\includegraphics[width=0.65\linewidth]{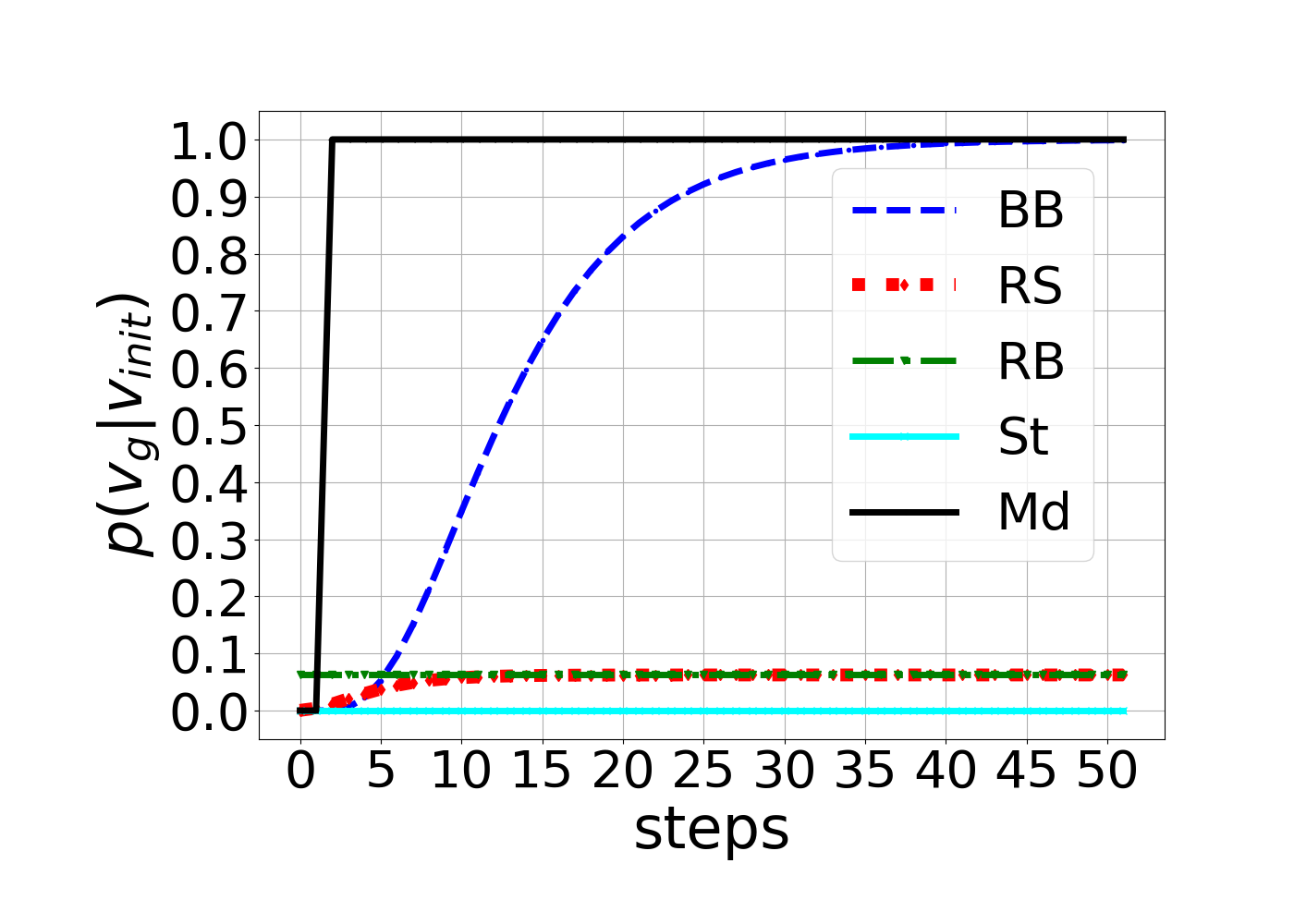}
\caption{Probabilities to reach the goal}
\label{fig:t-d:probas-toGoal-0to1-calculated}
\end{minipage}
\begin{minipage}{0.5\textwidth}
\centering
\includegraphics[width=0.65\linewidth]{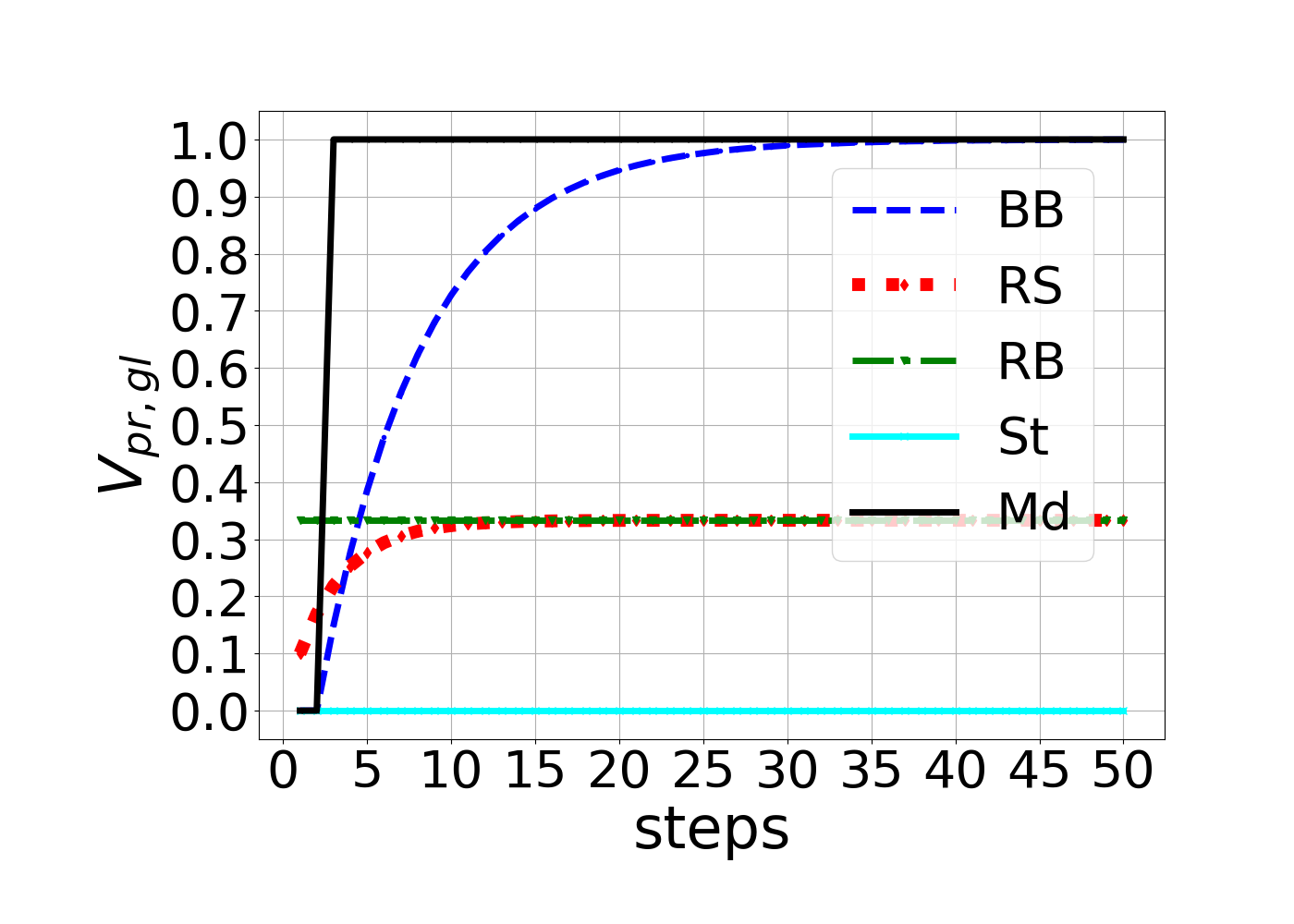}
\caption{Pragmatic goal values with correct info}
\label{fig:t-d:pv-goal_0to1_0-025-075-1}
\end{minipage}
\newline
\begin{minipage}{0.5\textwidth}
\centering
\includegraphics[width=0.65\linewidth]{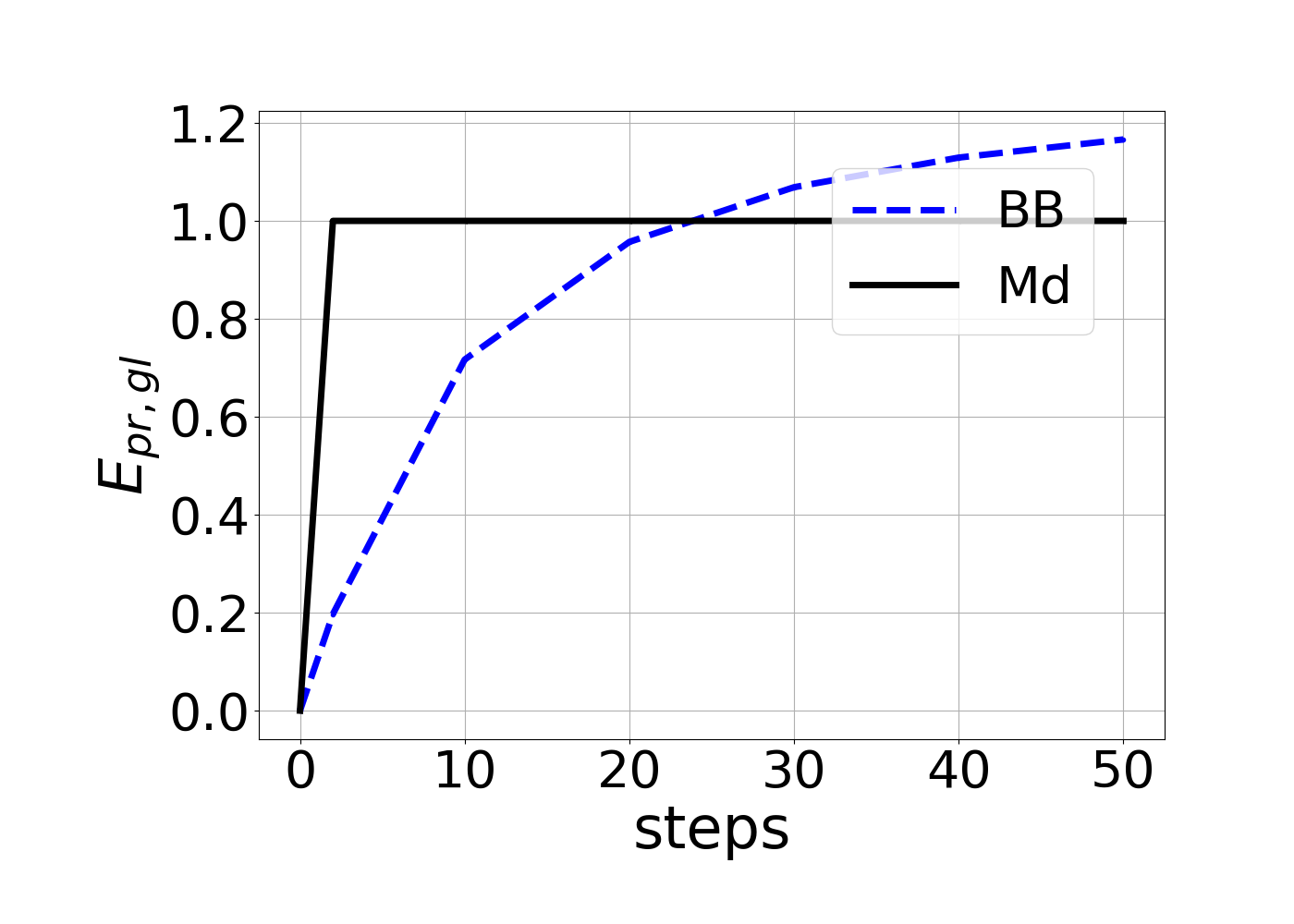}
\caption{Efficiency of the pragmatic goal value 
}
\label{fig:t-d:pv-effi-res-nrml}
\end{minipage}
\begin{minipage}{0.5\textwidth}
\centering
\includegraphics[width=0.65\linewidth]{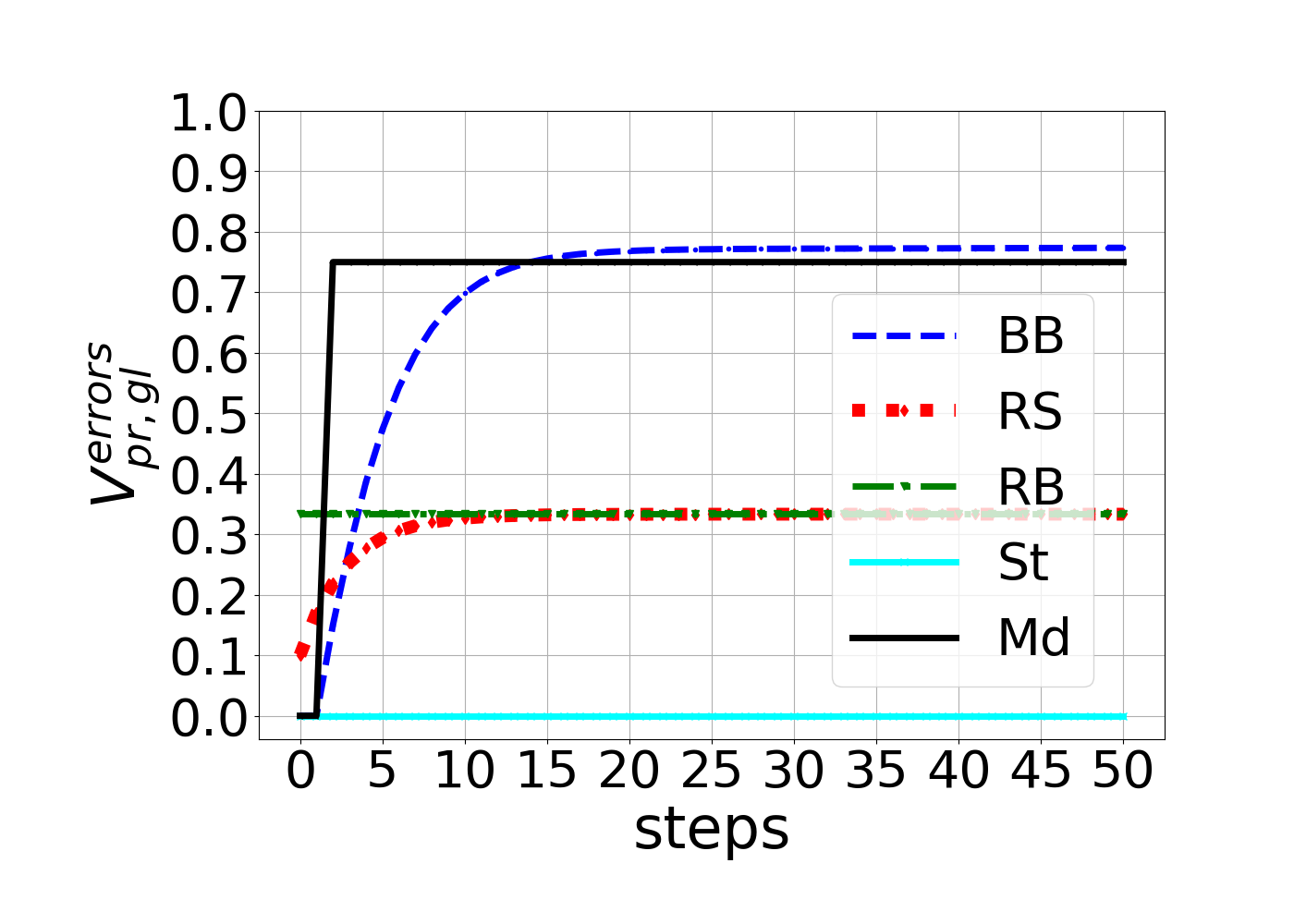}
\caption{Pragmatic goal values with error info}
\label{fig:t-d:pv-goal_0to1_0-025-075-1_withError}
\end{minipage}
\setcounter{subfigure}{-1}
\caption{Probabilities of goal convergence and pragmatic goal values within 50 steps; from the initial state $v_{S_0}^{t_0}=0000$ to the goal state $v_g=1111$, for different adaptation strategies. \\
To calculate $V_{pr,gl}^{t_0 \rightarrow t_{49}}$  we considered state values $SV_{gl}(S_0)=\{0,0.25,0.5,0.75,1\}$.}
\label{fig:t-d:PV-goal-all}
\end{subfigure}

\subsection{Hierarchical Oscillators}

\subsubsection{Model Description}
This case study is based on the model of coupled biochemical oscillators in \cite{Kim2010}, which was extended to a hierarchy of oscillators in \cite{Mellodge2021}.  Coupled biochemical oscillators are observed throughout many systems in nature (e.g., cellular processes involving circadian rhythms). A single oscillator consists of two interacting components $X$ and $Y$ (e.g., mRNA and protein) in which $X$ inhibits its own synthesis while promoting that of $Y$ and $Y$ inhibits its own and $X$’s synthesis, resulting in a feedback relationship within a single oscillator that induces oscillations in the concentrations of both under appropriate conditions.  The interaction is modeled using differential equations that describe the concentrations in the $X$ and $Y$ components. For coupled biochemical oscillators, the pair of oscillators are coupled by a connection between their $X$ components.  There are two types of coupling: double-positive (PP) in which each $X$ promotes the synthesis of the other, and double-negative (NN) in which each $X$ inhibits the synthesis of the other.  The interactions have two parameters: coupling strength $F$ between the two $X$’s and communication time delay $\tau$ associated with their interaction.  It was shown in \cite{Kim2010} that the different types of coupling and the different values of $F$ and $\tau$ result in different behaviors (synchronized oscillation, unsynchronized oscillation, and no oscillation).

\begin{figure}
    \centering
    \includegraphics[width=0.5\textwidth]{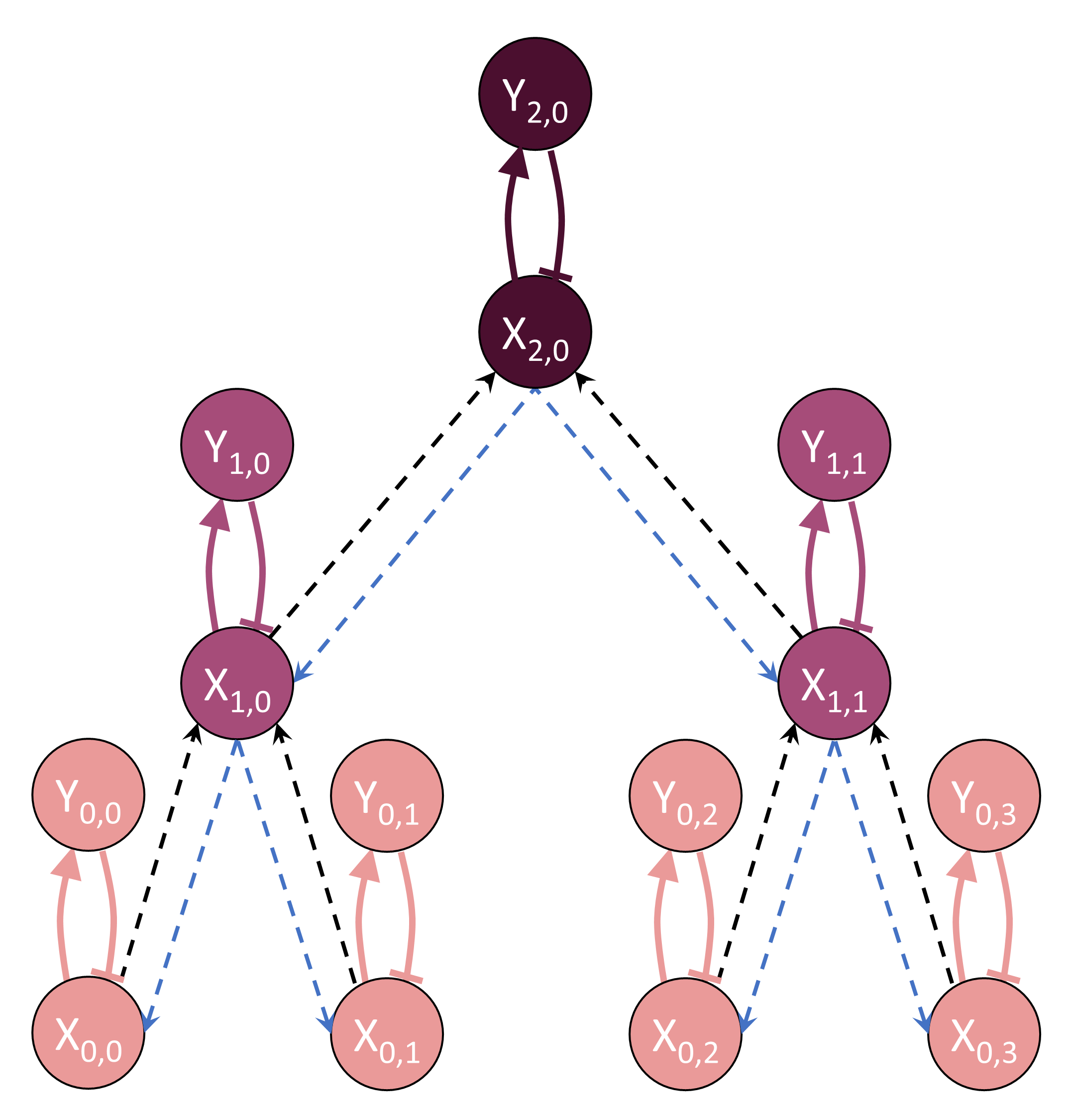}
    \caption{The hierarchical oscillator system with three scales.  Each oscillator is an ($X,Y$) pair and communication across scales is by means of $X$ concentration values.}
    \label{fig:HO_model_detail}
\end{figure}

For the hierarchical oscillator system, the differential equation model from \cite{Kim2010} was extended to form multiple scales of oscillators that communicate their $X$ concentration levels to each other as shown in Figure \ref{fig:HO_model_detail}.  The interactions are modeled by a system of coupled differential equations as follows:
\begin{align}
\frac{dX_{m,i}}{dt} = &\frac{1+PP \left( F_m \Gamma_{m,i}(t-\tau_m) \right)^3}{1+ \left( F_m \Gamma_{m,i}(t-\tau_m) \right)^3 + \left( \frac{Y_{m,i}(t-2)}{0.5} \right)^3} - 0.5X_{m,i}(t) + 0.1 \label{eqn:middle_X} \\
\frac{dY_{m,i}}{dt} = &\frac{\left( \frac{X_{m,i}(t-2)}{0.5} \right)^3}{1+ \left( \frac{X_{m,i}(t-2)}{0.5} \right)^3} - 0.5Y_{m,i}(t) + 0.1
\end{align}
where
\begin{align}
\Gamma_{m,i}(\cdot) = &W X_{m+1,p_{m,i}}(\cdot) + (1-W)\bar{X}_{m,i}(\cdot) \label{eqn:gamma}
\end{align}
and $\bar{X}_{m,i}(\cdot)$ denotes the mean $X$ concentration taken over the children of $X_{m,i}$; $m = 0..M-1$, where $M$ is the number of scales; $i = 0..N_m-1$, where $N_m$ is the number of oscillators at scale $S_m$; $p_{m,i}$ is the position of the parent of $X_{m,i}$; $F_m$ is the coupling strength; $\tau_m$ is the time delay; and $W \in [0,1]$ is a factor that sets the weight given to information from above or below.  Parameters $F_m$ and $\tau_m$ are constant across a scale and $W$ is constant across the system.  $PP$ is set to 1 for PP type coupling and 0 for NN type coupling.

In this system, communication occurs across scales only (i.e., oscillators at a given scale do not communicate directly with each other).  The abstracted information contained in oscillator $i$ at scale $S_m$ is the average $X$ concentration of its children oscillators in $S_{m-1}$ and is given by $\bar{X}_{m,i}(\cdot)$.  The control information received by oscillator $i$ in $S_m$ is the $X$ concentration of its parent oscillator in $S_{m+1}$, which impacts the $X$ concentration of oscillator $i$ through (\ref{eqn:middle_X}).  The function $\Gamma_{m,i}(t-\tau_m)$ in (\ref{eqn:gamma}) shows how both the abstracted and reified information impact the behavior of oscillator $i$ in $S_m$, with $W$ controlling the relative importance of both types of information.  For the top and bottom scales, $\Gamma$ is modified to eliminate the first and second terms, respectively, due to lack of reified information from above and lack of aggregated information from below.  Communication between oscillators happens through multiple feedback loops in which the $X$ concentration is collected from the immediate lower scale as averages, combined with the average from the immediate higher scale, and sent back down as $X$ concentration with an associated time delay $\tau_m$.  As in the case with a single pair of coupled oscillators, the behavior of the hierarchical system can achieve three steady-state conditions: synchronized oscillation, unsynchronized oscillation, and no oscillation.  The overall goal is for the system to synchronize oscillators at the bottom scale.

\subsubsection{Comparison strategies}
We focus on the comparison of two HO systems using NN type coupling with two different scales (2-scale and 3-scale).  Both systems have four oscillators at the bottom scale $S_0$.  The 2-scale system has one oscillator at $S_1$ which receives the average of the four oscillators in $S_0$.  The 3-scale system is as shown in Figure \ref{fig:HO_model_detail} which contains two oscillators in $S_1$ and one oscillator in $S_2$ and each oscillator receives the average $X$ concentration of its two children.  Thus the 3-scale system breaks down the averaging into two steps, resulting in more delay in its structure since each oscillator at $S_1$ and $S_2$ passes down reified information with delay $\tau_m$.

The system is simulated using a time-delay Ordinary Differential Equations (ODE) solver (see details in Supplementary Material). 
One simulation is run for each system for 300 seconds.  The initial $X$ concentrations of the bottom scale oscillators are set to 0.2, 0.4, 0.6, and 0.8.  The remaining oscillators are initialized to be the average values of their children.  Parameters $F_m$ and $\tau_m$ are set to values known to result in synchronized oscillation.  The cases of unsynchronized or no oscillation are not considered.

To calculate $C_{syn}$, we assume that oscillators in $S_0$ have uniformly distributed $X$ concentrations on the interval $[0,1]$ which represents the most random situation.  The probability distributions at the higher scales are calculated using the Bates distribution.  Integration for $C{syn}$ is carried out numerically using the trapezoidal rule (see details in the Supplementary Material).

\subsubsection{Information measures}

\paragraph{Calculations}

Since the HO system is modeled using differential equations which are continuous-time descriptions of the system behavior, the measures which use comparisons between $t$ and $t-\theta$ are calculated using derivatives in this case study. 

\subparagraph{Syntactic information}

Due to the continuous nature of $X$ concentrations, 
syntactic information content cannot be calculated 
as in the other case studies using Shannon entropy or memory usage.  Instead, we use a common entropy measure for continuous valued random variables, Jensen-Shannon (JS) Divergence between probability distributions $f_1$ and $f_2$ as defined in (\ref{eqn:JSdivergence}) \cite{Lin1991}\cite{Nielsen2019}.
\begin{equation}
D_{JS}(f_1||f_2) = \frac{1}{2} \left( D_{KL}(f_1||\bar{f}) + D_{KL}(f_2||\bar{f}) \right)
\label{eqn:JSdivergence}
\end{equation}
where $\bar{f} = \frac{f_1+f_2}{2}$ and $D_{KL}$ is the Kullback-Leibler (KL) Divergence defined as \cite{kullback1951information}
\begin{equation}
D_{KL}(p||q) = \int_{x \in \chi} p(x) \log \frac{p(x)}{q(x)} dx,\label{eqn:KLdivergence}  
\end{equation}
where $\chi$ is the support of $X$.

The JS Divergence provides a metric to measure the difference between probability distributions.  $D_{JS} \in [0,\log(2)]$ with $D_{JS} = 0$ indicating that the two distributions are identical and higher values indicating that they are more dissimilar.  
We use $D_{JS}(f_{m,i}||U)$ to calculate the amount of information contained in oscillator $i$ at scale $S_m$ compared to uniform distribution $U$.  That is, the syntactic information for any given oscillator is calculated as
\begin{equation}
C_{syn_{m,i}} = 1 - \frac{D_{JS}(f_{m,i}||U)}{\log(2)}
\end{equation}
Using this definition, $C_{syn_{m,i}} \in [0,1]$ with values of 0 and 1 indicating the minimum and maximum amount of information, respectively.  \textbf{Syntactic information content} for the system $C_{syn}$ is calculated as the sum of $C_{syn_{m,i}}$ over all oscillators at all scales in the hierarchy.
\begin{equation}
    C_{syn} = \sum_{m=0}^{M-1} \sum_{i=0}^{N_m-1} C_{syn_{m,i}}
\end{equation}

\subparagraph{Semantic information}

This case study does not involve changes to the differential equation model itself, so the \textbf{semantic delta} measures how much change there is to the knowledge $\Gamma_{m,i}$ being used by oscillator $i$ at scale $S_m$.  For the lowest scale of oscillators, $\Gamma_{0,i} = X_{1,p_{0,i}}$, the average $X$ concentration passed down to oscillator $i$ from its parent $p_{0,i}$ with a delay of $\tau_0$.

\begin{equation}
    \Delta_{sm}(t) = \frac{1}{N_0} \sum_{i=0}^{N_0-1} \frac{d X_{1,p_{0,i}}(t-\tau_0)}{dt}
\end{equation}

The \textit{ground truth} $X_{th}$ for the system is the average $X$ concentration among all oscillators at the lowest scale, which is calculated as
\begin{equation}
    X_{th}(t) =  \frac{1}{N_0} \sum_{i=0}^{N_0-1} X_{0,i}(t)
\end{equation}
The \textbf{semantic truth value} is then calculated as
\begin{equation}
    V_{sm,th}(t) = -\frac{1}{N_0} \sum_{i=0}^{N_0-1} \frac{d}{dt} \left| X_{th}(t) - X_{1,p_{0,i}}(t-\tau_0) \right|
\end{equation}
The efficiency is calculated by dividing the semantic truth value by $C_{syn}$.
 \begin{equation}
    E_{sm,th}(t) = \frac{V_{sm,th}(t)}{C_{syn,tot}}
\end{equation}

\subparagraph{Pragmatic information}

An action taken by an oscillator is to change its $X$ concentration based on micro and/or macro information it receives about $X$ concentrations in the overall system, with the scope of this information determined by its location in the hierarchy.  Thus, the numerical value of the action is calculated using (\ref{eqn:middle_X}) and the \textbf{pragmatic delta} $\Delta_{pr}$ is calculated as

\begin{equation}
    \Delta_{pr}(t) = \frac{1}{N_0} \sum_{i=0}^{N_0-1} \frac{d}{dt} \left( \frac{dX_{0,i}}{dt} \right)
\end{equation}


Since the goal is to achieve synchronized oscillation at the lowest scale, the result is identical concentrations of $X$ among these oscillators at every moment $t$.  The distance from this goal, \textit{delta of goal} $\Delta_{gl}(t)$, is the variance of $X$ and its negative derivative is the \textbf{pragmatic goal value} $V_{pr,gl}(t)$.
\begin{equation}
    \Delta_{gl}(t) = \frac{1}{N_0} \sum_{i=0}^{N_0-1} \left( X_{0,i}(t) - \mu(t) \right)^2        
\end{equation}
where $\mu(t) = \frac{1}{N_0} \sum_{i=0}^{N_0-1} X_{0,i}(t)$
\begin{equation}
    V_{pr,gl}(t) = \frac{d \Delta_{gl}(t)}{dt}    
\end{equation}
The efficiency is calculated by dividing the pragmatic goal value by $C_{syn}$.
\begin{equation}
    E_{pr,gl}(t) = \frac{V_{pr,gl}(t)}{C_{syn}}
\end{equation}

\paragraph{Results}

\begin{figure}
    \centering
    \includegraphics[width=\textwidth]{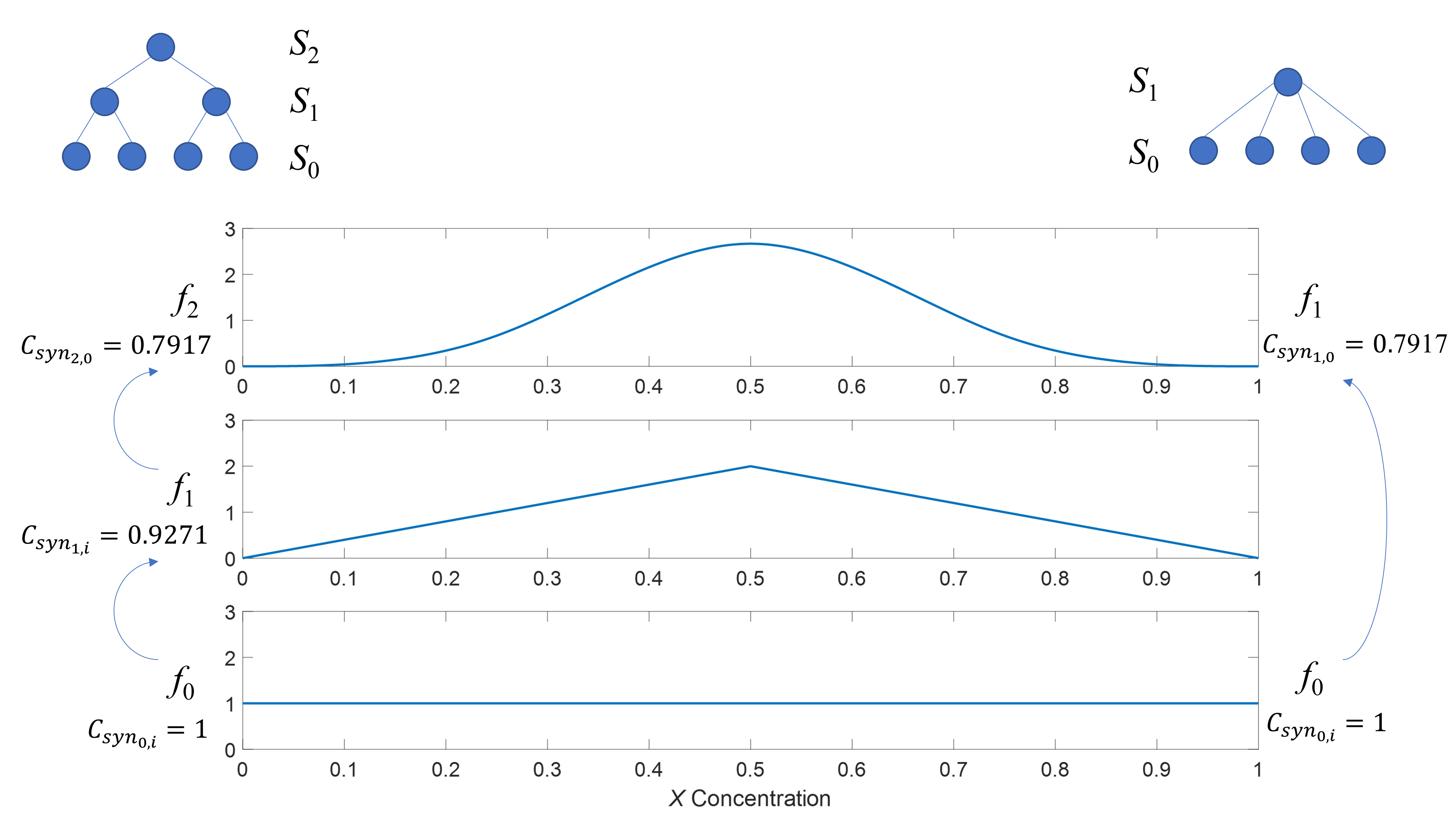}
    \caption{Calculated values for syntactic information.  The left side of the figure depicts the 3-scale system while the right side depicts the 2-scale system.}
    \label{fig:DJSfigure}
\end{figure}

Figure \ref{fig:DJSfigure} shows $C_{syn}$ results for the two HO systems under study.  The three plots in the middle represent the probability density functions at each scale.  For the 3-scale system, the oscillators at the middle scale receive the average of the two children oscillators below it, resulting in a triangle distribution as shown in the middle plot.  The highest scale oscillator receives the average of the two oscillators below it with triangle distributions, which is equivalent to the average of the four uniform distributions at the bottom scale.  The values for $C_{syn_{m,i}}$ are shown for each of the three scales.  $C_{syn}$ is obtained by summing the values.  Thus for the 3-scale system, $C_{syn} = (4 \times 1) + (2 \times 0.9271) + 0.7917 = 6.6459$, while for the 2-scale system $C_{syn} = (4 \times 1) + 0.7917 = 4.7917$. These values indicate that more information is contained in the 3-scale system compared to the 2-scale system, making 2-scale system a more efficient structure for achieving the same result.

Figure \ref{subfig:Xconcentrations} shows the $X$ concentrations for the four bottom scale oscillators for each system.  It is clear that the 2-scale system synchronizes much faster than the 3-scale system.

Figures \ref{subfig:semanticDelta} to \ref{subfig:pragmaticGoalValue} show the information measures for the two HO system plotted together for comparison.  As information is passed through the system in the form of $X$ concentrations, which are oscillatory in nature, most information measures themselves oscillate.  When viewed together, such information measures for each system have a small time offset due to oscillators in the 2-scale system being slightly out of phase with those in the 3-scale system.  In all the information measures, there is a distinct change in behavior when the systems transition from unsynchronized to synchronized oscillations.  During synchronization, all information measures (except efficiencies) are the same in both systems.

For the semantic delta, positive values indicate that the knowledge used by the oscillators to act is growing.  Larger magnitudes indicate faster knowledge change.  The 3-scale system exhibits a smaller $|\Delta_{sm}|$ than the 2-scale system during the pre-synchronization period.  This relationship indicates that the feedback in the 2-scale system has a larger impact on the knowledge used by the oscillators to adapt their $X$ concentrations.  Once synchronization is achieved, both systems have identical $\Delta_{sm}(t)$, meaning that both systems achieve the same impact on an oscillator's knowledge.

For the semantic truth value, positive values indicate that the average knowledge among the oscillators is getting closer to the ground truth.  As with the $\Delta_{sm}$, $|V_{sm,th}|$ is smaller for the 3-scale system during the pre-synchronization period leading to the conclusion that the 2-scale system achieves more impact in bringing an oscillators' knowledge closer to the ground truth.  Also, during synchronized operation, $V_{sm,th}$ for both is identical, but $E_{sm,th}$ emphasizes that the 2-scale achieves noticeably larger efficiency values (maximum of 0.03 vs. 0.02 for the 3-scale system) due to less information flowing in the system.

For the pragmatic delta, larger magnitudes indicate a larger change in the oscillator's adaptation.  The 2-scale system achieves more change in adaptation during the pre-synchronization period, while the two systems have identical $\Delta_{pr}$ once synchronized.  Since the distance from the goal is measured as the variance in the $X$ concentrations among oscillators, a positive $V_{pr,gl}$ indicates that the system is moving toward the goal of synchronization.  Larger positive magnitudes indicate that it is moving toward the goal at a faster rate.  In comparing the two systems, the 3-scale system moves faster toward the goal, but this is offset by faster movement away from the goal in nearly equal amounts due to larger time lags in each oscillators knowledge.  Thus the 3-scale system shows a greater degree of overshoot and undershoot on its way toward the goal, while the 2-scale system is more direct and faster in its approach toward synchronization.  As with $E_{sm,th}$, $E_{pr,gl}$ indicates the 2-scale system is more efficient in achieving synchronization.

\begin{subfigure}
    \centering
    \begin{minipage}[b]{0.3\textwidth}
        \includegraphics[width=1\textwidth]{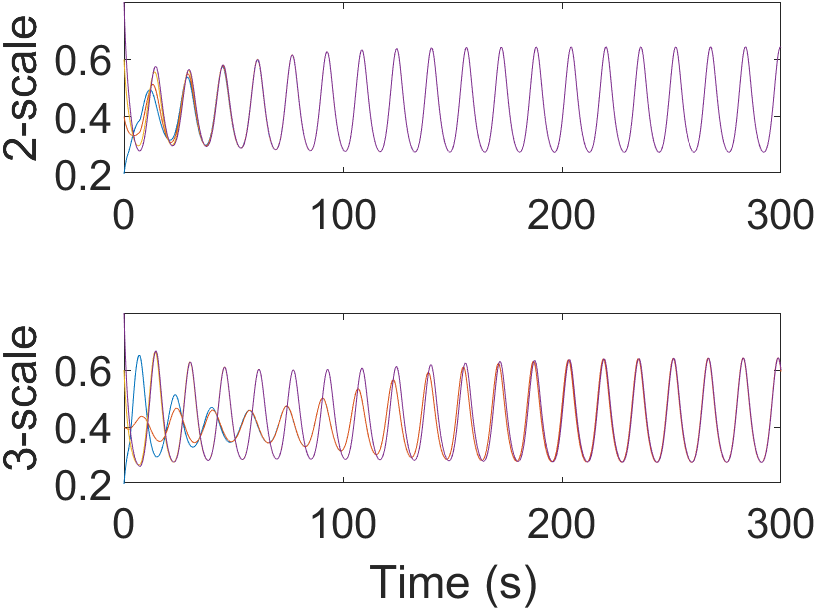}
        \caption{$X$ concentrations}
        \label{subfig:Xconcentrations}
    \end{minipage}
    \begin{minipage}[b]{0.3\textwidth}
        \includegraphics[width=1\textwidth]{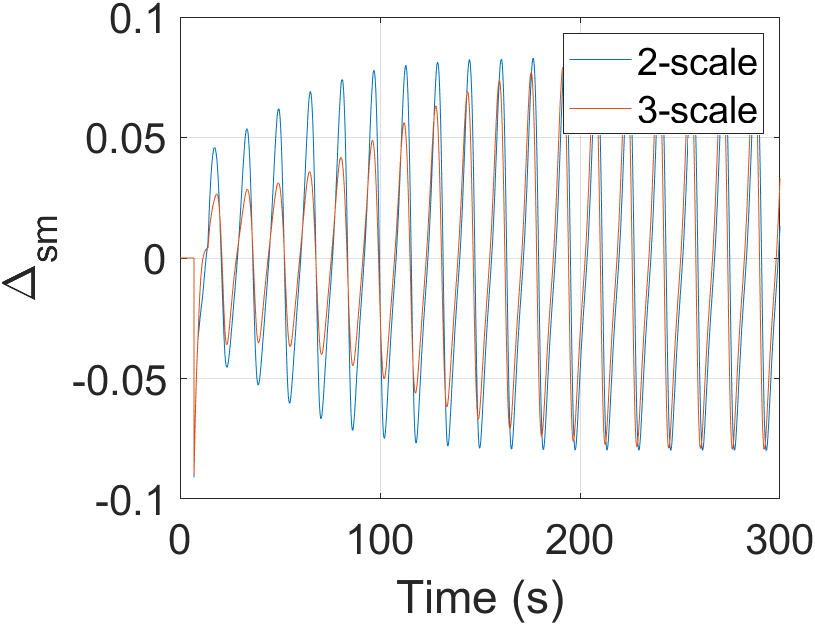}
        \caption{Semantic delta}
        \label{subfig:semanticDelta}
    \end{minipage}
    \begin{minipage}[b]{0.3\textwidth}
        \includegraphics[width=1\textwidth]{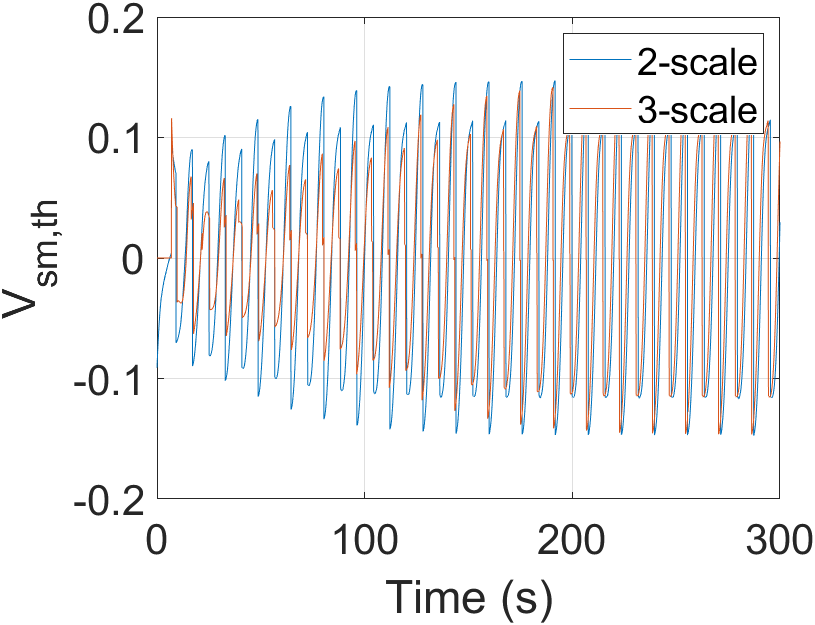}
        \caption{Semantic truth value}
        \label{subfig:semanticTruthValue}
    \end{minipage}
    \begin{minipage}[b]{0.3\textwidth}
        \includegraphics[width=1\textwidth]{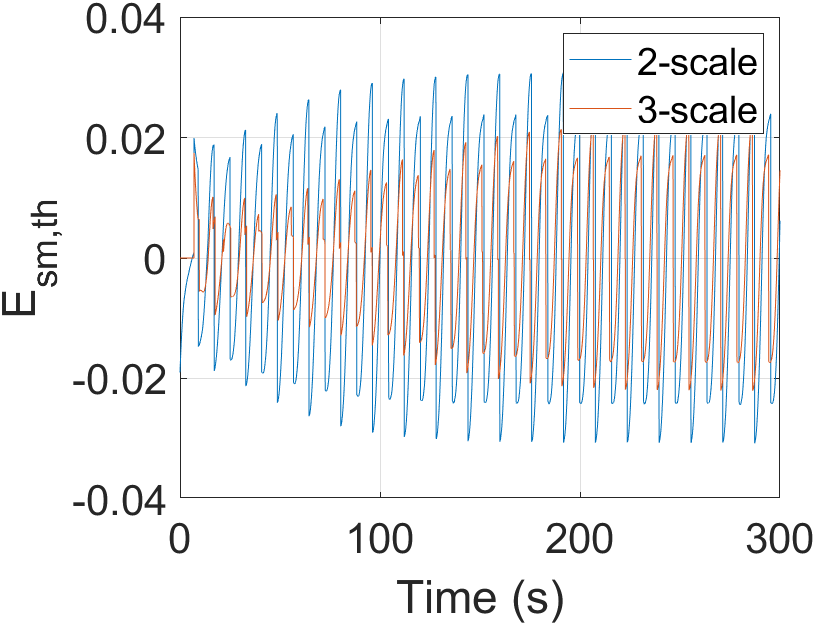}
        \caption{Semantic truth value efficiency}
        \label{subfig:semanticTruthValueEfficiency}
    \end{minipage}
   \begin{minipage}[b]{0.3\textwidth}
        \includegraphics[width=1\textwidth]{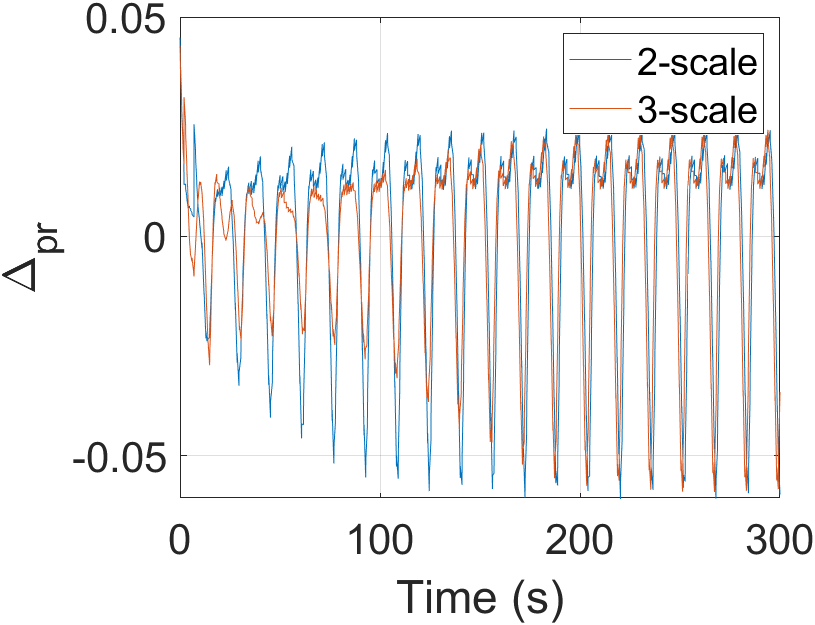}
        \caption{Pragmatic delta}
        \label{subfig:pragmaticActionDelta}
    \end{minipage}
    \begin{minipage}[b]{0.3\textwidth}
        \includegraphics[width=1\textwidth]{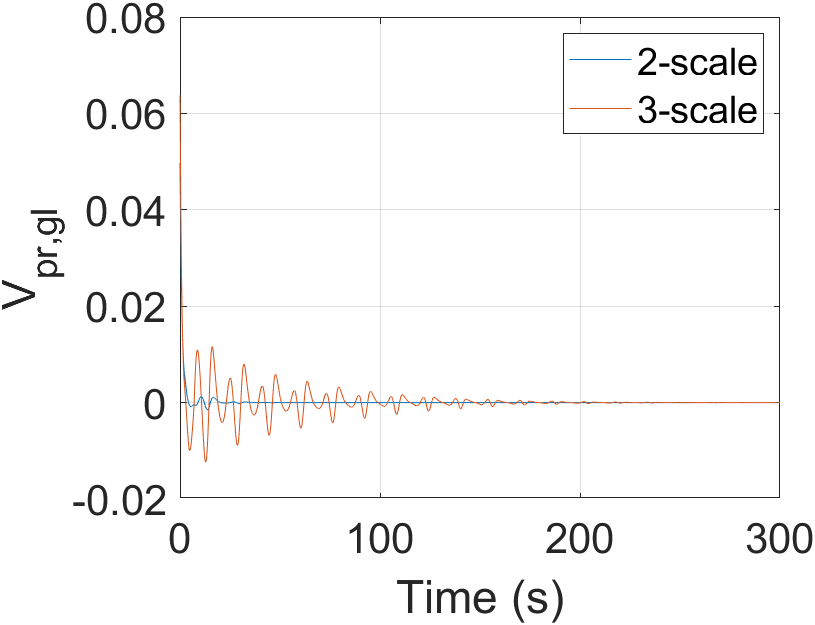}
        \caption{Pragmatic goal value}
        \label{subfig:pragmaticGoalValue}
    \end{minipage}

\setcounter{subfigure}{-1}
    \caption{Results for the two hierarchical oscillator systems.  The blue line shows the 2-scale system and the red line shows the 3-scale system.}
    \label{fig:HOresults}
\end{subfigure}


\section{Discussion \& Conclusions}
 
CAS typically consist of many components that interact with each other across at least two scales (\cite{ahl1996hierarchy} \cite{salthe1993development}), adapting through feedback loops to reach a shared goal \cite{flack2013timescales}. In such MSFS, there is always a loss of syntactic information when abstracting from the micro- to the macro-scale. Conversely, control information flowing from the macro- to the micro-scale can be either further abstracted (e.g., different macro-states may issue the same micro-controls); stay the same (e.g., through top-down forwarding); or gain information (e.g., through extra micro-scale context). 
Syntactic information measures, including Shannon entropy, can be used to quantify the size of these information flows, and the amount of information abstracted across scales (as shown in the hierarchical oscillator and task distribution case studies). As information becomes more symbolic, the relevance of syntactic measures decreases, calling for value-oriented measures.

We proposed a non-exhaustive set of information measures within the syntactic, semantic and pragmatic categories, and exemplified their usage through four cases of MSFS. Feedback is central to the behavior of adaptive information-processing systems, including single-scale ones. Whenever there is a feedback cycle there is a system that adapts, and an information flow controlling that system. The inherent delays and uncertainties of the adapting system and its environment play a key part in the efficacy and efficiency of the feedback cycle.  When the feedback spans across multiple scales, affecting components that function at different granularities of time and information, adjusting its operation to the multitude of delays and uncertainties becomes increasingly difficult. The presence of multiple feedback cycles that operate simultaneously between different scales exacerbates this difficulty. To deal with such increased complexity, parts of the system become quasi-independent \cite{simon2012architecture}, with minimal information flows coordinating them to reach a shared goal. Tracking information flows and their mutual influences throughout such MSFS becomes essential for understanding their behavior and/or adjusting their design.   
Using different types of information measures, which can be calculated for the whole system or at single scales, can help in understanding such increased complexity in MSFS. This requires identifying a goal.
Goals can be determined at different scales and can be exogenous or endogenous, depending on where the observer is placed. 
Observer identification is necessary for all types of information measures, including syntactic ones, while the goal choice is needed to measure informational values (semantic and pragmatic). 

Such information measures are interrelated and best understood when considered simultaneously. Disturbances on the information flow (measured syntactically) have an impact on knowledge (measured semantically) and on adaptation (measured pragmatically). In our examples, these disturbances were due to time delays or partial (and/or imperfect) information collection. \textbf{Time delays} can occur in information communication and processing or in system adaptation. Within feedback cycles, they can be tracked by comparing changes 
in the actual state measures, and in their semantic and pragmatic values, over time, at different scales. 
Depending on the system 
characteristics, delays can be beneficial in achieving the goal. This is the case when the delay in the feedback cycle matches a slow and inaccurate adaptation process. 

In the robotic collective, for example, adaptation 
delays caused by the robots' spatial dynamics compensated for inaccuracies in their collective state information influencing their movement. In the collective decision-making case, inaccurate information (i.e. an inaccurate collective opinion $O_{coll}$) leading to fast decisions performed better than accurate information that caused a longer feedback cycle. In the task distribution example, delays caused outdated state information that adaptation strategies had to compensate for. In \cite{Mellodge2021}, we showed how delays could lead to adaptation towards different attractors, or goals. This includes the hierarchical oscillator example, where delays are essential to system synchronization.

\textbf{Partial or imperfect information} leads to inaccurate collective knowledge. Studying the delta of truth $\Delta_{th}$ together with the semantic delta value $V_{sm,th}$ over time can show whether the system's knowledge acquisition accumulates inaccuracies, learns to diminish them, or maintains them within a constant range (this latter case applying to all examples shown here). Trade-offs in knowledge acquisition can also be introduced by the abstraction process. In the robotics case, although the macro-scale information is derived by abstracting meso-scale information, comparing measures across scales did not result in the same performance ranking across different strategies, showing how a strategy that improves knowledge precision at one scale may reduce precision at a higher scale. Knowledge inaccuracies propagate through multi-scale systems, impacting adaptation towards the goal. This is observed by considering the semantic truth value $V_{sm,th}$ together with the pragmatic goal value $V_{pr,gl}$. In the task distribution case, perfect knowledge (as implemented by the Model strategy) led to maximum efficacy in reaching the goal, even if the knowledge is outdated. In comparison, the BB strategy achieved the same $V_{pr,gl}$ with a lower accuracy of knowledge on the management scales, but over a longer period. On the short term, the random strategy (RS) used no collective knowledge and performed similarly to BB.
These insights can help tune the system configuration and strategy depending on the targeted timescale and goal -- e.g., RS may be efficient in highly fluctuating or unpredictable environments. The robotic collective example showed a different relation between semantic and pragmatic measures. Accurate knowledge hindered efficacy, due to adaptive overreactions and execution delays. The ground truth strategy gave robots perfect knowledge of their collective state, but their individual reactions led to large oscillations around the goal state when compared to the partial knowledge scenario. 

In the collective decision-making model, semantic and pragmatic measures tracked the impact of varying system parameters (the number of agents $N$ and of information sources $R$) on collective knowledge and adaptation. The $random_{CN}$ strategy led to a higher pragmatic goal value $V_{pr,gl}$ for low $R$, compared with the $consensus$ strategy, despite having a higher average delta of truth $\Delta_{th}$ -- showing how this performance is a result of a fast feedback cycle, rather than accurate knowledge. This leads to fast and tight oscillations and variations of $O_{coll}$, which could be beneficial or not depending on the chosen system goal. Tuning the $consensus$ strategy to environmental conditions, on the other hand, can lead to smoother parameter transitions, enhancing resilience and scalability. 
In the hierarchical oscillator model, oscillating semantic measures lead to oscillating pragmatic measures, both in the two-scale and three-scale designs. This highlights the underlying oscillations of adaptations, which are essential to maintaining the biochemical oscillators while synchronizing them. Since perfect information does not always lead to the best adaptation, the semantic goal value (not calculated in the examples) could be used to find the most effective and efficient information flows for a given system. Still, tailoring information flows to a specific configuration may hinder transferability, so the semantic truth value and semantic goal value are best considered simultaneously -- assuming that the ground truth can provide a common denominator across configuration and parameter ranges. 

These examples provide initial insights into how coupled information measures can be used to understand the behavior of CAS. This can help system analysts and designers study the value of information flows for system properties, such as reactivity, stability and scalability. More broadly, identifying how different patterns of information flows correlate with the behavior and evolution of CAS across scales can help characterize existing CAS through a unified informational lens \cite{krakauer2020information}, and understand fundamental regularities across such systems.

\section*{Conflict of Interest Statement}
The authors declare that the research was conducted in the absence of any commercial or financial relationships that could be construed as a potential conflict of interest.

\section*{Author Contributions}
LJDF coordinated the project. All authors contributed to: conceptualization, data curation, formal analysis, investigation, methodology, visualization, writing -- original draft, and writing -- review \& editing.

\section*{Funding}
LJDF acknowledges funding by the Margarita Salas program of the Spanish Ministry of Universities, funded by the European Union-NextGeneraionEU.


\section*{Supplemental Data}
 A separate Supplementary Material file is available.


\bibliographystyle{Frontiers-Harvard} 

\begin{thebibliography}{59}
\providecommand{\natexlab}[1]{#1}
\expandafter\ifx\csname urlstyle\endcsname\relax
  \providecommand{\doi}[1]{doi:\discretionary{}{}{}#1}\else
  \providecommand{\doi}{doi:\discretionary{}{}{}\begingroup
  \urlstyle{rm}\Url}\fi
\providecommand{\selectlanguage}[1]{\relax}
\providecommand{\bibAnnoteFile}[1]{%
  \IfFileExists{#1}{\begin{quotation}\noindent\textsc{Key:} #1\\
  \textsc{Annotation:}\ \input{#1}\end{quotation}}{}}
\providecommand{\bibAnnote}[2]{%
  \begin{quotation}\noindent\textsc{Key:} #1\\
  \textsc{Annotation:}\ #2\end{quotation}}

\bibitem[{Ahl and Allen(1996)}]{ahl1996hierarchy}
Ahl, V. and Allen, T.~F. (1996).
\newblock \emph{Hierarchy theory: a vision, vocabulary, and epistemology}
  (Columbia University Press)
\bibAnnoteFile{ahl1996hierarchy}

\bibitem[{Atmanspacher(1991)}]{atmanspacher1991information}
Atmanspacher, H. (1991).
\newblock \emph{Information Dynamics} (Plenum Press)
\bibAnnoteFile{atmanspacher1991information}

\bibitem[{Bellman et~al.(2021)Bellman, Diaconescu, and
  Tomforde}]{selfIntegrationMastering2021}
Bellman, K.~L., Diaconescu, A., and Tomforde, S. (2021).
\newblock Special issue on "self-improving self integration".
\newblock \emph{Future Gener. Comput. Syst.} 119, 136--139.
\newblock \doi{10.1016/J.FUTURE.2021.02.010}
\bibAnnoteFile{selfIntegrationMastering2021}

\bibitem[{Bellman and Goldberg(1984)}]{bellman-lang&mov1984}
Bellman, K.~L. and Goldberg, L.~J. (1984).
\newblock Common origin of linguistic and movement abilities.
\newblock \emph{American Journal of Physiology-Regulatory, Integrative and
  Comparative Physiology} 246, R915--R921.
\newblock \doi{10.1152/ajpregu.1984.246.6.R915}
\bibAnnoteFile{bellman-lang&mov1984}

\bibitem[{Brillouin(1953)}]{brillouin1953negentropy}
Brillouin, L. (1953).
\newblock The negentropy principle of information.
\newblock \emph{Journal of Applied Physics} 24, 1152--1163
\bibAnnoteFile{brillouin1953negentropy}

\bibitem[{Brillouin(1962)}]{brillouin1962science}
Brillouin, L. (1962).
\newblock Science and information theory
\bibAnnoteFile{brillouin1962science}

\bibitem[{Dessalles(2010)}]{jldBadLuckEmotion2010}
Dessalles, J.-L. (2010).
\newblock Emotion in good luck and bad luck: predictions from simplicity
  theory.
\newblock In \emph{Proceedings of the 32nd Annual Conference of the Cognitive
  Science Society}, eds. S.~Ohlsson and R.~Catrambone (Austin, TX: Cognitive
  Science Society), 1928--1933
\bibAnnoteFile{jldBadLuckEmotion2010}

\bibitem[{Dessalles(2013)}]{jldRelevance2013}
Dessalles, J.-L. (2013).
\newblock Algorithmic simplicity and relevance.
\newblock In \emph{Algorithmic probability and friends - LNAI 7070}, ed. D.~L.
  Dowe (Berlin, D: Springer Verlag), 119--130.
\newblock \doi{10.1007/978-3-642-44958-1_9}
\bibAnnoteFile{jldRelevance2013}

\bibitem[{Di~Felice and Zahadat(2022)}]{difelice2022agent}
Di~Felice, L.~J. and Zahadat, P. (2022).
\newblock An agent-based model of collective decision-making in correlated
  environments.
\newblock \emph{Proceedings of the 4th International Workshop on Agent-Based
  Modelling of Human Behaviour (ABMHuB’22)}
\bibAnnoteFile{difelice2022agent}

\bibitem[{Diaconescu et~al.(2019)Diaconescu, Di~Felice, and
  Mellodge}]{diaconescu2019multi}
Diaconescu, A., Di~Felice, L.~J., and Mellodge, P. (2019).
\newblock Multi-scale feedbacks for large-scale coordination in self-systems.
\newblock In \emph{2019 IEEE 13th International Conference on Self-Adaptive and
  Self-Organizing Systems (SASO)} (IEEE), 137--142
\bibAnnoteFile{diaconescu2019multi}

\bibitem[{Diaconescu et~al.(2021{\natexlab{a}})Diaconescu, Di~Felice, and
  Mellodge}]{diaconescu2021exogenous}
Diaconescu, A., Di~Felice, L.~J., and Mellodge, P. (2021{\natexlab{a}}).
\newblock Exogenous coordination in multi-scale systems: How information flows
  and timing affect system properties.
\newblock \emph{Future Generation Computer Systems} 114, 403--426
\bibAnnoteFile{diaconescu2021exogenous}

\bibitem[{Diaconescu et~al.(2021{\natexlab{b}})Diaconescu, Di~Felice, and
  Mellodge}]{diaconescu2021information}
Diaconescu, A., Di~Felice, L.~J., and Mellodge, P. (2021{\natexlab{b}}).
\newblock An information-oriented view of multi-scale systems.
\newblock In \emph{2021 IEEE International Conference on Autonomic Computing
  and Self-Organizing Systems Companion (ACSOS-C)} (IEEE), 154--159
\bibAnnoteFile{diaconescu2021information}

\bibitem[{Feistel and Ebeling(2016)}]{feistel2016entropy}
Feistel, R. and Ebeling, W. (2016).
\newblock Entropy and the self-organization of information and value.
\newblock \emph{Entropy} 18, 193
\bibAnnoteFile{feistel2016entropy}

\bibitem[{Fetzer(2004)}]{fetzer2004information}
Fetzer, J.~H. (2004).
\newblock Information: Does it have to be true?
\newblock \emph{Minds and Machines} 14, 223--229
\bibAnnoteFile{fetzer2004information}

\bibitem[{Flack(2021)}]{flack2021complexity}
Flack, J. (2021).
\newblock Complexity begets complexity.
\newblock In \emph{Artificial Life Conference Proceedings 33} (MIT Press One),
  vol. 2021, 7
\bibAnnoteFile{flack2021complexity}

\bibitem[{Flack(2017)}]{flack2017coarse}
Flack, J.~C. (2017).
\newblock Coarse-graining as a downward causation mechanism.
\newblock \emph{Philosophical Transactions of the Royal Society A:
  Mathematical, Physical and Engineering Sciences} 375, 20160338
\bibAnnoteFile{flack2017coarse}

\bibitem[{Flack et~al.(2013)Flack, Erwin, Elliot, and
  Krakauer}]{flack2013timescales}
Flack, J.~C., Erwin, D., Elliot, T., and Krakauer, D.~C. (2013).
\newblock Timescales, symmetry, and uncertainty reduction in the origins of
  hierarchy in biological systems.
\newblock \emph{Evolution cooperation and complexity} , 45--74
\bibAnnoteFile{flack2013timescales}

\bibitem[{Floridi(2005)}]{floridi2005semantic}
Floridi, L. (2005).
\newblock Is semantic information meaningful data?
\newblock \emph{Philosophy and phenomenological research} 70, 351--370
\bibAnnoteFile{floridi2005semantic}

\bibitem[{Floridi(2008)}]{floridi2008defence}
Floridi, L. (2008).
\newblock A defence of informational structural realism.
\newblock \emph{Synthese} 161, 219--253
\bibAnnoteFile{floridi2008defence}

\bibitem[{Frank(2003)}]{frank2003pragmatic}
Frank, A.~U. (2003).
\newblock Pragmatic information content—how to measure the information in a
  route.
\newblock \emph{Foundations of geographic information science} , 47
\bibAnnoteFile{frank2003pragmatic}

\bibitem[{Gernert(2006)}]{gernert2006pragmatic}
Gernert, D. (2006).
\newblock Pragmatic information: Historical exposition and general overview.
\newblock \emph{Mind and Matter} 4, 141--167
\bibAnnoteFile{gernert2006pragmatic}

\bibitem[{Gould(1974)}]{gould1974risk}
Gould, J.~P. (1974).
\newblock Risk, stochastic preference, and the value of information.
\newblock \emph{Journal of Economic Theory} 8, 64--84
\bibAnnoteFile{gould1974risk}

\bibitem[{Grunwald and Vitanyi(2008)}]{AIT_Vitanyi2008}
Grunwald, P.~D. and Vitanyi, P.~M. (2008).
\newblock Algorithmic information theory.
\newblock In \emph{Philosophy of Information} (Elsevier)
\bibAnnoteFile{AIT_Vitanyi2008}

\bibitem[{Haken and Portugali(2016)}]{haken2016information}
Haken, H. and Portugali, J. (2016).
\newblock Information and self-organization.
\newblock \emph{Entropy} 19, 18
\bibAnnoteFile{haken2016information}

\bibitem[{Jablonka(2002)}]{jablonka2002information}
Jablonka, E. (2002).
\newblock Information: Its interpretation, its inheritance, and its sharing.
\newblock \emph{Philosophy of science} 69, 578--605
\bibAnnoteFile{jablonka2002information}

\bibitem[{Kephart and Chess(2003)}]{kephartAC2003}
Kephart, J. and Chess, D. (2003).
\newblock The vision of autonomic computing.
\newblock \emph{Computer} 36, 41--50.
\newblock \doi{10.1109/MC.2003.1160055}
\bibAnnoteFile{kephartAC2003}

\bibitem[{Kim et~al.(2010)Kim, Shin, Jung, Heslop-Harrison, and Cho}]{Kim2010}
Kim, J.-R., Shin, D., Jung, S.~H., Heslop-Harrison, P., and Cho, K.-H. (2010).
\newblock A design principle underlying the synchronization of oscillations in
  cellular systems.
\newblock \emph{Journal of Cell Science} 123, 537--543.
\newblock \doi{10.1242/jcs.060061}
\bibAnnoteFile{Kim2010}

\bibitem[{Kolchinsky and Wolpert(2018)}]{kolchinsky2018semantic}
Kolchinsky, A. and Wolpert, D.~H. (2018).
\newblock Semantic information, autonomous agency and non-equilibrium
  statistical physics.
\newblock \emph{Interface focus} 8, 20180041
\bibAnnoteFile{kolchinsky2018semantic}

\bibitem[{Kounev et~al.(2017)Kounev, Lewis, Bellman, Bencomo, C{\'{a}}mara,
  Diaconescu et~al.}]{selfAwareNotion2017}
Kounev, S., Lewis, P.~R., Bellman, K.~L., Bencomo, N., C{\'{a}}mara, J.,
  Diaconescu, A., et~al. (2017).
\newblock The notion of self-aware computing.
\newblock In \emph{Self-Aware Computing Systems}, eds. S.~Kounev, J.~O.
  Kephart, A.~Milenkoski, and X.~Zhu (Springer International Publishing).
  3--16.
\newblock \doi{10.1007/978-3-319-47474-8\_1}
\bibAnnoteFile{selfAwareNotion2017}

\bibitem[{Krakauer et~al.(2020)Krakauer, Bertschinger, Olbrich, Flack, and
  Ay}]{krakauer2020information}
Krakauer, D., Bertschinger, N., Olbrich, E., Flack, J.~C., and Ay, N. (2020).
\newblock The information theory of individuality.
\newblock \emph{Theory in Biosciences} 139, 209--223
\bibAnnoteFile{krakauer2020information}

\bibitem[{Kullback and Leibler(1951)}]{kullback1951information}
Kullback, S. and Leibler, R.~A. (1951).
\newblock On information and sufficiency.
\newblock \emph{The annals of mathematical statistics} 22, 79--86
\bibAnnoteFile{kullback1951information}

\bibitem[{Lalanda et~al.(2013{\natexlab{a}})Lalanda, McCann, and
  Diaconescu}]{lalanda2013autonomic}
Lalanda, P., McCann, J., and Diaconescu, A. (2013{\natexlab{a}}).
\newblock \emph{Autonomic Computing: Principles, Design and Implementation}.
\newblock Undergraduate Topics in Computer Science (Springer London)
\bibAnnoteFile{lalanda2013autonomic}

\bibitem[{Lalanda et~al.(2013{\natexlab{b}})Lalanda, McCann, and
  Diaconescu}]{autonomicLalanda2013}
Lalanda, P., McCann, J.~A., and Diaconescu, A. (2013{\natexlab{b}}).
\newblock \emph{Autonomic Computing - Principles, Design and Implementation}.
\newblock Undergraduate Topics in Computer Science (Springer).
\newblock \doi{10.1007/978-1-4471-5007-7}
\bibAnnoteFile{autonomicLalanda2013}

\bibitem[{Lewis(1930)}]{lewis1930symmetry}
Lewis, G.~N. (1930).
\newblock The symmetry of time in physics.
\newblock \emph{Science} 71, 569--577
\bibAnnoteFile{lewis1930symmetry}

\bibitem[{Lin(1991)}]{Lin1991}
Lin, J. (1991).
\newblock Divergence measures based on the shannon entropy.
\newblock \emph{IEEE Transactions on Information Theory} 37, 145--151.
\newblock \doi{10.1109/18.61115}
\bibAnnoteFile{Lin1991}

\bibitem[{Mellodge et~al.(2021)Mellodge, Diaconescu, and
  Di~Felice}]{Mellodge2021}
Mellodge, P., Diaconescu, A., and Di~Felice, L.~J. (2021).
\newblock { Timing configurations affect the macro-properties of multi-scale
  feedback systems }.
\newblock In \emph{2021 IEEE International Conference on Autonomic Computing
  and Self-Organizing Systems (ACSOS)} (Los Alamitos, CA, USA: IEEE Computer
  Society), 100--109.
\newblock \doi{10.1109/ACSOS52086.2021.00032}
\bibAnnoteFile{Mellodge2021}

\bibitem[{Ming~Li(2019)}]{kolmogorovIntro2019}
Ming~Li, P.~V. (2019).
\newblock \emph{An Introduction to Kolmogorov Complexity and Its Applications}.
\newblock Texts in Computer Science (Springer Cham).
\newblock \doi{https://doi.org/10.1007/978-3-030-11298-1}
\bibAnnoteFile{kolmogorovIntro2019}

\bibitem[{Morris(1938)}]{morris1938foundations}
Morris, C. (1938).
\newblock Foundations of the theory of signs.
\newblock \emph{Foundations of the Theory of Science} 1
\bibAnnoteFile{morris1938foundations}

\bibitem[{M{\"u}ller-Schloer et~al.(2011)M{\"u}ller-Schloer, Schmeck, and
  Ungerer}]{müller2011organic}
M{\"u}ller-Schloer, C., Schmeck, H., and Ungerer, T. (2011).
\newblock \emph{Organic Computing — A Paradigm Shift for Complex Systems}.
\newblock Autonomic Systems (Springer Basel)
\bibAnnoteFile{müller2011organic}

\bibitem[{M{\"u}ller-Schloer and Tomforde(2017)}]{OrganicCMllerSchloer2017}
M{\"u}ller-Schloer, C. and Tomforde, S. (2017).
\newblock Organic computing – technical systems for survival in the real
  world.
\newblock In \emph{Autonomic Systems}
\bibAnnoteFile{OrganicCMllerSchloer2017}

\bibitem[{Nehaniv(1999)}]{nehaniv1999meaning}
Nehaniv, C.~L. (1999).
\newblock Meaning for observers and agents.
\newblock In \emph{Proceedings of the 1999 IEEE International Symposium on
  Intelligent Control Intelligent Systems and Semiotics (Cat. No. 99CH37014)}
  (IEEE), 435--440
\bibAnnoteFile{nehaniv1999meaning}

\bibitem[{Nielsen(2019)}]{Nielsen2019}
Nielsen, F. (2019).
\newblock On the jensen–shannon symmetrization of distances relying on
  abstract means.
\newblock \emph{Entropy} 21
\bibAnnoteFile{Nielsen2019}

\bibitem[{Pattee(1973)}]{Pattee1973}
Pattee, H.~H. (1973).
\newblock \emph{Hierarchy theory; the challenge of complex systems} (New York:
  G. Braziller)
\bibAnnoteFile{Pattee1973}

\bibitem[{Roederer et~al.(2005)}]{roederer2005information}
Roederer, J.~G. et~al. (2005).
\newblock \emph{Information and its Role in Nature} (Springer)
\bibAnnoteFile{roederer2005information}

\bibitem[{Salthe(1993)}]{salthe1993development}
Salthe, S.~N. (1993).
\newblock \emph{Development and evolution: complexity and change in biology}
  (Mit Press)
\bibAnnoteFile{salthe1993development}

\bibitem[{Shannon(1948)}]{shannon1948mathematical}
Shannon, C.~E. (1948).
\newblock A mathematical theory of communication.
\newblock \emph{The Bell system technical journal} 27, 379--423
\bibAnnoteFile{shannon1948mathematical}

\bibitem[{Simon(1991)}]{Simon1991}
Simon, H.~A. (1991).
\newblock \emph{The Architecture of Complexity} (Boston, MA: Springer US).
\newblock \doi{10.1007/978-1-4899-0718-9_31}
\bibAnnoteFile{Simon1991}

\bibitem[{Simon(2012)}]{simon2012architecture}
Simon, H.~A. (2012).
\newblock The architecture of complexity.
\newblock In \emph{The Roots of Logistics} (Springer). 335--361
\bibAnnoteFile{simon2012architecture}

\bibitem[{Sowinski et~al.(2023)Sowinski, Carroll-Nellenback, Markwick,
  Pi{\~n}ero, Gleiser, Kolchinsky et~al.}]{sowinski2023semantic}
Sowinski, D.~R., Carroll-Nellenback, J., Markwick, R.~N., Pi{\~n}ero, J.,
  Gleiser, M., Kolchinsky, A., et~al. (2023).
\newblock Semantic information in a model of resource gathering agents.
\newblock \emph{arXiv preprint arXiv:2304.03286}
\bibAnnoteFile{sowinski2023semantic}

\bibitem[{Theraulaz et~al.(1998)Theraulaz, Bonabeau, and
  Deneubourg}]{Theraulaz98}
Theraulaz, G., Bonabeau, E., and Deneubourg, J.-L. (1998).
\newblock Response threshold reinforcement and division of labour in insect
  societies.
\newblock In \emph{In Proc. Royal Society of London}. vol. 265 of \emph{5},
  327-- 332
\bibAnnoteFile{Theraulaz98}

\bibitem[{Timpson(2013)}]{timpson2013quantum}
Timpson, C.~G. (2013).
\newblock \emph{Quantum information theory and the foundations of quantum
  mechanics} (OUP Oxford)
\bibAnnoteFile{timpson2013quantum}

\bibitem[{Tribus and McIrvine(1971)}]{tribus1971energy}
Tribus, M. and McIrvine, E.~C. (1971).
\newblock Energy and information.
\newblock \emph{Scientific American} 225, 179--190
\bibAnnoteFile{tribus1971energy}

\bibitem[{Von~Uexk{\"u}ll(2013)}]{von2013foray}
Von~Uexk{\"u}ll, J. (2013).
\newblock \emph{A foray into the worlds of animals and humans: With a theory of
  meaning}, vol.~12 (U of Minnesota Press)
\bibAnnoteFile{von2013foray}

\bibitem[{Walker(2014)}]{walker2014top}
Walker, S.~I. (2014).
\newblock Top-down causation and the rise of information in the emergence of
  life.
\newblock \emph{Information} 5, 424--439
\bibAnnoteFile{walker2014top}

\bibitem[{Weinberger(2002)}]{weinberger2002theory}
Weinberger, E.~D. (2002).
\newblock A theory of pragmatic information and its application to the
  quasi-species model of biological evolution.
\newblock \emph{Biosystems} 66, 105--119
\bibAnnoteFile{weinberger2002theory}

\bibitem[{Weizs{\"a}cker and
  Weizs{\"a}cker(1972)}]{weizsacker1972wiederaufnahme}
Weizs{\"a}cker, E. U.~v. and Weizs{\"a}cker, C.~v. (1972).
\newblock Wiederaufnahme der begrifflichen frage: Was ist information.
\newblock \emph{Nova Acta Leopoldina} 37, 535--555
\bibAnnoteFile{weizsacker1972wiederaufnahme}

\bibitem[{Weyns(2021)}]{sasWeynsBook2021}
Weyns, D. (2021).
\newblock \emph{Basic Principles of Self-Adaptation and Conceptual Model} (John
  Wiley \& Sons, Ltd), chap.~1.
\newblock 1--15.
\newblock \doi{https://doi.org/10.1002/9781119574910.ch1}
\bibAnnoteFile{sasWeynsBook2021}

\bibitem[{Wong et~al.(2022)Wong, Wagner, and Treude}]{SASreview2022}
Wong, T., Wagner, M., and Treude, C. (2022).
\newblock Self-adaptive systems: A systematic literature review across
  categories and domains.
\newblock \emph{Information and Software Technology} 148, 106934.
\newblock \doi{https://doi.org/10.1016/j.infsof.2022.106934}
\bibAnnoteFile{SASreview2022}

\bibitem[{Zahadat(2023)}]{Zahadat2023ACSOS}
Zahadat, P. (2023).
\newblock Local estimation vs global information: the benefits of slower
  timescales.
\newblock \emph{4th IEEE International Conference on Autonomic Computing and
  Self-Organizing Systems (ACSOS)}
\bibAnnoteFile{Zahadat2023ACSOS}

\end{thebibliography}

\end{document}